\documentclass[apj,twocolumn, floatfix]{aastex63}
\usepackage{amsmath}
\usepackage{longtable}

\newdimen\figwd
\newcount\figno
\newcount\figcol
\newcount\figrow
\newcount\maxrows
\def\fighgap{10pt}
\def\figvgap{2pt}
\def\gridoffigs#1,#2\relax{%
\def\figtemp{#2}%
\advance\figcol1 \includegraphics[width=\figwd]{#1}%
\ifnum\figcol=\figno 
   \ifnum\figrow=\maxrows
   \gdef\savedfigs{#2}%
   \def\figtemp{}%
   \else
   \\[\figvgap]%
   \advance\figrow1 \figcol=0
   \fi
\else
\hskip\fighgap
\fi
\if\relax\figtemp\relax \hfill \else
\expandafter\expandafter\expandafter\gridoffigs \expandafter\figtemp\expandafter\relax
\fi
}
\newcommand{\figgrid}[2][5]{\global\figno=#1%
\figrow=1
\setlength{\figwd}{\hsize}%
\figcol=1
\loop
\advance\figcol1
\advance\figwd-\fighgap
\ifnum\figcol<\figno \repeat
\figcol=0
\divide\figwd\figno
\gridoffigs#2,\relax
}
\def\stripcomma#1,!{\def\fixedfigs{#1}}%
\newcommand{\contfiggrid}{%
\addtocounter{figure}{-1}%
\def\thefigure{\arabic{figure}, continued}%
\expandafter\stripcomma\savedfigs!
\expandafter\expandafter\expandafter\figgrid\expandafter[\expandafter\the\expandafter\figno\expandafter]\expandafter{\fixedfigs}%
}

\shorttitle{Realistic Ages in the bulge}
\shortauthors{Joyce et al.}

\graphicspath{{./}{figures/}}

\begin{document}

\title{The Ages of Galactic bulge Stars with Realistic Uncertainties}

\correspondingauthor{Meridith Joyce}
\email{meridith.joyce@csfk.org}

\author[0000-0002-8717-127X]{Meridith Joyce}
\affiliation{Space Telescope Science Institute \\
3700 San Martin Drive \\
Baltimore, MD 21218, USA}
\altaffiliation{Lasker Fellow}
\affiliation{Kavli Institute for Theoretical Physics, University of Santa Barbara, California, 93106, USA}
\affiliation{Konkoly Observatory, Research Centre for Astronomy and Earth Sciences, H-1121 Budapest Konkoly Th. M. \'ut 15-17., Hungary}
\affiliation{CSFK, MTA Centre of Excellence, Budapest, Konkoly Thege Mikl\'os \'ut 15-17., H-1121, Hungary}

\author[0000-0002-8878-3315]{Christian I. Johnson}
\affiliation{Space Telescope Science Institute \\
3700 San Martin Drive \\
Baltimore, MD 21218, USA}

\author[0000-0003-0064-0692]{Tommaso Marchetti}
\affiliation{European Southern Observatory\\
Karl-Schwarzschild-Strasse 2\\
85748 Garching bei München, Germany}

\author[0000-0003-0427-8387]{R. Michael Rich}
\affiliation{Department of Physics and Astronomy, UCLA\\
430 Portola Plaza, Box 951547\\
Los Angeles, CA 90095-1547, USA}9

\author[0000-0001-8889-0762]{Iulia Simion}
\affiliation{Shanghai Key Lab for Astrophysics, Shanghai Normal University, 100 Guilin Road, Shanghai, 200234, China}

\author[0000-0002-4958-5404]{John Bourke}
\affiliation{Mathematical Sciences Publishers, 798 Evans Hall No.\ 3840, Berkeley, CA 94720, USA}

\begin{abstract}
Using modern isochrones with customized physics and carefully considered statistical techniques, we recompute the age distribution for a sample of 91 micro-lensed dwarfs in the Galactic bulge presented by \citet{Bensby17} and do not produce an age distribution consistent with their results. In particular, our analysis finds that only 15 of 91 stars have ages younger than 7 Gyr, compared to their finding of 42 young stars in the same sample. While we do not find a constituency of very young stars, our results do suggest the presence of an $\sim8$ Gyr population at the highest metallicities, thus contributing to long-standing debate about the age--metallicity distribution of the Galactic bulge. We supplement this with attempts at independent age determinations from two sources of photometry, BDBS and \textit{Gaia}, but find that the imprecision of photometric measurements prevents reliable age and age uncertainty determinations. 
Lastly, we present age uncertainties derived using a first-order consideration of global modeling uncertainties in addition to standard observational uncertainties. The theoretical uncertainties are based on the known variance of free parameters in the 1D stellar evolution models used to generate isochrones, and when included, result in age uncertainties of $2$--$5$ Gyr for this spectroscopically well-constrained sample. These error bars, which are roughly twice as large as typical literature values, constitute realistic lower limits on the true age uncertainties. 
\end{abstract}


\section{Introduction} \label{sec:intro}
The star formation history of the Galactic bulge remains a challenge, with age constraints on significant mass components of the population still ranging from 1 to $>10$ Gyr. The underlying cause of this controversy is that a range in stellar metallicity, the large intrinsic depth of the bulge, 
severe crowding, high reddening, spatial distribution, and the intervening disk population all conspire to make it extremely difficult to constrain the age and age--metallicity relationship for the bulge. Even the advent of \textit{Gaia} parallaxes has not made a significant impact: while \textit{Gaia} data can remove stars in the foreground, the parallaxes are not sufficiently precise to give good age constraints on stars at the bulge distance of $\sim 8$ kpc.
 
Unlike the case for globular clusters, the bulge also has significant physical depth, as much of the mass is in a bar with major axis $\sim 20-30^\circ$ from the line of sight toward $(l,b)=0,0$ \citep[e.g.,][]{Stanek97,Wegg13,Simion21}. While a detailed discussion of the history of the age controversy for the bulge may be found in various review articles, e.g.\ \citet{Rich13} and \citet{Barbuy18}, we offer a brief overview here.

The star formation history of the bulge remains an important problem even in the era of high redshift direct observations of galaxy formation and evolution.  No high redshift galaxy's evolution can be traced to the present day, but the Milky Way preserves all the stars that ever formed in our Galaxy.  Metallicities derived from deep photometry  \citep{Johnson20} and spectroscopy (e.g.\ \citealt{Zoccali08,Johnson13_offaxis,Ness13_mdf}) show a strong tendency for the metal-rich stars to be concentrated toward the plane. Hence, if there is any age--metallicity relationship, there is an implied age gradient with presumably younger, more metal-rich stars concentrated toward the plane. Further, for ages $\gtrsim 8$ Gyr, the differences in main sequence turnoff (MSTO) morphology are sufficiently subtle that the spatial depth, reddening, metallicity spread, and contamination by the foreground population blur color--magnitude diagrams. Yet, this age regime likely traces the processes that led to the formation of the bar and might therefore reveal the progression of the formation of the bulge. Given the enhanced $\alpha$-elements measured for bulge giants \citep[e.g.,][]{Gonzalez11,Johnson14,Bensby17,Rojas20} and the rapid---if not explosive---star formation history this implies, it is possible that the strong gradient of abundance with latitude lacks a corresponding age gradient.
It is just as possible that there is an age gradient, if the bulge/bar formed more slowly, or if other dynamical processes contributed to its formation. A fundamental tension is that age determinations based on photometry nearly always find a uniformly old bulge, while those based on spectroscopy find evidence of a broad age dispersion \citep{Clarkson08,Bensby17,Renzini18}. However, current ages are simply too coarse to tell definitively.

Despite 30 years of HST observations, the age and star formation history of the bulge remain open problems. For the reasons stated above, and due also to significant trends in abundance with Galactic latitude, HST photometry at the main sequence turnoff still fails to give the kind of age precision that characterizes, e.g., the local disk, or globular clusters. Overall, there is currently a debate between those who analyse color--magnitude diagrams of the bulge (often foreground-subtracted or proper motion-cleaned, e.g.\ \citealt{Clarkson08,Renzini18}) giving a $\sim 10$ Gyr age, and analyses of individual micro-lensed dwarfs, which argue for a metal rich component of the bulge as young as 1--5 Gyr \citep{Bensby17}. As stellar evolution models have advanced and observations improve, we now revisit the ages computed for micro-lensed dwarfs to explore whether this schism in age/star formation history derived using these different methods remains. 


The deep imaging of the Hubble space telescope offered compelling evidence of a predominantly old stellar population. 
\citet{Ortolani95} compared the luminosity function of stars in Baade’s window against those of the bulge globular clusters NGC 6528 and NGC 6553 and found that the bulge was compatible in age with these clusters while possessing an age spread no larger than about 10$\%$.  Further results by \citet{Zoccali03}, \citet{Valenti13}, and more recently by \citet{Renzini18} with HST treasury data, seemed to confirm that the bulge is largely composed of old stars with little evidence supporting a young component.

As was first shown by \citet{Kuijken02}, one may use the $l-$ proper motion to veto members of the foreground disk. This method was elegantly applied in \citet{Clarkson08,Clarkson11} to argue that the bulge is a stellar population dominated by a roughly Solar metallicity, $>10$ Gyr component. This analysis of color--magnitude diagrams would appear to settle the matter of the age, but that is not the case.

A series of papers has advanced the case for a young/intermediate age population in the bulge based on analysis of dwarf spectra serendipitously experiencing micro-lensing \citep{bensby11,bensby13,bensby14,Bensby17}. Such events can brighten a source by upwards of an order of magnitude, making possible high-resolution spectroscopy of dwarfs otherwise too faint to observe. As summarized in \citet{Bensby17}, the derived [Fe/H], $\log g$, and $T_\text{eff}$ result in strikingly young ages---around $1$--$5$ Gyr---especially for metal-rich stars. Their work finds $\sim 70\%$ of the stars with [Fe/H]$>0$ are $<8$ Gyr. The broad implication of these age estimates is that much of the population that constitute the bar, with $-0.5 < \rm{[Fe/H]} < +0.5$, is younger than 8 Gyr, and that non-negligible portions are younger than 5 Gyr. As these ages are in direct conflict with the results from main sequence turnoff studies, it is of the utmost importance to carefully evaluate these conclusions using the most current isochrones, the most appropriate statistical methods, and  
meticulous consideration of all sources of uncertainty (and, when available, actual photometry).

In this study, we perform an independent re-analysis of the spectroscopic parameters presented in \citet{Bensby17} using modern, customized stellar models and a combination of statistical techniques including maximum likelihood methods, $\chi^2$ frequentist approaches, and Monte Carlo simulations. For the bulk of our analysis, we use modern isochrones taken directly from the publicly available MIST database (MESA Isochrones and Stellar Tracks; \citealt{Choi2016}). We supplement this with explorations using isochrones computed with the Yale Rotating Stellar Evolution Code  (YREC; \citealt{YY}), to more closely mimic \citet{Bensby17}'s method, and custom isochrones built with the Modules for Experiments in Stellar Astrophysics (MESA) 1D stellar evolution program. In all cases, our analysis of the spectroscopic parameters produces an age--metallicity distribution consistent with the presence of an age gradient in the bulge, but notably inconsistent with the age results of \citet{Bensby17}.   

As an independent check, we also attempt to perform age derivations from photometric sub-samples of the same population using the Blanco DECam Bulge Survey \citep[BDBS;][]{Rich20,Johnson20} and \textit{Gaia} EDR3 \citep{GaiaEDR3}. While we can produce an age--metallicity distribution visually consistent with the absence of an age gradient---and almost totally bereft of young stars---in both cases, our attempts to compute observational age uncertainties reveal that the photometric parameters are simply too imprecise to provide meaningful estimates of either the ages or their uncertainties. Nonetheless, 
we include detailed photometric analysis (see Appendix \ref{sec:photometric_fits}) with the hope that it serves as both a cautionary tale and useful guide to others performing photometric age determinations.

We discuss the details of our age determination algorithm at length, and while our explicitness in this regard is not typical of similar analyses, our transparency serves to underscore the dangers of ``black box'' modeling results. Though machine learning and similar techniques are undoubtedly powerful, over-interpretation of results is more common when the internal workings of the algorithm or method are obscured from the user.

We conclude our analysis by performing age determinations that account for observational as well as theoretical (i.e.\ modeling) sources of uncertainty, where first-order estimates of the theoretical uncertainty contributions are determined by quantifying the effects of physically justified variations in free parameter choices. While studying the impact of, for example, variations in degrees of $\alpha$-element enhancement in stellar models is common in more careful analyses, this is the first time that variations in the convective mixing length, $\alpha_{\text{MLT}}$, have been propagated into MIST isochrones. The age uncertainties derived using this $\alpha_{\text{MLT}}$-adaptive grid of isochrones are state-of-the-art and constitute a significant advancement, in terms of realism and accuracy, in the arena of global age uncertainty estimates. 

This paper is organized as follows: 
In Section \ref{sec:micro-lensed_sample}, we discuss the sample of the micro-lensed dwarfs and their potential to resolve bulge age tensions. 
In Section \ref{sec:isochrones}, we discuss our choice of isochrones, the key differences between MIST and YREC, and the effects of $\alpha$-element enhancement. 
In Section \ref{sec:statistical_analysis}, we detail our age determination algorithm and present the resulting age and age--metallicity distributions for the spectroscopic data, supplemented by age uncertainties discussed in Section~\ref{sec:error_analysis}.
In Section \ref{sec:comparison_of_age_determinations}, we discuss the spectroscopic results in context, and in 
Section \ref{sec:modeling_uncertainty}, we present realistic age error estimates based on modeling uncertainties. We compare our age--metallicity distribution to HST results in Section \ref{sec:hst_comparison}.
We summarize our results and discuss their implications in Sections \ref{sec:implications} and \ref{sec:summary}.
For completeness, we apply our age determination technique to BDBS and \textit{Gaia} photometry in Appendix \ref{sec:photometric_fits}, taking care to discuss the shortcomings of these results and the reasons why these, and all, photometric age determinations should be approached with caution. 

\section{Ages of the micro-lensed sample}
\label{sec:micro-lensed_sample}

The micro-lensed sample described in detail by \citet{Cohen08,Cohen10} and \citet{Bensby10,bensby11,bensby13,bensby14,Bensby17} is a rare and powerful tool. Micro-lensing produces brightness magnification of 10 to 1000x, enabling the inference of spectroscopic parameters for sub-dwarf stars that would normally be much too faint for most instruments. In micro-lensing conditions, it is possible to perform high-resolution spectroscopy with 8m-class telescopes on these targets, thus providing measurements of their physical parameters---effective temperate and surface gravity---directly and independently of distance. A sample of micro-lensed bulge stars, then, has the potential to provide robust fundamental parameters for stars in this region. Given the implications the bulge age distribution carries for its star formation history, the origin of the bulge, and the formation history of the Galaxy itself, stellar ages are of particular importance. 

Recently, age determinations for the sample of 91 micro-lensed stars in the bulge were conducted by \citet{Bensby17}. Their analysis concluded that the Galactic bulge has complex age and abundance properties; in particular, that the peaks in the bulge metallicity distribution, star formation episodes, and the abundance trends suggest the Galactic bulge/bar may be an inner extension of the Galactic thin/thick disk. They also found that the star formation rate appears to have been faster in the bulge than in the local thick disk, which indicates a denser stellar environment closer to the Galactic center. 

In sum, their results support the idea of a secular origin for the Galactic bulge, suggesting the bulge was built up slowly from existing stellar populations in the central region of the Galaxy. The primary piece of evidence for these assertions is \citet{Bensby17}'s finding of a large numbers of young stars, which implies prolonged star formation in the region. 

The bulge, however, has long been understood to be old, and an overabundance of young stars in this region calls into question our understanding of the formation history of the Galaxy and of galaxy evolution mechanisms more generally. An old bulge implies that the bulge assembled first, which is consistent with several chemical enrichment model fits to the observed [Fe/H] and [alpha/Fe] distributions \citep[e.g.,][]{Ballero07,Grieco12,Matteucci19}.
Further, our knowledge of the chemical distribution of the Galaxy is not consistent with recent star formation episodes in this region, which is in conflict with the fact that younger stars must have been formed more recently. Alternatively, if the young stars did not form there, we must then explain how younger stars moved into the region.

There is a known tension in the literature between age determinations from photometry, which produce a uniformly old bulge \citep[e.g.,][]{Ortolani95,Zoccali03,Clarkson08,Valenti13,Renzini18, Bovy19, Hasselquist20}, and age determinations from micro-lensing, which produce a very broad age distribution \citep[e.g.,][]{bensby13,Bensby17}.
However, age determinations from stellar models are difficult and sometimes unreliable. Often, they are plagued by inappropriate statistical techniques, out-of-date physical assumptions in the models, and inadequate treatment of uncertainties in both observations and models. The first step towards resolving this tension, then, is to examine rigorously the age determination process itself. 

\section{Choice of Isochrones}
\label{sec:isochrones}
For the bulk of our analysis, we use non-rotating isochrones from the modern and publicly available MIST database (MESA Isochrones and Stellar Tracks; \citealt{Choi2016}). The stellar tracks underlying these models are computed with the Modules for Experiments in Stellar Astrophysics software instrument (MESA; \citealt{MESAI, MESAII, MESAIII, MESAIV, MESAV}). The models adopt the solar abundance scale of \citet{Asplund09}, the OPAL opacities of \citet{IglesiasRogers1996}, a default value for the convective mixing length parameter of $\alpha_{\text{MLT}}=1.82$ with the MLT (Mixing Length Theory; \citealt{Boehm-Vitense79}) prescription of \citet{Henyey1964}, photosphere tables based on the 1D model atmospheres of \citet{Kurucz1970} as surface boundary conditions, and a small degree of core convective overshoot, $f_{\text{ovs}}=0.016$. The models include the effects of diffusion on the main sequence, where diffusion is modeled via five representative elemental species \citep{Thoul94}. The key physical assumptions made in MIST are summarized in Table 1 of \citet{Choi2016} and discussed in further detail throughout.   

\begin{figure*}
\begin{center}
\includegraphics[width=\textwidth]{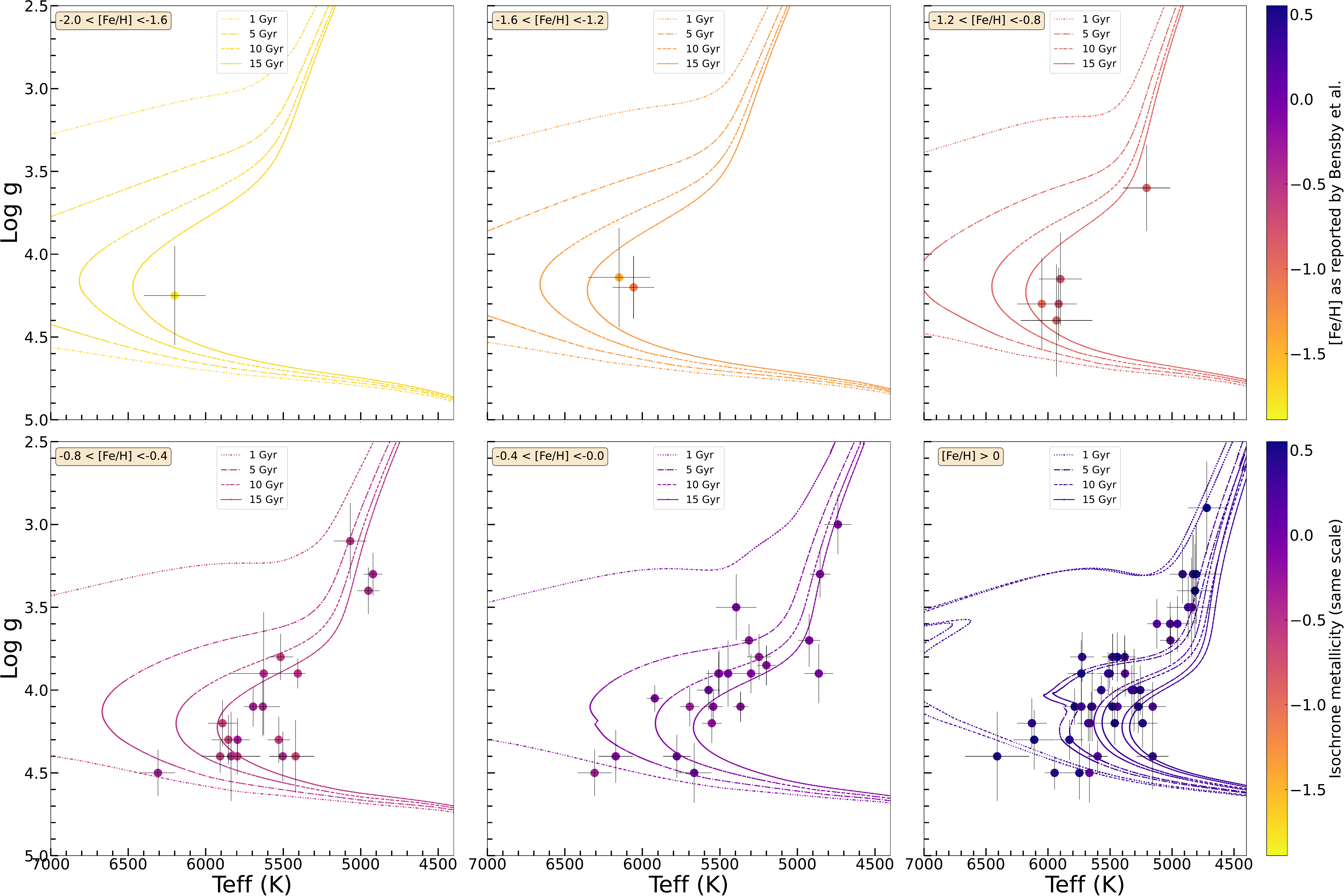}
 \caption{Reproduction of \citet{Bensby17}'s Figure 6 but using MIST isochrones. Metallicity bins are the same as in \citet{Bensby17}. }
\label{fig:6panel}
\end{center}
\end{figure*}

We begin by re-fitting \citet{Bensby17}'s sample directly. Figure \ref{fig:6panel} provides a reproduction of \citet{Bensby17}'s Figure 6, with the panels divided according to the same metallicity bins, as indicated in the upper left-hand caption. Each panel contains four MIST isochrones with ages 1, 5, 10, and 15 Gyr shown in dotted, dot-dashed, dashed, and solid lines, respectively. The metallicities assigned to the isochrones are the median of each bin: from metal-poor to metal-rich, [Fe/H]$_{\text{iso}}= -1.8, -1.4, -1.0, -0.6, -0.2$, and the most metal-rich bin shows [Fe/H]= $+0.2$ and $+0.3$.   
Darker colored tracks indicate higher metallicites (purple for highest, yellow for lowest), as shown on the color bar. 

A side-by-side comparison of Figure \ref{fig:6panel} with \citet{Bensby17}'s Figure 6 already suggests some discrepancy. In general, the MIST isochrones are hotter (left-shifted) relative to the observations than are the isochrones in \citet{Bensby17}. If all physical assumptions in both sets of models were identical, this would translate to the prediction of older ages for the same stars with MIST. The choices of physics are, of course, not identical. Differences in the models' treatment of abundances, in particular, complicates the physical picture.

We control for the impact of differences in the choice of modeling physics in two ways: First, we examine the choice of isochrone basis by comparing results computed with MIST to results computed with the Yonsei--Yale (hereafter abbreviated YY; \citealt{YY}) isochrones, the models used in \citet{Bensby17}. Second, we compare results from MIST isochrones with simulated $\alpha$-element enhancement to those without.  

\subsection{YY vs MIST}
\label{YY_vs_MIST}
The version of the YY isochrones used in \citet{Bensby17} (hereafter abbreviated B17) date back to 2003, making them nearly 15 years older than MIST. They also constitute a sparser theoretical data set, as the number of parameter variations covered by the publicly available YY grid is smaller, and the metallicity resolution lower, than in MIST's case. Despite this, the YY isochrones were among the first to incorporate the effects of helium and $\alpha$-element enhancement carefully \citep{YY}. 

\begin{figure*}
\begin{center}
\includegraphics[width=0.49\textwidth]{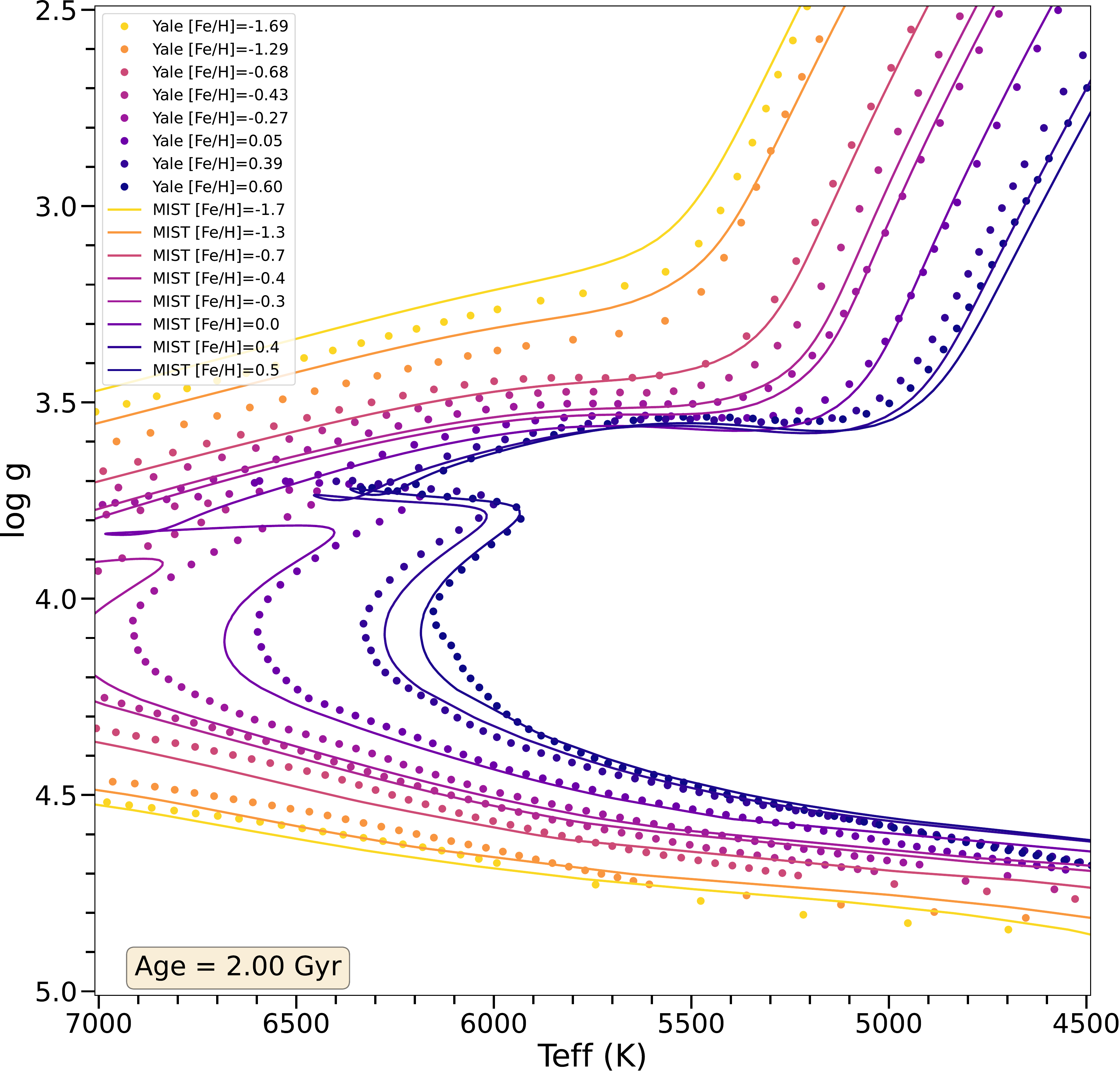}
\includegraphics[width=0.49\textwidth]{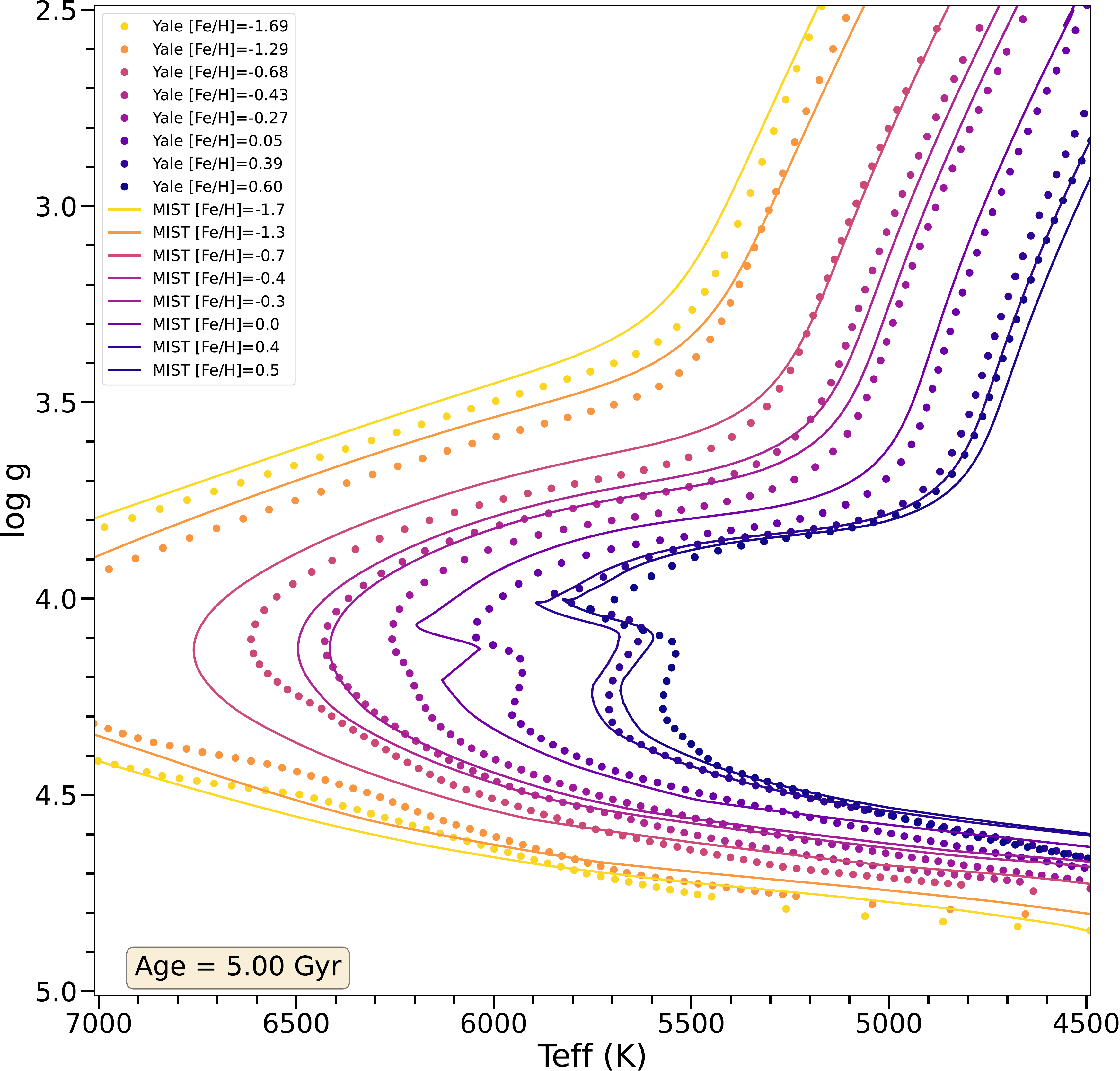}
\includegraphics[width=0.49\textwidth]{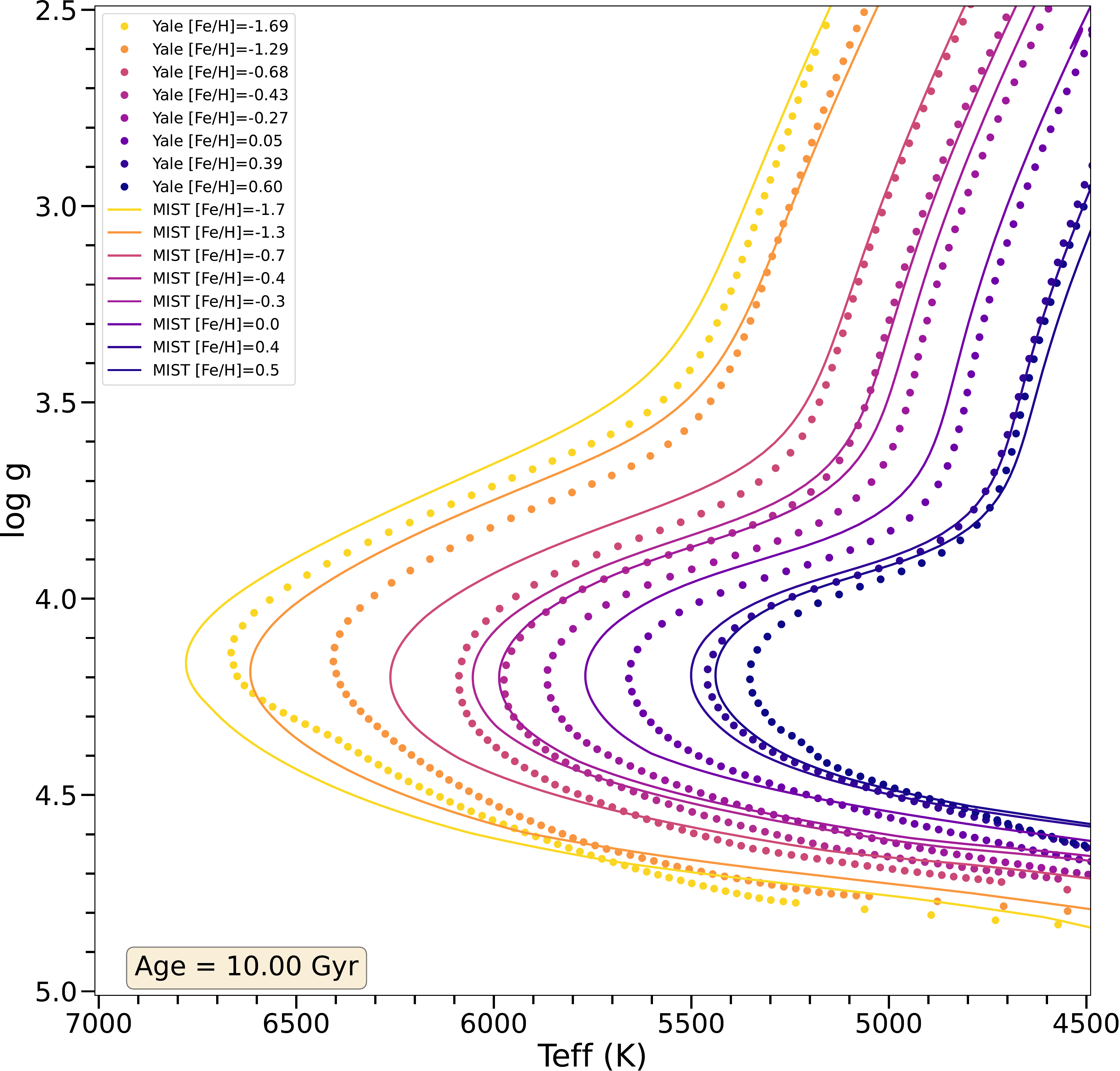}
\includegraphics[width=0.49\textwidth]{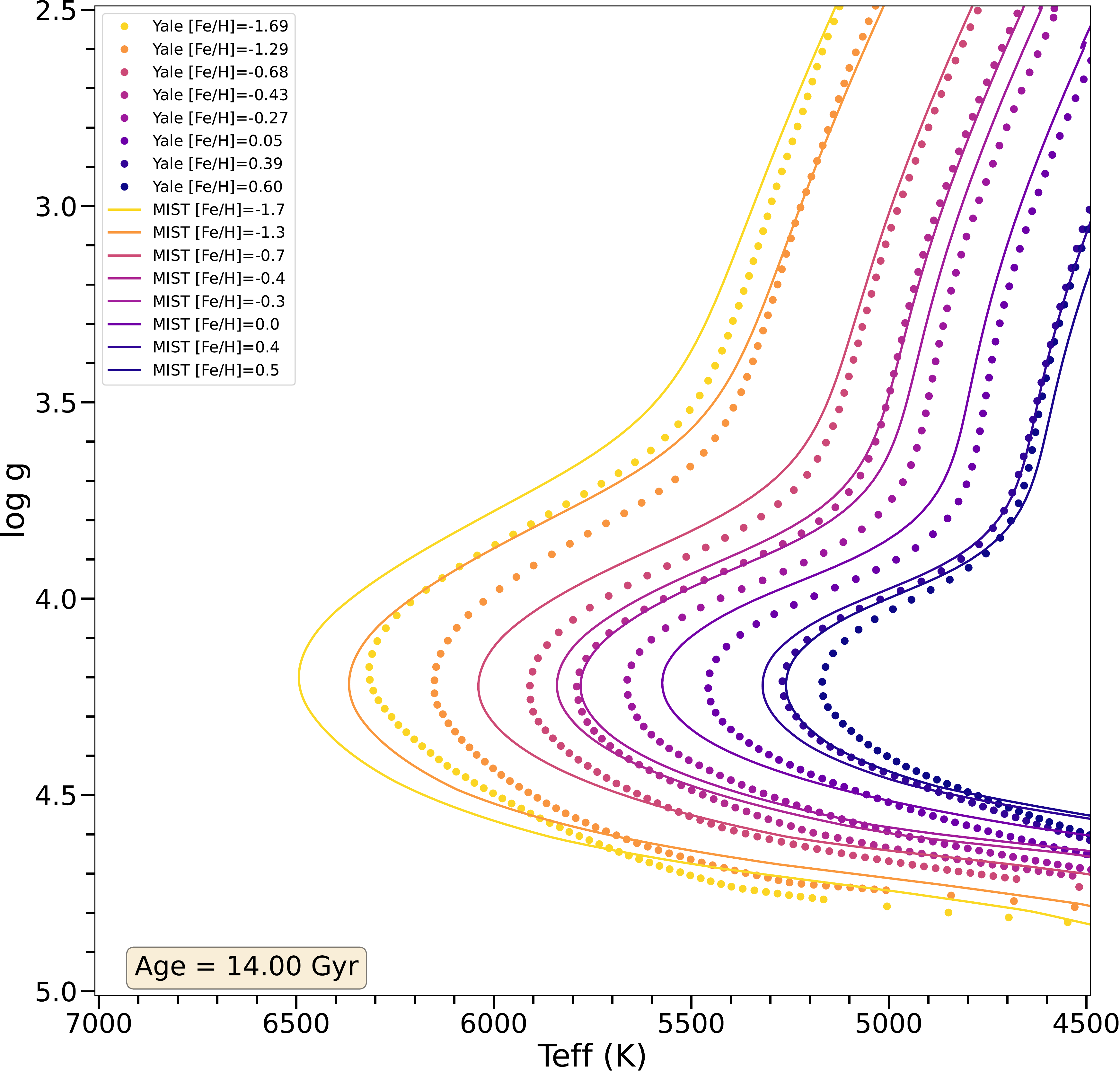}
\caption{A selection of MIST and Yale isochrones are shown for a selection of ages: 2, 5, 10, and 14 Gyr. MIST models are shown in dots, Yale with solid lines. Models of similar color have similar metallicities.}
\label{fig:yale_cmds}
\end{center}
\end{figure*}

Figure \ref{fig:yale_cmds} shows a selection of the 2004 YY isochrones (dotted lines) overlaid on a selection of MIST isochrones (solid lines) with similar metallicities for a range of ages. Curves with similar colors are similar metallicities, with the same color scale used in Figure \ref{fig:6panel}.

For fixed age and metallicity, the Yale isochrones generally predict cooler temperatures compared to MIST isochrones---which is what we should expect based on the comparison of Figure \ref{fig:6panel} to B17's Figure 6---but this relationship inverts, or becomes ambiguous, near the base of the red giant branch. 

There is noticeable discrepancy in the locations of the main sequence turn--off (MSTO) regions in the 2 and 5 Gyr panels. Since the majority of stars in the B17 sample fall exactly in this region, especially those stars identified as young, we should expect a non-trivial difference between the ages predicted by MIST and the ages predicted by Yale, all else being equal. 

It is well known that the choices of modeling physics have a large effect on the MSTO and base of the red giant branch (e.g. \citealt{Joyce2018aNotAll}), as the isochrones demonstrate. The key differences in modeling physics between MIST and YY can be summarized as follows:
\begin{enumerate}
    \item choice of solar abundance scale: YY uses the abundances of \citet{GS98}, which place the solar metallicity at Z = 0.020, whereas MIST uses the scale of \citet{Asplund09}, which places the solar metallicity at Z= 0.014;
    
    \item atmospheric boundary conditions: YY uses a grey model atmosphere, whereas MIST uses photosphere tables based on 3D atmosphere simulations;
    
    \item treatment of diffusion: YY includes the effects of diffusion throughout, whereas MIST only uses diffusion on the main sequence; and
    
    \item mixing length: the YY stellar tracks adopt a value of $\alpha_{\text{MLT}}= 1.743$ for the convective mixing length, whereas the MESA models use $\alpha_{\text{MLT}} =1.82$.
\end{enumerate}

Among the first three, the difference in solar abundance scale will dominate. Regarding point 3, we note that the incorporation of diffusion will not affect the evolution beyond the main sequence, but it is worth mentioning this difference in the isochrone databases given that many of our targets are not necessarily more evolved than the upper main sequence. We defer a discussion of point four until Section \ref{sec:modeling_uncertainty}. For a more comprehensive comparison of the assumptions made in these and other stellar evolution programs, see Table 1 of \citet{Tayar2022}.

\subsection{Effects of alpha--element enhancement}
\label{sec:alpha}
Although there is no way to request isochrones with a particular degree of $\alpha$-enhancement, or [$\alpha$/Fe], from the publicly available MIST database directly, $\alpha$-enhancement can be approximated in the models by rescaling the baseline abundance, [Fe/H]$_0$, to a new value [Fe/H]$_{\text{rescaled}}$ that encapsulates the relative increase in global metallicity imparted by larger numbers of $\alpha$-capture elements. This is done by computing an array of global metal mass fractions, $Z$, associated with a densely sampled grid of ([Fe/H], [$\alpha$/Fe]) combinations. This dictionary can then be used to find an appropriate pair ([Fe/H]$_{\text{rescaled}}$, [$\alpha$/Fe]$=0$) for any input combination of ([Fe/H],[$\alpha$/Fe]). Since this relation defines parameter pairs as equivalent according to their production of the same $Z$, the choice of solar scale will affect this relationship. 

We adopt an $\alpha$-element enhancement distribution for the isochrones according to the functions
\begin{itemize}
\item[] [alpha/Fe] = $+0.3$ for [Fe/H] $< -0.3$,
\item[] [alpha/Fe] = $-$[Fe/H] for $-0.3 <$ [Fe/H] $< 0$, and
\item[] [alpha/Fe] = 0 for [Fe/H] $> 0$,
\end{itemize}
which are based on the consistently observed [$\alpha$/Fe] versus [Fe/H] trends reported by numerous authors \citep[e.g.,][]{Gonzalez11,Johnson14,Bensby17,Rojas20}. This produces the rescaled [Fe/H] values given in Table \ref{table:alpha_rescale}.
\begin{table} 
\centering 
\caption{Re-scaled isochrone metallicities accounting for $\alpha$-element enhancement.}
\begin{tabular}{l l l  }  
\hline\hline
[Fe/H] & [$\alpha$/Fe] & rescaled [Fe/H]  \\
\hline
$-2.0$ & $+0.3$ & $-1.787$ \\
$-1.9$ & $+0.3$ &  $-1.687$\\
$-1.8$ & $+0.3$ & $-1.587$ \\
$-1.7$ & $+0.3$ & $-1.487$ \\
$-1.6$ & $+0.3$ & $-1.387$ \\
$-1.5$ & $+0.3$ & $-1.287$ \\
$-1.4$ & $+0.3$ & $-1.187$ \\
$-1.3$ & $+0.3$ & $-1.087$ \\
$-1.2$ & $+0.3$ & $-0.987$ \\
$-1.1$ & $+0.3$ & $-0.887$ \\
$-1.0$ & $+0.3$ & $-0.787$ \\
$-0.9$ & $+0.3$ &  $-0.687$\\
$-0.8$ & $+0.3$ & $-0.588$ \\
$-0.7$ & $+0.3$ &  $-0.486$\\
$-0.6$ & $+0.3$ & $-0.384$ \\
$-0.5$ & $+0.3$ &  $-0.285$ \\
$-0.4$ & $+0.3$ & $-0.184$ \\
$-0.35$ & $+0.3$ & $-0.137$\\
$-0.32$ & $+0.3$ &  $-0.108$\\
$-0.3$ & $+0.3$ &  $-0.088$\\
$-0.26$ & $+0.26$& $-0.077$\\
$-0.2$ & $+0.2$ & $-0.060$ \\
$-0.1$ & $+0.1$ & $-0.037$\\
0 to $+0.5$ & $0.0$ & original [Fe/H] \\
\hline
\end{tabular}
\label{table:alpha_rescale}
\end{table}

\begin{figure*}
\begin{center}
\includegraphics[width=0.49\textwidth]{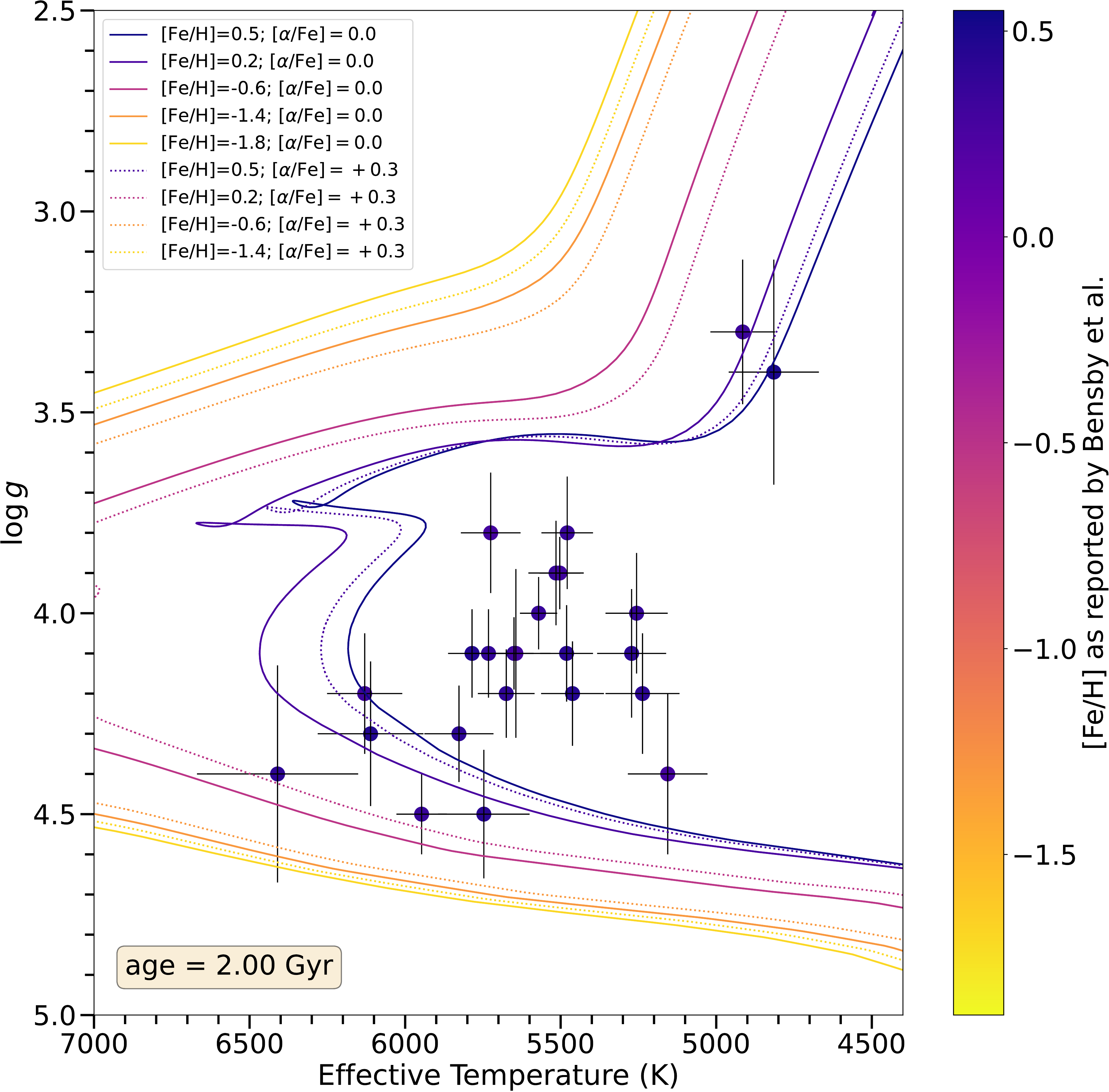}
\includegraphics[width=0.49\textwidth]{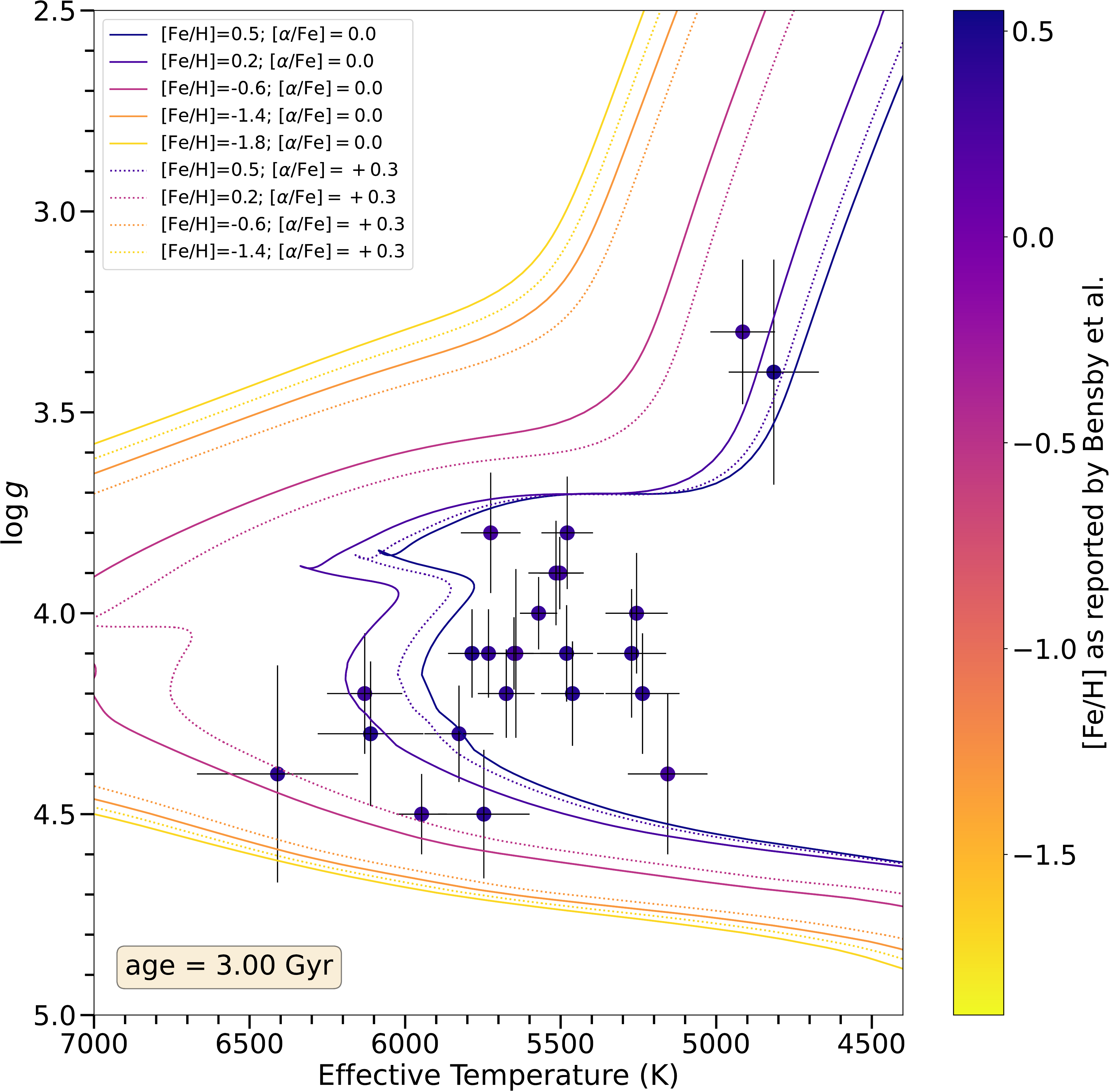}
\includegraphics[width=0.49\textwidth]{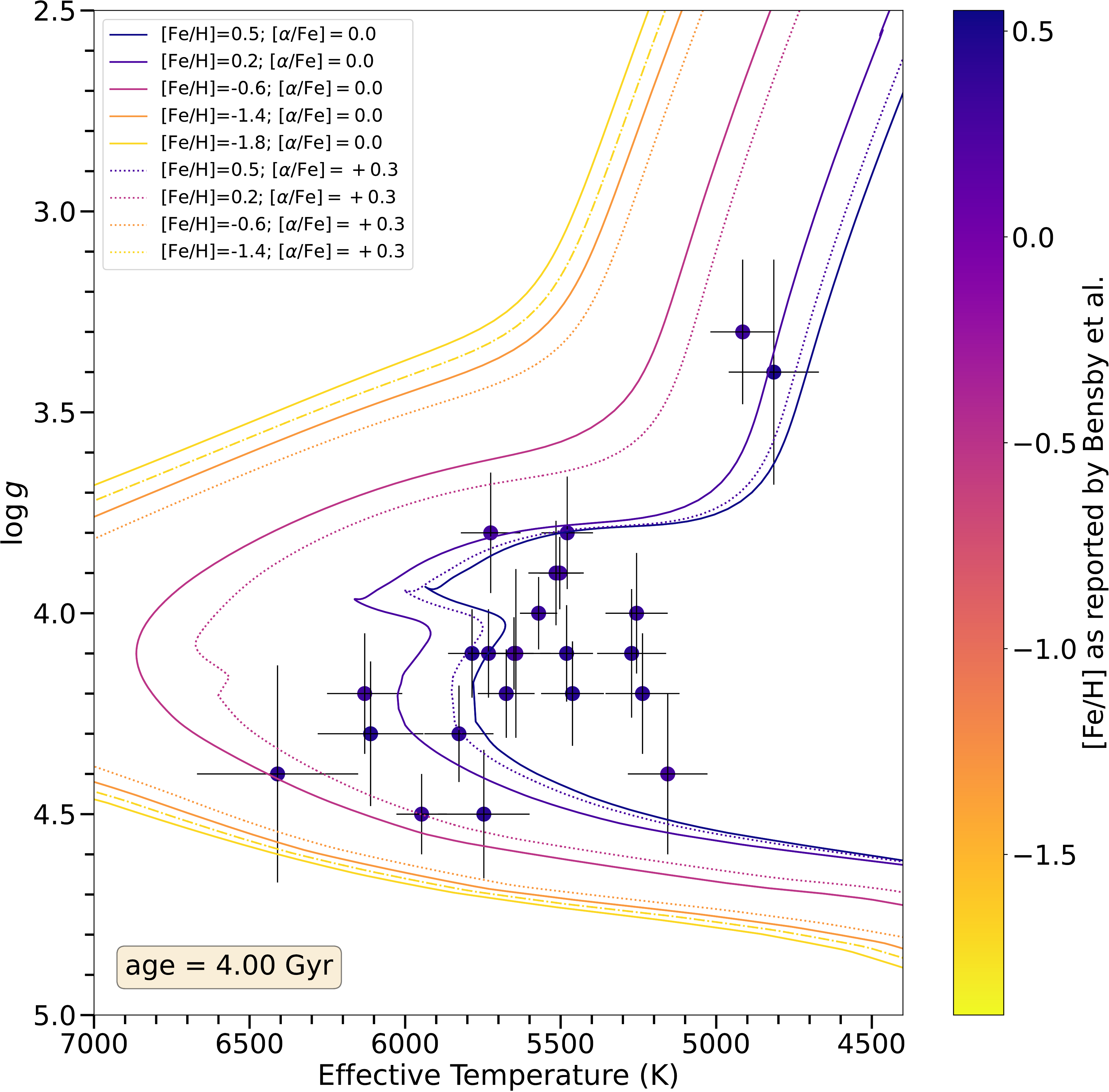}
\includegraphics[width=0.49\textwidth]{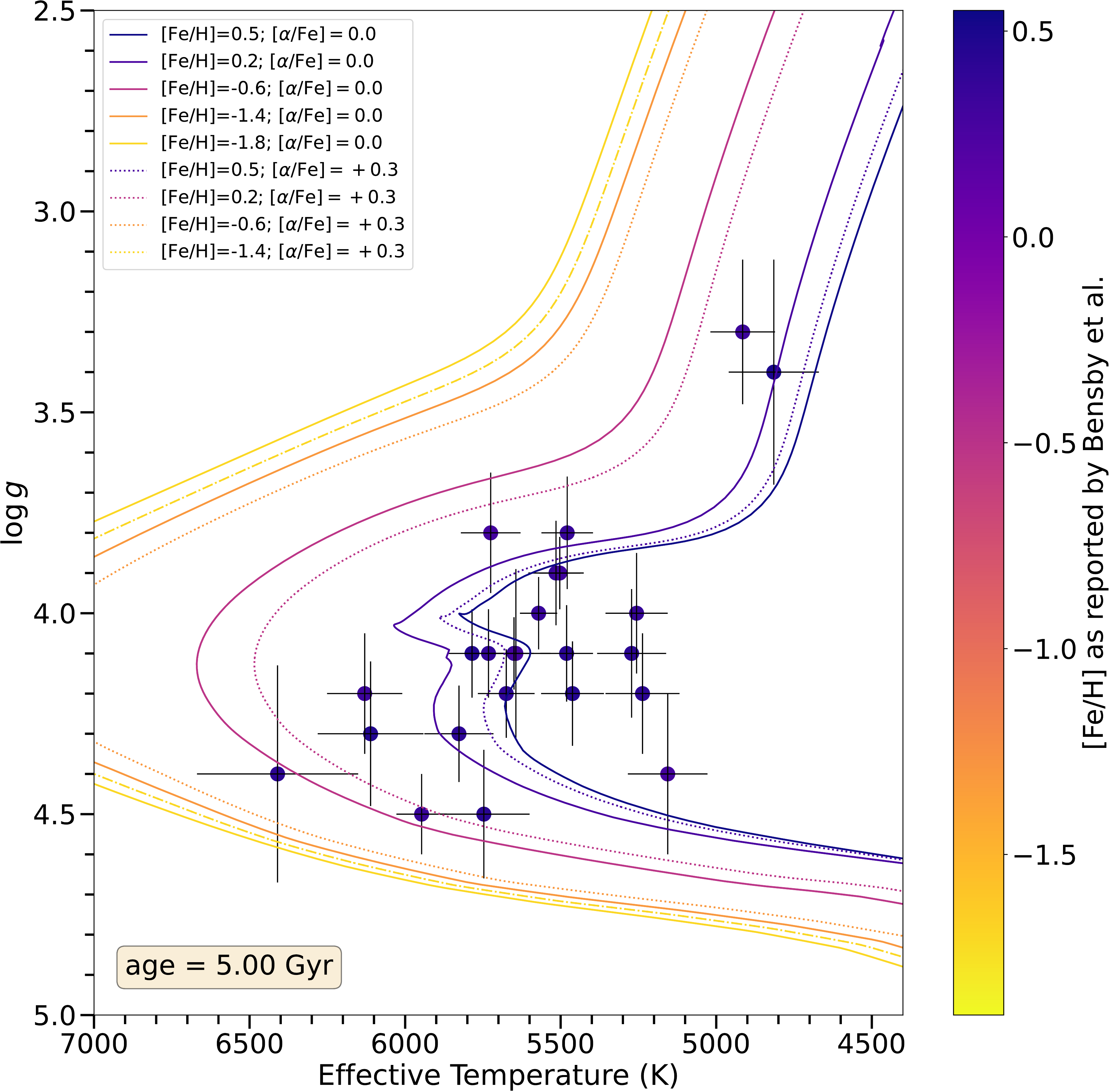}
 \caption{Isochrones at a small range of metallicities and their $\alpha$-scaled counterparts are shown for the ages of $2$--$5$ Gyr, as indicated. Isochrones that take into account $\alpha$-element enhancement are indicated on these four panels by dashed lines. A solid line of the same color indicates the same isochrone without $\alpha$-enhancement. In the case of the isochrone with [Fe/H] $= 0.5$, no $\alpha$-enhanced counterpart exists due to the global metallicity upper limit of $+0.5$ imposed by the MIST database. The most visible effect of including $\alpha$-element enhancement in the isochrones is to shift the isochrones towards cooler temperatures, but the net effect of increased metallicity is that stars with equivalent temperatures and luminosities have lower masses, translating to slower evolution and an increase in inferred age as a function of [Fe/H].}
\label{fig:young_alpha_shift}
\end{center}
\end{figure*}

The difference between $\alpha$-enhanced and standard MIST isochrones can be seen in Figure \ref{fig:young_alpha_shift}. The primary effect of including $\alpha$-element enhancement is to shift the isochrones rightward towards cooler temperatures (redder colors) for equivalent ages and baseline metallicities. 
This appears similar to the difference between MIST and YY isochrones. However, the reason for the difference in this case is that, as metallicity increases, stars maintain the same reaction rates at lower temperatures. The number of donor electrons increases in the atmosphere, which causes an increase in the continuous opacity and subsequent decrease in the surface temperature. The isochrones constructed from such models will therefore skew cooler, but their evolution also lags behind isochrones without $\alpha$-enhancement. When $\alpha$-enhancement is handled properly, age predictions will increase. 

\begin{figure*}
\includegraphics[width=\textwidth]{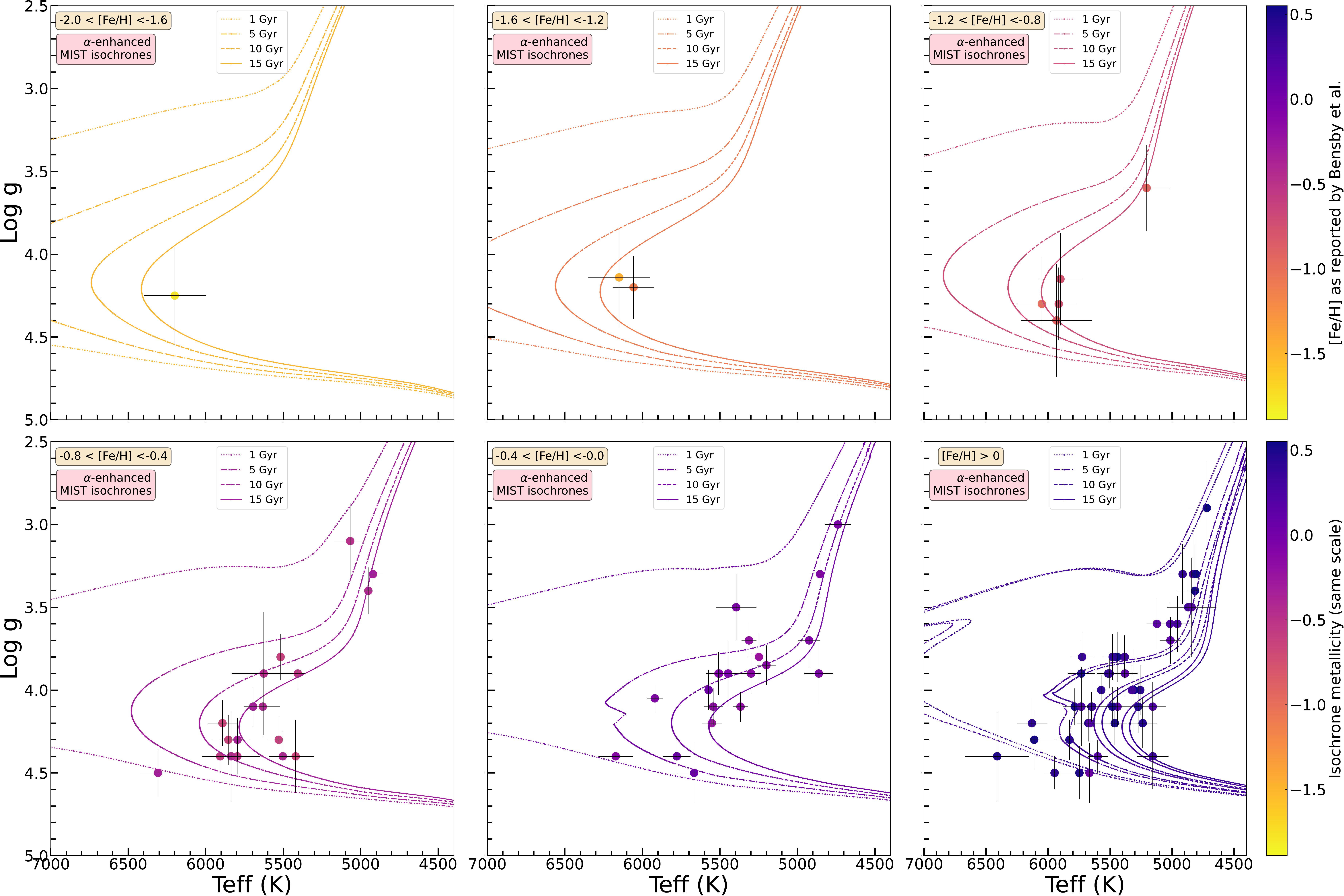}
 \caption{Same as Figure \ref{fig:6panel} but using $\alpha$-enhanced MIST isochrones.}
\label{fig:6panel_alpha}
\end{figure*}

Figure \ref{fig:6panel_alpha} is a reproduction of Figure \ref{fig:6panel} using $\alpha$-enhanced MIST isochrones. They provide a slightly better visual fit to the B17 sample and are more reflective of real $\alpha$ abundance trends with metallicity. They are also most similar in composition to the abundances assumed in the YY isochrones used in \citet{Bensby17}'s analysis, while still being up-to-date and adopting the most modern physical conventions. We proceed using the $\alpha$-enhanced MIST isochrones as the basis for our quantitative age determinations.

%
%
\section{Age Determinations and Statistical Analysis} 
\label{sec:statistical_analysis}
Though it is straightforward to fit an isochrone to an observation ``by eye,'' it is much more difficult to construct a mathematically rigorous definition of a best-fitting model. This is especially true in a situation where: 
\begin{enumerate}
    \item many observations are plausibly consistent with a large number of isochrones (i.e., the isochrone falls within the star's $1 \sigma$ uncertainties);

    \item the isochrones are discretely spaced in age and metallicity and thus limited in resolution; and 
    
    \item the models contain their own intrinsic uncertainties that are both difficult to quantify and highly variant over different parameter regimes---a point to which we will return in later discussion.
    \end{enumerate}
To deal with these issues, we apply a combination of known statistical techniques to estimate the age and observational age uncertainty for each star according to an appropriate basis of isochrones. Our method incorporates elements of maximum--likelihood analysis, Monte Carlo simulations, and a frequentist approach to goodness-of-fit/$\chi^2$ (see, for instance, \citealt{Joyce2018balphaCen, MurphyJoyce2021, TangJoyce2021}).
The chosen isochrone basis constitutes a set of candidate hypotheses against which we test our stars. In all cases, the MIST isochrones range in age from $0.5$ to $20$ Gyr, sampled at a resolution of $0.5$ Gyr, and in metallicity from [Fe/H] = $-2.0$ to [Fe/H] = +0.5, sampled at a resolution of 0.1 dex. We permit ages above the formal age of the Universe \citep[13.5 Gyr;][]{Planck16} for consistency with \citet{Bensby17}'s lack of upper bound on their own age determinations, and the standard MIST database permits ages up to 20 Gyr.

For each of \citet{Bensby17}'s targets, we measure the goodness-of-fit between the star and every point along each isochrone. Restricting to the appropriate evolutionary regime (we may confidently exclude the pre-main sequence and the red giant branch above the Red Giant Branch Bump, for example), the number of points along a particular isochrone is 220\footnote{We note, however, that this restriction was removed when custom isochrones were used in the fits due to the modified evolutionary trajectories of the underlying tracks caused by, e.g., very low assumptions for the convective mixing length; see Section \ref{sec:modeling_uncertainty}. A test comparing age determinations when using a restricted evolutionary domain on the isochrones to those using an unrestricted domain produced a shift toward older ages of less than 1\%.}

Goodness-of-fit for each theoretical, observational pair is determined according to a chi-square statistic, given by
\begin{equation}
\chi^2_{\text{for } i \text{ DoF}} = \sum_i  \frac{ (o_i - t_i)^2 }{\sigma_i^2 } 
\label{eq:chisq}
\end{equation}
where $o_i$ are the values derived from observations, $t_i$ are the values assumed in the models (theory), and $\sigma_i$ are the observational uncertainties. The number of degrees of freedom (DoF) is equal to the number of terms $i$ in the summation, which determines the distribution of the $\chi^2$ scores.
This metric, which ultimately underlies our claim that an isochrone of particular age is more or less consistent with a measurement, depends strongly on the number of independent\footnote{Observations are not strictly independent, but we must make this assumption for the sake of practicality.} observations, the uncertainties in the observations themselves, and the uncertainties implicit in, e.g., distance when a transformation from apparent to absolute magnitude is required. It is therefore crucial that we carefully consider the definitions of the constituent uncertainty terms in each case.

Explicitly, in the case of \citet{Bensby17}'s sample, there are three data sources and their associated uncertainties: error in $\log g$, error in effective temperature $T_\text{eff}$, and error in [Fe/H] (abbreviated below as~$Z$, the convention for representing metallicity in stellar models). As these are already physical coordinates and the uncertainties are reported directly, this $\chi^2$ definition is straightforward:
\begin{equation}
\chi^2_{\text{B17}} = \frac{ (\log{g}_{o} - \log{g}_{t})^2 }{\sigma_{\log{g},o}^2 }  + \frac{ (T_{\text{eff},o} - T_{\text{eff},t})^2 }{\sigma_{T_{\text{eff}},o}^2 }  
  + \frac{ (Z_{o}  - Z_t)^2 }{\sigma_{Z,o}^2 }, 
\label{eq:chisq_Bensby}
\end{equation}
as are the definitions for $\sigma_{\log g,o}, \sigma_{T_{\text{eff}},o}$, and $\sigma_{Z,o}$. We note that this statistic applies equal weighting to uncertainties in $T_\text{eff}$, $\log g$, and metallicity.

Fits to photometric data, on the other hand, are more complicated; they must take into account not only the uncertainties in the apparent magnitudes for the star in each filter used to build the CMD, but uncertainties in the distance to the star, extinction, and observational instrumental systematics. Further, the relationship among these quantities and appropriate relative weightings are not obvious. Definitions of uncertainty terms appropriate for the photometric analyses conducted here are discussed further in Appendix \ref{sec:photometric_fits}. 

For each star $S$ in the B17 sample, we compute an agreement statistic, or ``score,'' comparing $T_{\text{eff},S}$, $\log{g}_S$, and $Z_S$ to each of 220 evolutionary points $\times 40$ ages $\times 29$ metallicities for a total of 255,200 instances of $T_{\text{eff},t}$ and $\log{g}_{t}$ in Equation \ref{eq:chisq_Bensby}. Like age, [Fe/H] (AKA $Z_t$) is fixed per isochrone, so $Z_t$ does not vary over the set of evolutionary points. This algorithm thus returns the relative likelihoods of all candidate age--metallicity hypotheses (isochrones) for $S$.

We then construct a definition of the \textit{weighted average age} through which each age hypothesis is penalized according to its likelihood: Each score $\chi^2_n, n=1,\ldots,255200$ has a relative likelihood, $p_n$, given by the probability density function (pdf) for the $\chi^2$ distribution with three degrees of freedom. We caution that this association implicitly assumes the three variables $\log{g}$, $T_{\text{eff}}$, and [Fe/H] are independent and normally distributed. Though this is not strictly true (we note, for instance, that $\log{g}$ may more closely follow a log normal distribution, and any parameters derived from the same observational source---in this case, spectroscopy---are correlated by definition), we confirm that this is a reasonable enough approximation via Monte Carlo simulations, discussed in detail in Section \ref{sec:error_analysis} and the Appendix.   

The weighted average age of star $S$ is given by 
\begin{equation}
 t_S =  \frac{ \sum_n t_n p_n}{ \sum_n p_n},
\label{eq:weighted_age}
\end{equation}
where $t_n$ is the age of the isochrone hosting $\log  g_{t}, T_{\text{eff},t}$ and $Z_t$, $p_n$ is  the value of the chi square pdf at the score for that hypothesis given by Equation~\ref{eq:chisq_Bensby}, and $n$ is the total number of candidate hypotheses considered. The MIST-based age  we report for a star therefore represents a weighted average over all age solutions,
with the most consistent age fits weighted highest. 

\subsection{Comparison to other methods in the literature}
\label{sec:alternative_stats}

\citet{bensby11} briefly describe how they form a probability distribution for the age estimates as a weighted average of ``probabilities,'' which are products of the normal density functions comparing the observed parameters for a star to those of a suite of isochrones. This is similar to our method: both assume the observed parameters come from independent normal random variables, for example, though in our case the weights are the values of the density function for the $\chi^2$ distribution with three degrees of freedom. However, \citet{bensby11} relegate a more detailed discussion of their method to Mel\'endez et al.\ (in prep), which we were unable to find, and \citet{Bensby17} refer only to their 2011 paper regarding their use of the same method. 
While we do find some discussion of a seemingly similar method in \citet{Melendez2012}, it is unclear how they define their $1\sigma$ uncertainties. We speculate that, rather than fitting their derived distribution to a normal distribution, they extend their estimates outwards to collect $1\sigma$'s worth of points assuming normality and do so in either direction independently, which could be the cause of their highly asymmetric error bars. However, this is not stated explicitly, so we are unable to make a more rigorous comparison at this time.

In some capacity, \citet{bensby11} derive their $1\sigma$ error estimates from the shapes of the distributions they find. We instead use Monte Carlo resampling for two reasons. First, this requires no additional assumptions, whereas they must make some assumptions about these shapes. Second, through random resampling, we are able to incorporate model uncertainties (see Section \ref{sec:modeling_uncertainty}), which 
they did not take into account.

Various other methods have been used to estimate stellar ages using a suite of isochrones. \citet{Angus2019} used Markov Chain Monte Carlo random walks based on \citet{Foreman2013} to find the parameter region with the best-fitting isochrones. We found this inappropriate for our data, as our larger uncertainties mean too broad a range of isochrones offer plausible fits, which would lead to random walkers ending up in local minima as an artifact of the discrete isochrones. 
Similarly, \citet{Creevey22} used a combination of Gaia DR3 photometry, spectroscopy, and BaSTI isochrones to derive ages for millions of stars, but their method is more suited to targets in the solar neighborhood and has limited utility for stars with non-solar metallicities (see also \citealt{Kordopatis2022}).
\citet{Sanders2018} obtained age estimates for $\sim 3$ million stars from \textit{Gaia} and spectroscopic data using a Bayesian method that calculates a weighted average over isochrones due to \citet{Burnett2010}. Our method is similar in principle, and it would be worthwhile to compare our method to theirs in future work. Their method more readily accounts for the non-uniform spread of their isochrone grid, but we do not expect our lack of consideration for this feature to make a significant difference given that we tested the sensitivity of our results to the input isochrone grid and increased its age and metallicity resolution as needed. 

Another machine-learning method is \emph{The Cannon} \citep{Ness2015}, used for example by \citet{Ho2017,Hasselquist20,Sit2020} to estimate ages for large numbers of stars based on training data derived from APOGEE and other surveys. Our target, however, is a small number of individual stars for which we can make direct inferences, so this method is not ideal for our situation.
We also prefer to avoid machine-learning techniques in general, as they can still suffer from the same sensitivities to input (i.e.\ training sets) as any other flawed method, but these sensitives are obscured by ``black box'' usage, leading to over-confidence in the results. Any age determinations for imperfect data will inevitably be biased, whether by the training sample, the method, the priors, or data incompleteness and other problems. Though our method is not without its shortcomings and questionable-but-necessary assumptions, it is clearly defined, reproducible, and its limitations are not obscured by a black box.

\subsection{Ages according to basic MIST isochrones}
\begin{figure}
\begin{center}
\includegraphics[width=\columnwidth]{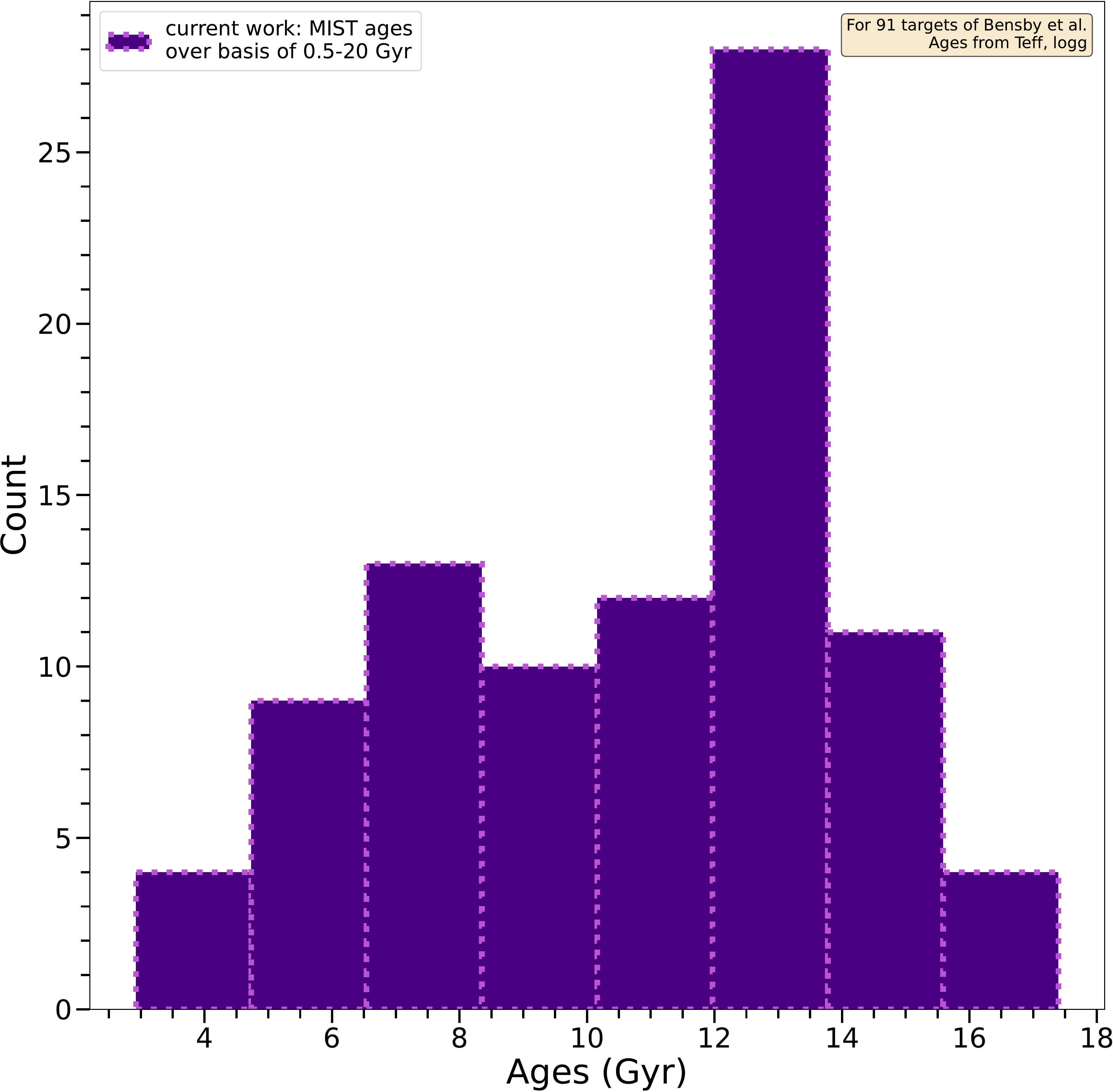}
\includegraphics[width=\columnwidth]{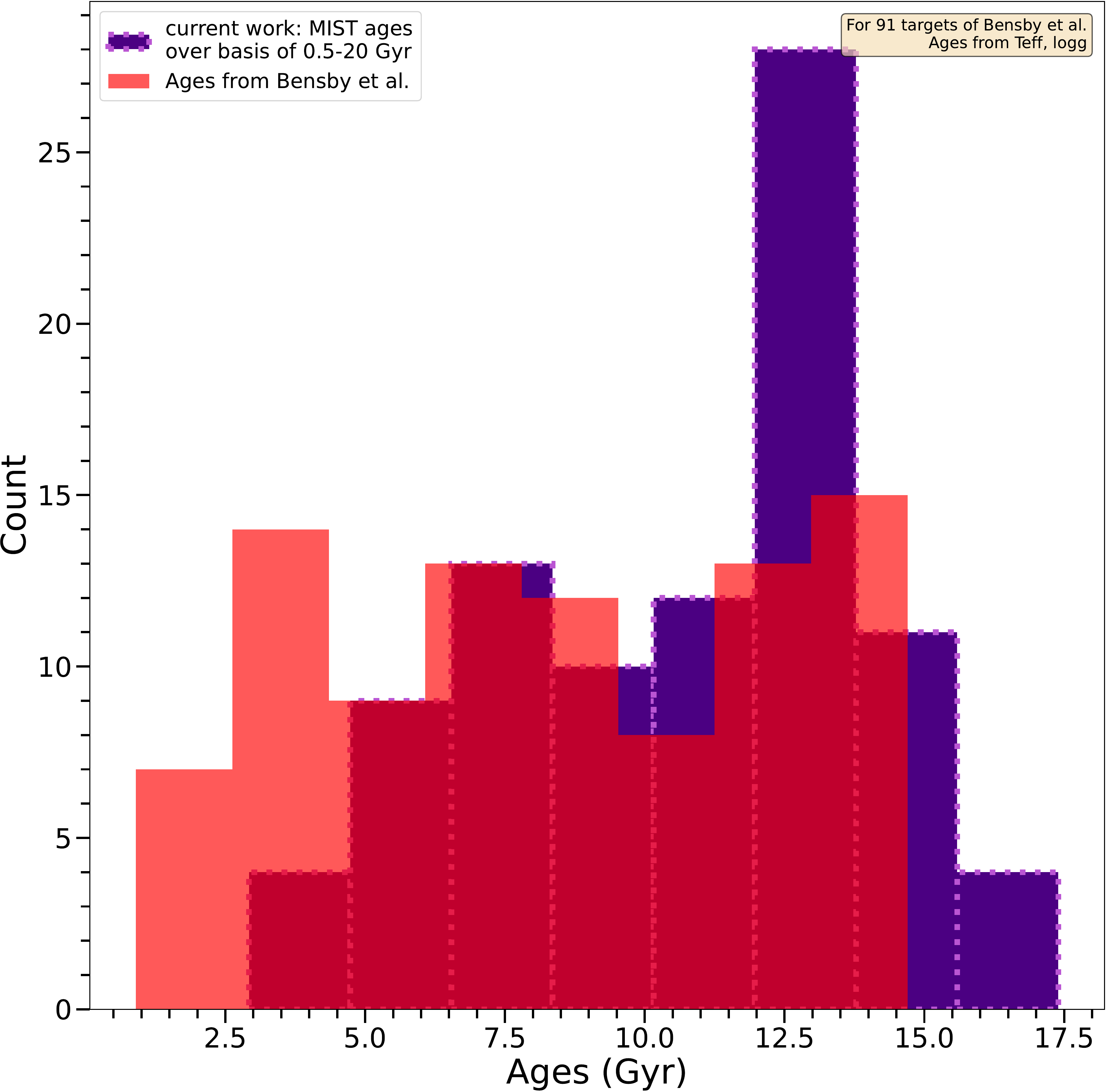}
\caption{\textbf{UPPER:} The age distribution computed using MIST isochrones and the algorithm described in section \ref{sec:statistical_analysis}.
\textbf{LOWER:} Two histograms overlaid, comparing the ages presented in \citet{Bensby17} (transparent red bars, foreground) to ages derived using MIST isochrones and the same logg and teff measurements (purple bars, background). }
\label{fig:age_histogram}
\end{center}
\end{figure}
The age distribution resulting from the application of our method to 91 B17 targets is given in Figure \ref{fig:age_histogram}.
The most striking feature of the distribution shown in the upper panel of Figure \ref{fig:age_histogram} is the strong peak at an age of approximately 13 Gyr. While some stars with ages below 8 Gyr are identified, only four stars are found to have ages below 5 Gyr, and we find none with ages less than 3 Gyr. On the contrary, \citet{Bensby17}'s age distribution, constructed from the same data and overlaid in red in the lower panel of Figure \ref{fig:age_histogram}, shows a nearly uniform distribution of ages between 0.5 Gyr and 14.5 Gyr.

\subsection{Ages according to Yale isochrones}
\begin{figure}
\begin{center}
\includegraphics[width=0.49\textwidth]{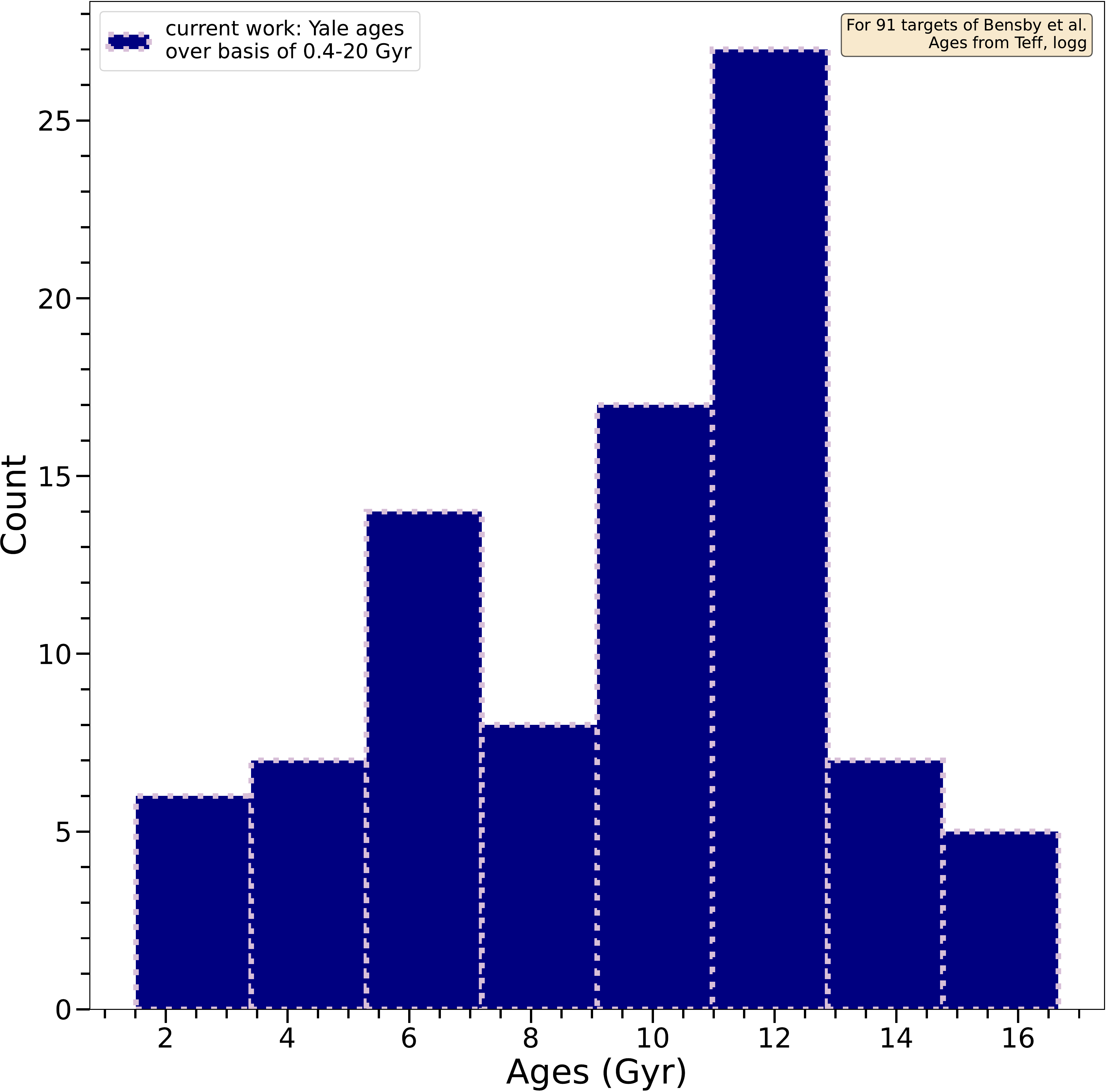}
\includegraphics[width=0.49\textwidth]{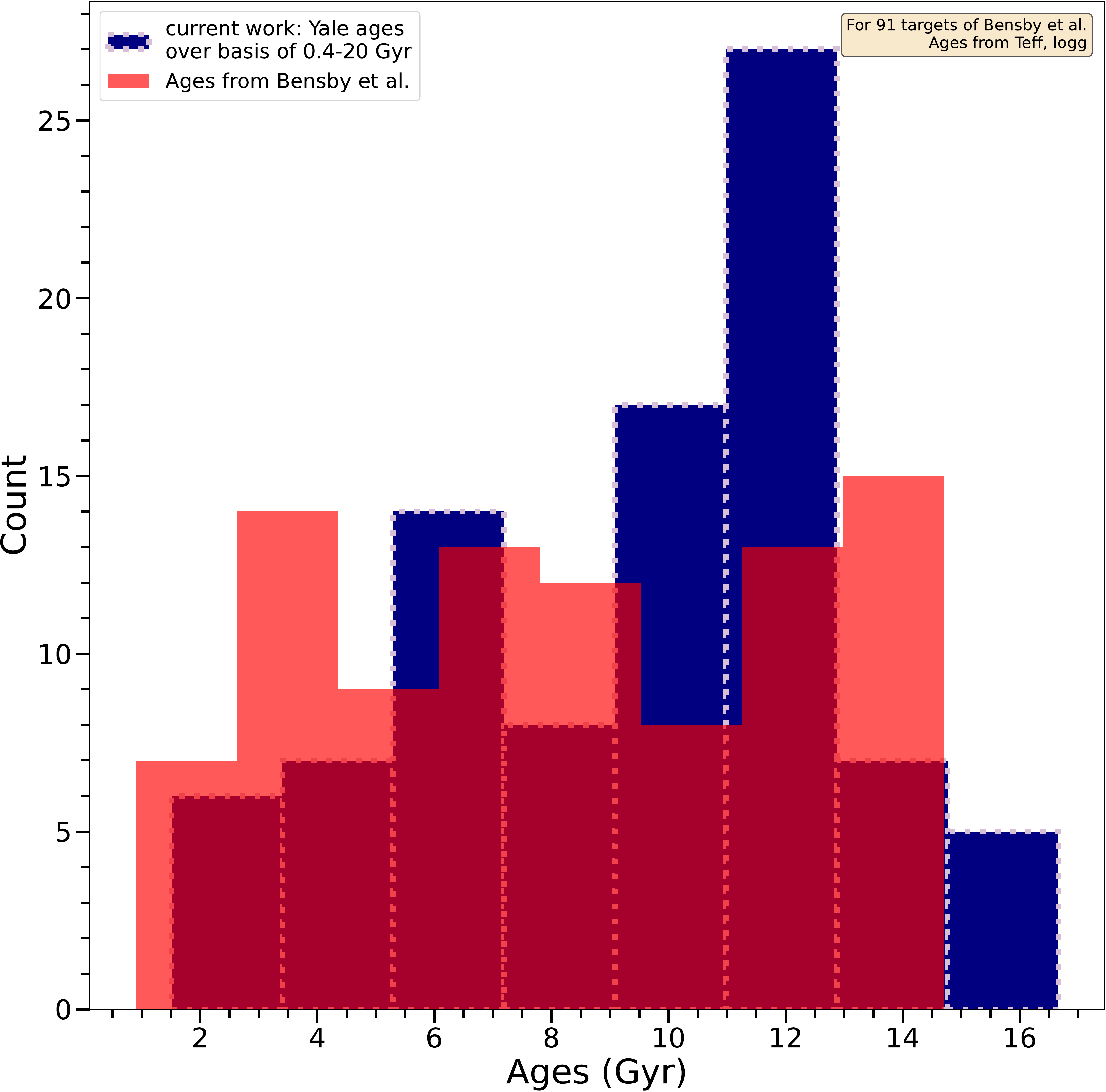}
\caption{\textbf{UPPER:} The panel shows the ages of 91 B17 targets computed using the age determination algorithm described in this manuscript, but instead adopting a basis of 2003 Yale isochrones instead of MIST isochrones. These were the isochrones used in \citet{Bensby17}'s original age determinations and adopt different physical assumptions than MIST. \textbf{LOWER:} Same as left, but with the B17 distribution overlaid, as in previous figures. Though the presence of a larger number of young stars is clear, the shape of the Yale-based age distribution is still not consistent with the distribution presented in B17.}
\label{fig:yale_hist}
\end{center}
\end{figure}

We test the hypothesis that the Yale isochrones should produce older ages, presented in Section \ref{sec:isochrones}, by applying the algorithm presented here to the publicly available set of 2003 Yale isochrones. In this case, the range of ages available is reduced to 0.5 to 20 Gyr in increments of 1 Gyr (half the resolution of MIST grid), and there are 8 metallicities available spanning [Fe/H]$=-1.7$ to +0.6. The resulting age distribution is shown in Figure~\ref{fig:yale_hist}.

First, we note that the mean of the Yale-based age distribution is younger than MIST's by a non-negligible 1.3 Gyr. Second, we observe a larger number of young stars (below 5 Gyr) detected and a significantly diminished high-age tail above 12 Gyr in the Yale distribution. There is also a more prominent peak around 6 Gyr than in Figure \ref{fig:age_histogram}. However, per the lower panel of Figure \ref{fig:yale_hist}, the shape of the Yale distribution still differs noticeably from the distribution presented by \citet{Bensby17}; it is much closer to the distribution found with MIST.

\subsection{Ages with alpha--enhanced MIST isochrones}
\begin{figure}
\begin{center}
\includegraphics[width=\columnwidth]{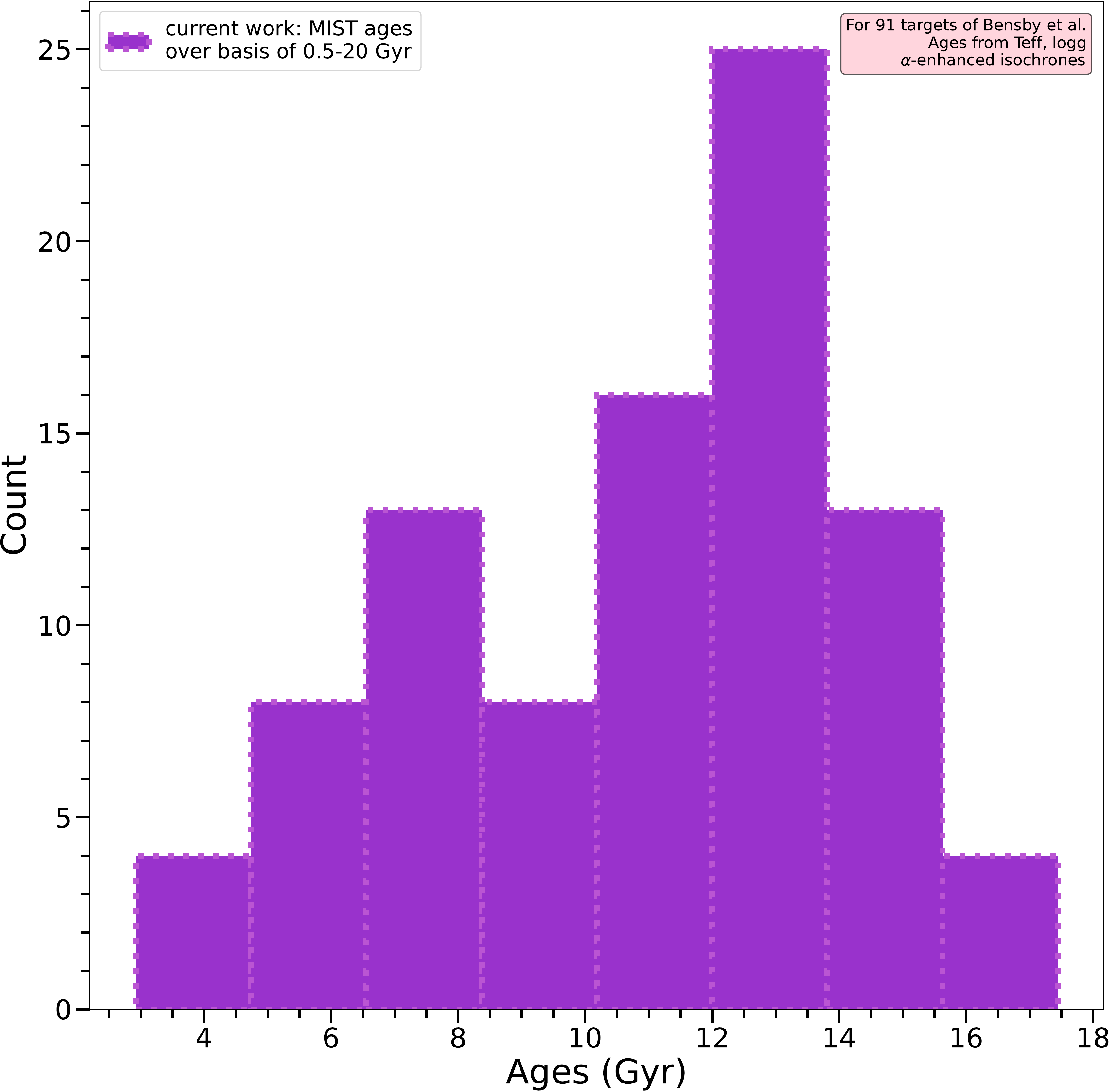}
\includegraphics[width=\columnwidth]{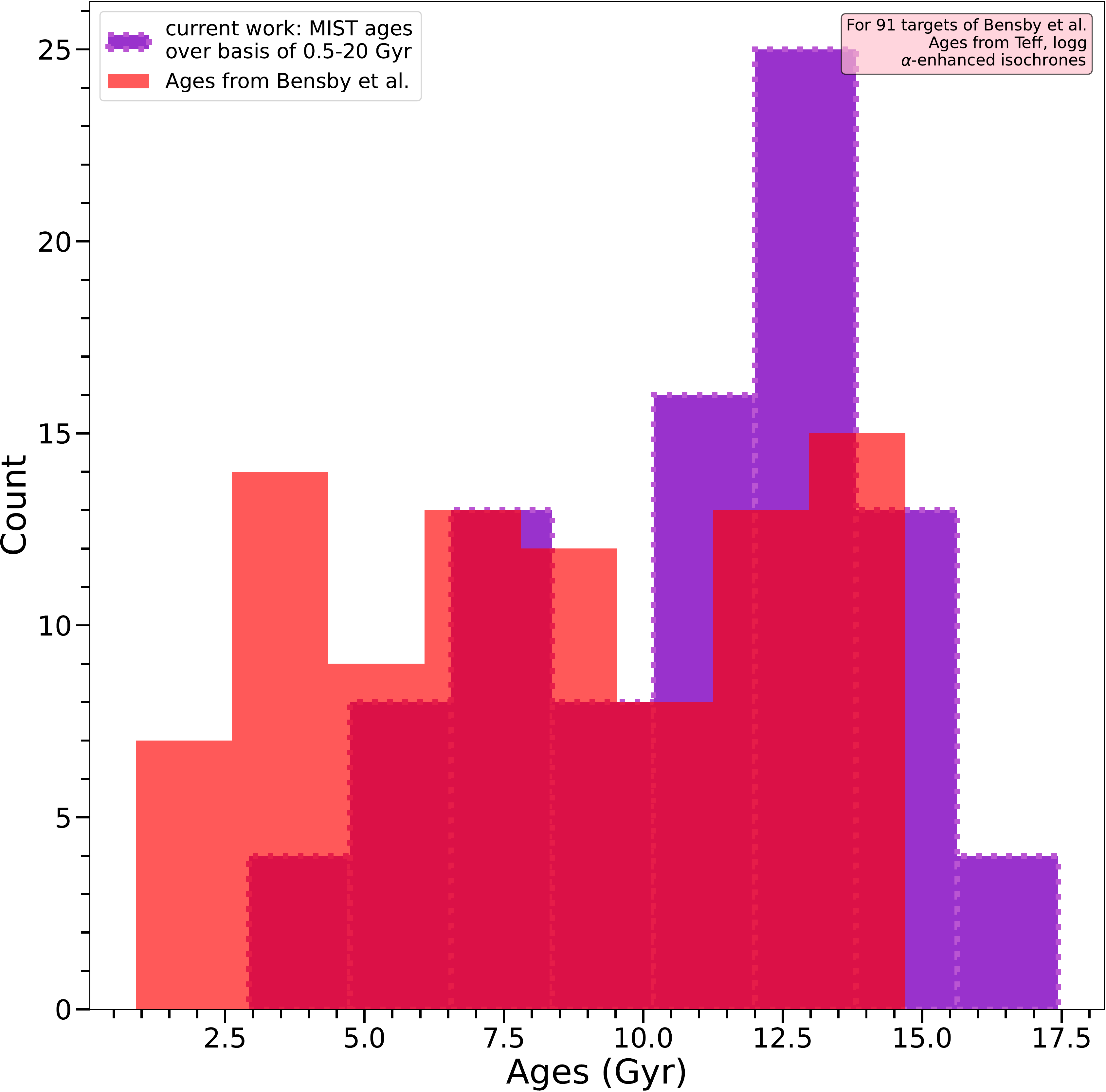}
\caption{\textbf{UPPER:} Same as Figure \ref{fig:age_histogram}, but assuming the $\alpha$-enhancement distribution of Table \ref{table:alpha_rescale} in the MIST isochrones. \textbf{LOWER:} The red foreground histogram shows results from \citet{Bensby17} overlaid on the histogram in the left panel.}
\label{fig:age_histogram_alpha}
\end{center}
\end{figure}

We quantify the effects of $\alpha$-element enhancement on stellar ages by recomputing the distribution presented in Figure \ref{fig:age_histogram} using the set of $\alpha$-enhanced MIST isochrones described in Section \ref{sec:alpha}. Figure \ref{fig:age_histogram_alpha} shows the resulting distribution.

The impact on the median age is small, producing an overall shift towards older ages of about 1\%: roughly 10.8 Gyr (no $\alpha$-enhancement) vs 10.9 Gyr (with $\alpha$-enhancement). However, the effect on the age bin near $8$--$9$ Gyr is substantial and makes more apparent one of the key differences between the age distribution derived in \citet{Bensby17} and the ones derived presently: Where \citet{Bensby17} find a three-peaked distribution---or, arguably, a near-uniform distribution\footnote{A Kolmogorov--Smirnov test confirms that \citet{Bensby17}'s distribution is in fact consistent with a uniform distribution: the null hypothesis cannot be rejected at the 5\% level.}--with an overabundance of young stars near 3 Gyr, fits with the age determination algorithm presented here reveal a distribution with one prominent peak around 13 Gyr, a subordinate peak around 8 Gyr (particularly in the fits of MIST isochrones with $\alpha$-enhancement and Yale isochrones) and a clear, monotonic decrease in count below $8$ Gyr. 

The shift of the distribution towards older ages, particularly for the oldest and most metal-poor stars that make up the most $\alpha$-enhanced population, is consistent with our expectations. We proceed with our analysis adopting the $\alpha$-basis MIST isochrones as the default physical configuration.

%
%
\subsection{The age--metallicity Distribution}

\begin{figure*}
\begin{center}
\includegraphics[width=\columnwidth]{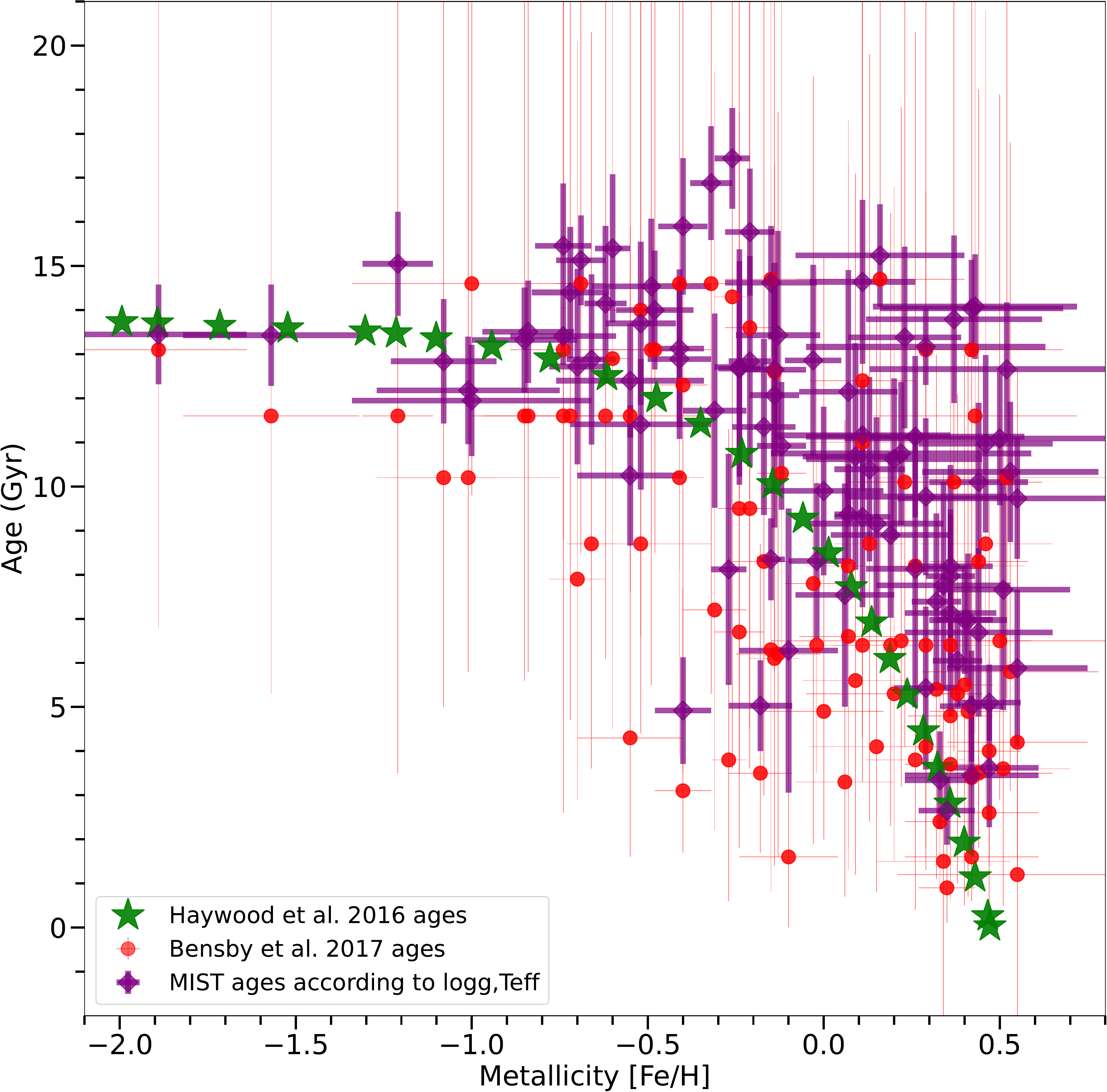}
\includegraphics[width=\columnwidth]{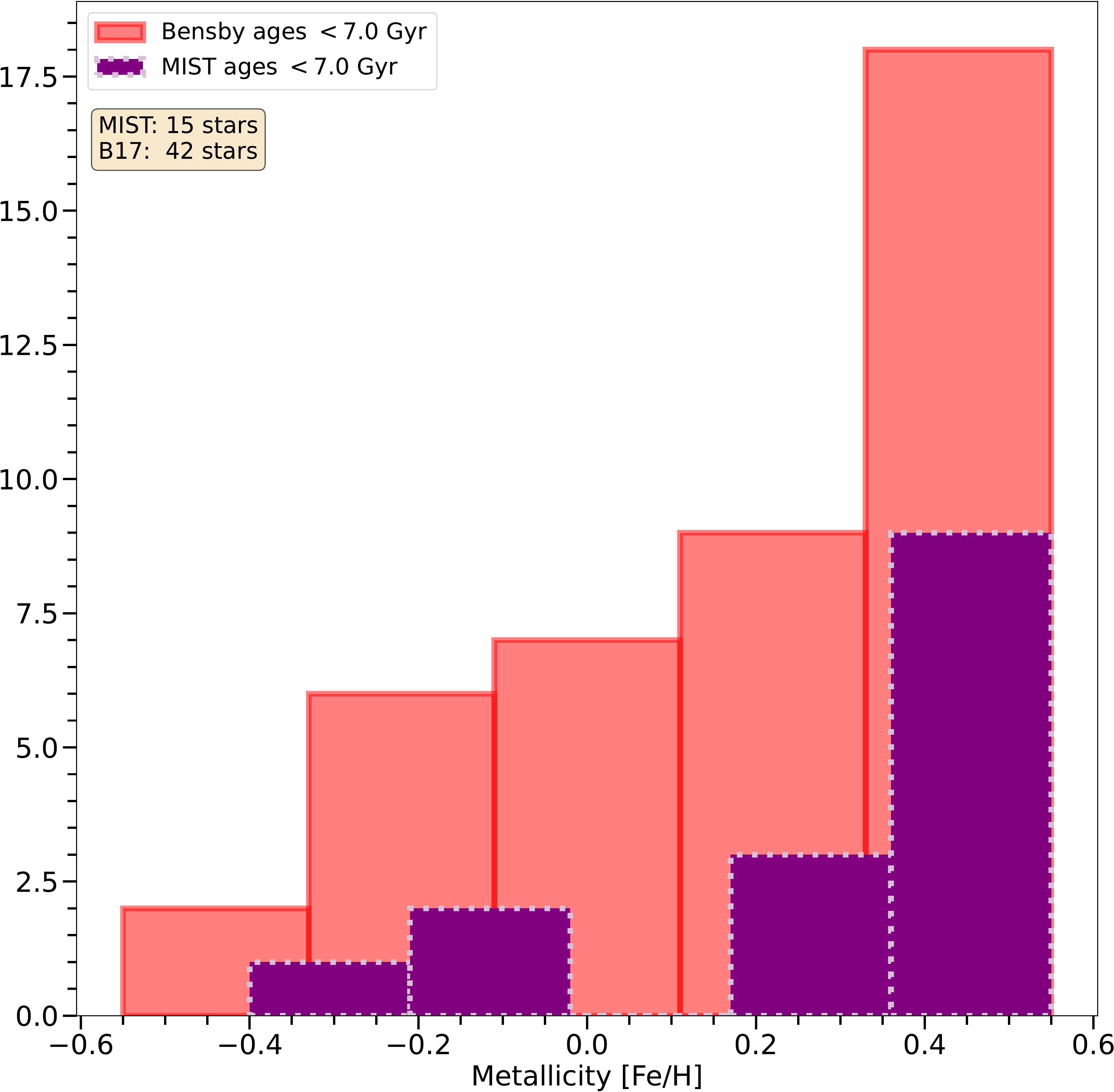}
\includegraphics[width=\columnwidth]{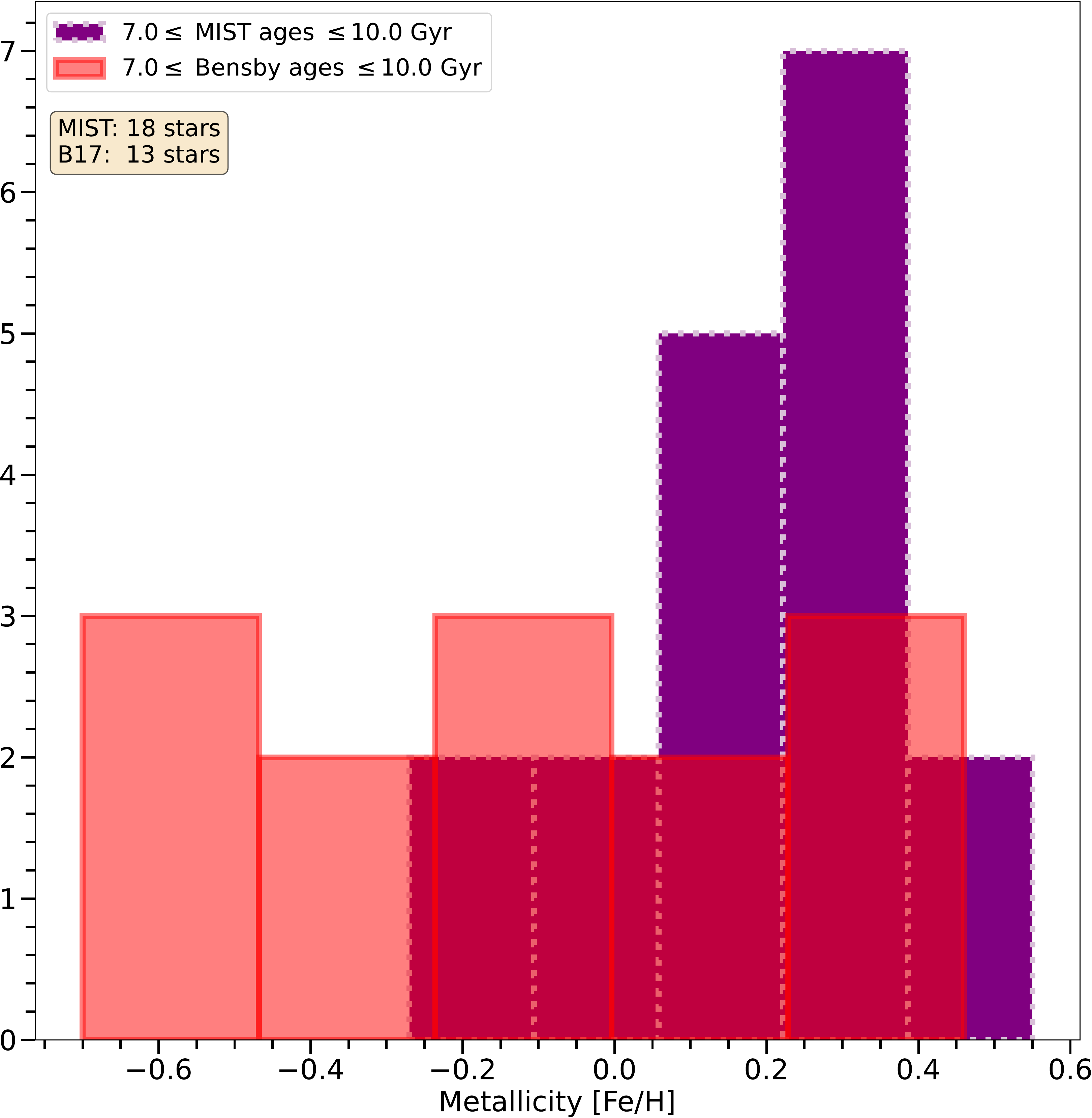}
\includegraphics[width=\columnwidth]{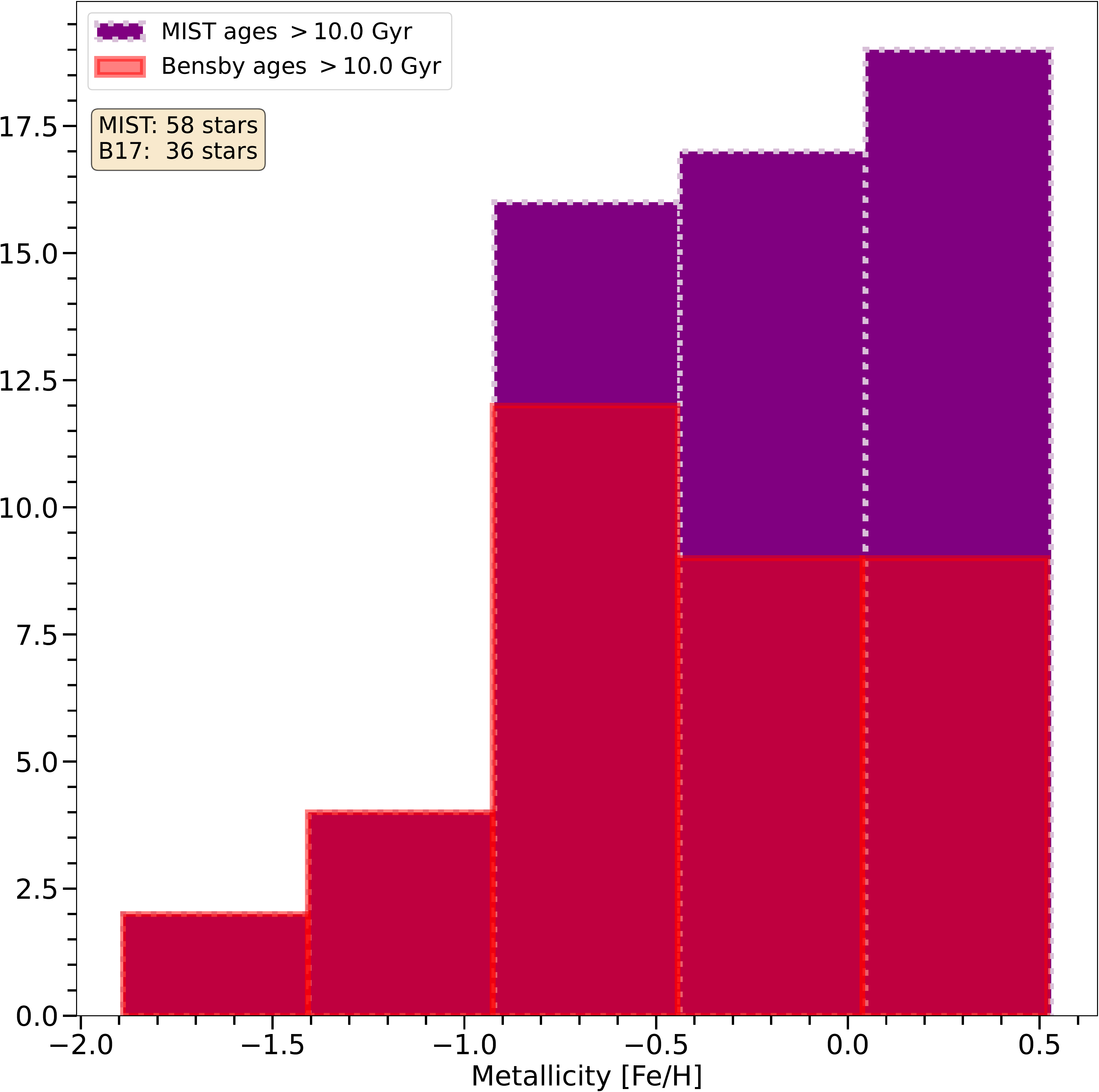}
\caption{
\textbf{UPPER LEFT} Age vs metallicity using logg, Teff and the $\alpha$-enhanced basis of isochrones, corresponding to the distribution shown in Figure \ref{fig:age_histogram_alpha}.
\textbf{UPPER RIGHT:} Histogram showing the distribution of metallicities [Fe/H] for the population of stars with ages less than 7.0 Gyr according to the ages determined in the current work (blue, solid borders) and \citet{Bensby17} (red, dashed borders), respectively. 
\textbf{LOWER LEFT:} Same as upper right, but for an intermediate age distribution as indicated.
\textbf{LOWER RIGHT:} Same as left, but for stars with ages greater than 10 Gyr according to the two age determinations, respectively. In all histogram panels, ``MIST ages" refers to the ages determined according to the $\alpha$-element enhanced basis of isochrones.
}
\label{fig:age_vs_met}
\end{center}
\end{figure*}

Figure \ref{fig:age_vs_met} compares the age--metallicity distribution found with MIST ($\alpha$-enhanced basis) to the distribution reported in \citet{Bensby17} as well as to the age-metallicity relation from \citet{Haywood16}. The upper left panel shows age vs [Fe/H] for the three data sets. The purple diamonds (MIST) suggest a characteristic broken power law, with ages roughly constant below [Fe/H]$=-0.5$ but declining with increasing metallicity above [Fe/H]$=-0.5$. The red circles (B17), on the other hand, show a large degree of age scatter at all but the lowest metallicities, and it would be difficult to claim a broken power law describes the data. A comparison between our new MIST-based age-metallicity relation and the relation from \citet{Haywood16} (green stars) shows that the present work extends the population of old ($>$ 10 Gyr) bulge stars to higher [Fe/H] than is predicted by the \citet{Haywood16} model.  Our results are in better agreement with the age-metallicity relation from \citet[][see also their Figure 10]{Bernard18}.

The remaining three panels break the distribution into metallicity histograms: the upper right shows the youngest age bin, $<7.0$ Gyr, the lower left shows $7 \le t \le 10$, and the lower right shows $t > 10$ Gyr. The youngest and intermediate panels provide a clear picture of where the distributions differ most: First, there is a large discrepancy in the number of young stars identified at all---this is most apparent in the $< 7$ \textcolor{blue}{Gyr} panel, where B17 report 42 of 91 stars with ages less than $7$ Gyr (nearly 50\%), as compared to our finding of 15 young stars (16\%). Second, though the intermediate-age panel shows similar numbers of stars identified in this age range by both MIST and B17, these are spread across a much larger range in metallicity in \citet{Bensby17}'s case. The distributions in the oldest panel are reasonably consistent, though MIST finds that fully two-thirds (66\%) of the stars have ages $>10$ Gyr compared to B17's 40\%.

\begin{figure}
\begin{center}
\includegraphics[width=\columnwidth]{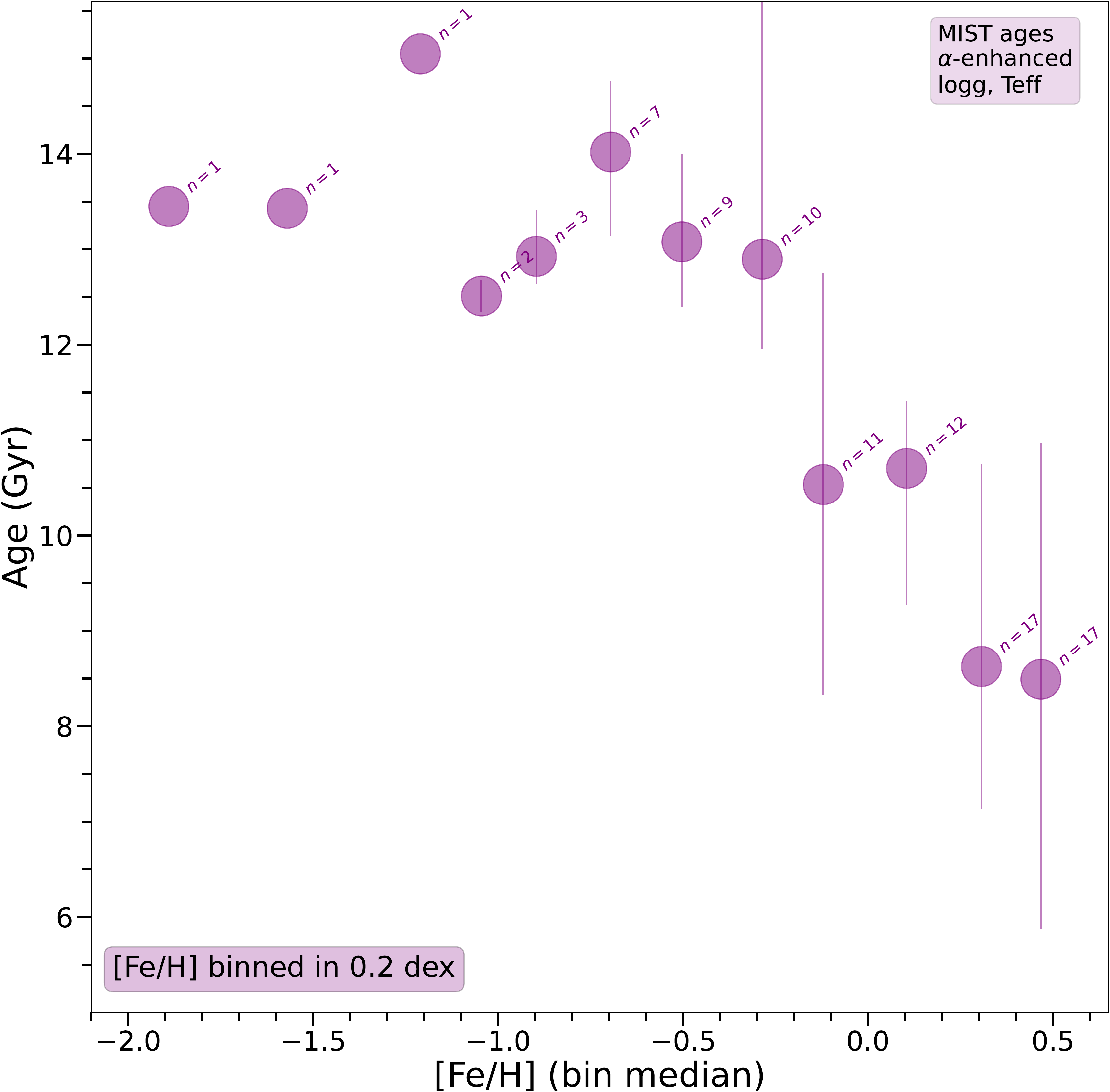}
\caption{Ages derived with MIST $\alpha$-enhanced isochrones are shown as a function of metallicity. Ages represent the average age within the corresponding metallicity bin, where each metallicity bin spans $0.2$ dex. The asymmetric vertical bars represent the age dispersion in a given bin: the first and fourth quartiles are the upper and lower segments, respectively. Points without dispersions correspond to bins in which there is only one age measurement. The number of age measurements in each bin is given by $n$ next to the corresponding marker.}
\label{fig:age_dispersions}
\end{center}
\end{figure}

Figure \ref{fig:age_dispersions} presents this relationship another way: binned average age is shown as a function of metallicity, with number of stars in each bin annotated. The data show an increase in age dispersion with increasing metallicity, which is not consistent with (primarily photometry--based) claims that the stellar ages display no metallicity dependence. However, this trend also does not suggest a large constituency of very young stars.
Another feature apparent in the upper left panel of Figure \ref{fig:age_vs_met} is the size of the age uncertainties reported by B17 relative to MIST's, with the MIST age uncertainties being significantly reduced relative to those presented in B17. We move now to a rigorous discussion of the calculation of the age uncertainties presented here.  

\section{Age Uncertainty Analysis}
\label{sec:error_analysis}
The method described in Section \ref{sec:statistical_analysis} produces an age estimate for a star according to a particular basis of isochrones. However, it does not provide a straightforward way of estimating the uncertainty in the $t_S$ of Equation \ref{eq:weighted_age} due to limitations in our knowledge of the observational data set.\footnote{Though formal statistics provides a standard definition for the variance in Equation \ref{eq:weighted_age}, the validity of that definition hinges on features that our algorithm and data set cannot satisfy: first, that $t_0,\ldots,t_n$ be normally distributed---they are not, as these are a set of uniformly spaced candidate ages. Second, all contributing uncertainties must be dimensionless and log-free. They are not, and in some cases we do not have enough information to convert them to the appropriate coordinates.}. Therefore, to estimate the variance ($1 \sigma$ error) in $t_S$, we use a Monte Carlo (MC) technique inspired by discussion in \citet{Angus2019} and thoroughly detailed in \citet{ErrorEstAstronomy}.

To find the observational age uncertainty for star $S$, we adopt as our starting assumption the effective temperature, gravity, metallicity, $1 \sigma$ temperature uncertainty, $1 \sigma \log g$ uncertainty, and $1 \sigma$ [Fe/H] uncertainty reported for $S$ by \citet{Bensby17}. From these, we construct three independent normal distributions with densities
\begin{equation}
\nonumber
f(x) = \frac1{\sigma\sqrt{2\pi}} e^{-(x-\mu)^2/2\sigma^2}
\end{equation}
defined by 
\begin{alignat}{2}
\nonumber
\mu &= T_{\text{eff},S},&\quad \sigma &= \sigma_{T_{\text{eff},S }},  
\\
\label{eq:normal_dist}
\mu &= \log g_{S}, &\quad \sigma &= \sigma_{\log g_S} ,
\\\mu &= Z_{S}, &\quad \sigma &= \sigma_{Z_S} 
\nonumber
\end{alignat}
for temperature, $\log g$, and [Fe/H], respectively. 

In MC resampling, we repeat the age determination 100 times. We calculate the weighted average for one star using 100 independent samples of the parameters $T_{\text{eff}}, \log g$, and [Fe/H] from the distributions of Equation \ref{eq:normal_dist} in place of $\log g_o, T_{\text{eff}, o}$, and $Z_{o}$, respectively, in Equation \ref{eq:chisq_Bensby}, but always use the B17 uncertainties. This gives a spread of age estimates that we could have seen if \citet{Bensby17}'s measurements had varied according to their uncertainties; an example is given for MOA-2013-BLG-402S in Figure~\ref{fig:MC_example}, and similar histograms for all stars can be found in the Appendix. The distribution in Figure \ref{fig:MC_example} has a standard deviation (root variance) of $\sigma = 1.83$~Gyr for MOA-2013-BLG-402S. In general, this technique gives our estimate of the standard error ($1 \sigma$ uncertainty) in our age estimate, $t_S$.

We note that as the number of MC trials increases, the distribution of the
age estimates should approach a normal distribution with mean equal to our age
estimate,\footnote{If the variables in the $\chi^2$ definition are indeed normally distributed, then the mean of the Monte Carlo age distribution, $\mu_{MC}$, should approach the predicted age, $t_S$, as the number of trials increases.} $t_S = 10.62$ in the case of MOA-2013-BLG-402S. We see good agreement ($\mu_{\text{simulated}} = 10.66$ vs $t_S = 10.62$, within 0.5\%) at 100 trials.

\begin{figure}
\begin{center}
\includegraphics[width=\columnwidth]{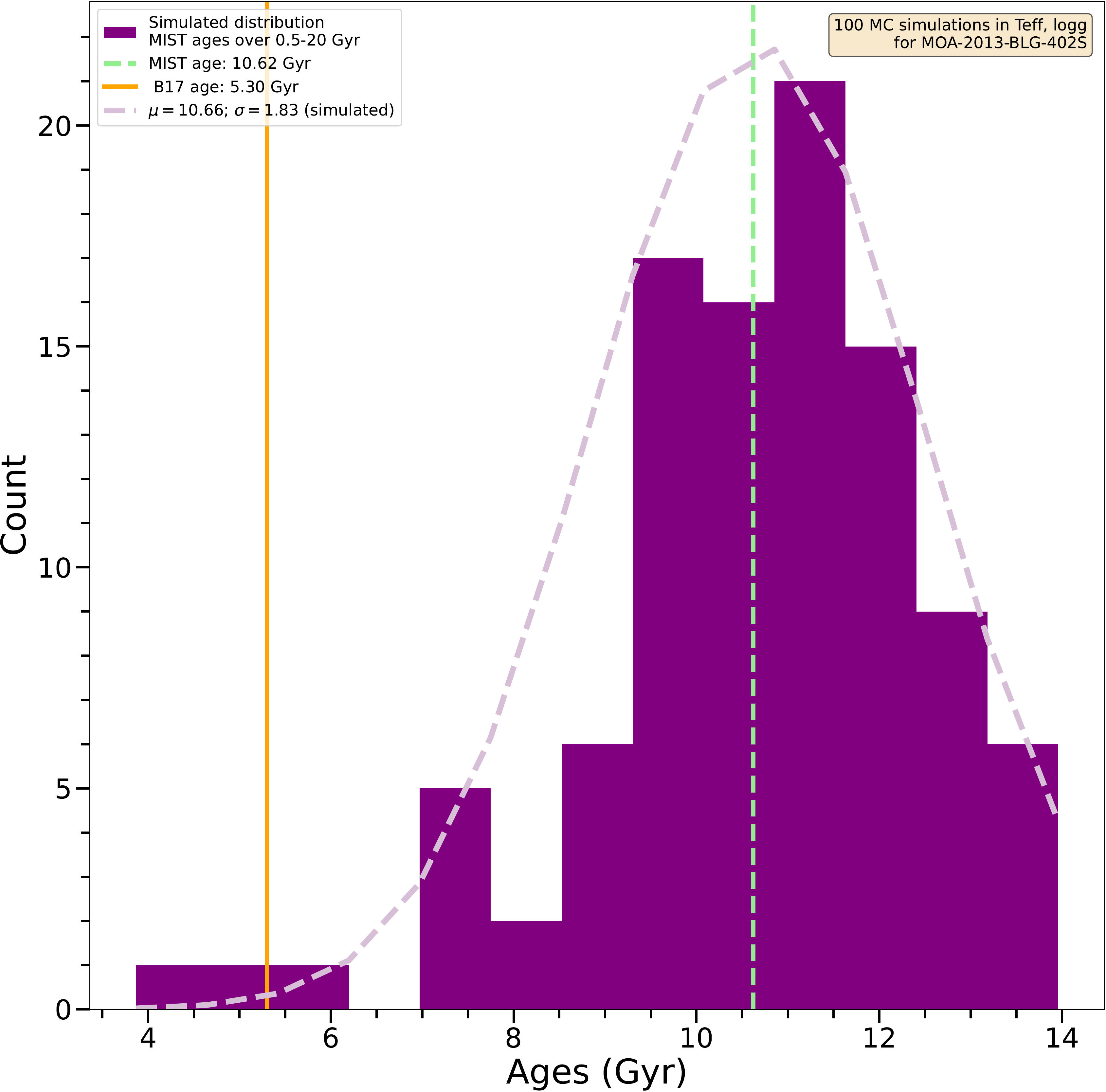}
\caption{A MIST-based age distribution simulated from 100 Monte Carlo trials in physical coordinate (logg, Teff) for the target MOA-2013-BLG-402S is shown. The variance (second moment, $\sigma^2$) of this distribution provides a reasonable estimate for the age uncertainty. }
\label{fig:MC_example}
\end{center}
\end{figure}

It is important to note that a valid age determination technique must be basis-invariant; that is, it must work independently of the color--magnitude or brightness--temperature system in which the isochrone fits are performed. Our technique satisfies this requirement, as demonstrated through its application to photometry, described in Appendix \ref{sec:photometric_fits}.  

\pagebreak
\section{Comparison of age determinations}
\label{sec:comparison_of_age_determinations}
Table \ref{table:mega_age_results} summarizes all age determinations: the MIST-based ages according to the default and $\alpha$-enhanced isochrone bases as well as the BDBS-based and \textit{Gaia}-based photometric ages discussed in Appendix \ref{sec:photometric_fits}. The age determinations from \citet{Bensby17} are reproduced for comparison. All ages and uncertainties are in Gyr; [Fe/H] and its uncertainty (in dex) are taken from \citet{Bensby17} and adopted in all age determinations for all data sets, physical and photometric. Where observations for a particular star are not available in a given photometric systems, the entry is marked ``NA.'' Membership in either or both of the BDBS and {\it Gaia} gold samples is indicated with markers at the end of the star's name. In all four systems, age uncertainties are estimated via the Monte Carlo technique described in Section \ref{sec:modeling_uncertainty} using 100 trials.

The two most important caveats on results thus far are the following:
\begin{enumerate}
    \item In all cases discussed previously, the age uncertainties are computed using \textit{observational sources of error only}; and
    \item the error bars are only physically meaningful in the spectroscopic cases. 
\end{enumerate}
Point (2) is discussed in the Appendix; we elaborate on point (1) in the next section.

Among the spectroscopic determinations, the ages derived according to default and $\alpha$-enhanced isochrone bases are consistent with each other. For a given star, the differences between them range from about 5\% - 10\%. Though this difference is not necessarily negligible, it is eclipsed by nearly a factor of two in the observational age uncertainty, which hovers around 1-2 Gyr, or roughly 20\%. This finding suggests that careful consideration of abundances in the models may not be worthwhile given existing spectroscopic precision, but such considerations could become important in light of modest observational uncertainty reductions. Age uncertainties of this order are consistent with other estimates for sub-giants and early RGB stars in the literature (e.g.\ \citealt{Tayar2022}), but they are not comparable to the upper and lower age limits quoted by \citet{Bensby17}. However, the relationship between the age and uncertainty estimation method presented here and the one used to derive the upper and lower age bounds in \citet{Bensby17} is unclear (per discussion in Section~ \ref{sec:alternative_stats}).

\begin{center}
{\footnotesize \begin{table*}
\centering 
\caption{Age Summary} 
\begin{tabular}{l l c c c cccc  c }  
\hline\hline 
 &	Name &	[Fe/H]	& B17 Age	& MIST Age & .. $\alpha$-enhanced  & ...model err & BDBS Age & Gaia Age  \\ \hline 
1 & MOA-2009-BLG-174S &  $0.11 \pm 0.08$ &  $6.4^{+9.7}_{-3.1}$ &  $8.9 \pm 2.0$ &  $9.3 \pm 2.0$ &  $9.3 \pm 4.0$ &  $9.3 \pm 1.4$  &  $10.0\pm 1.2$  \\
2 & MOA-2009-BLG-259S &  $0.34 \pm 0.19$ &  $1.5^{+8.9}_{-99.9}$ &  $7.2 \pm 2.4$ &  $7.8 \pm 2.4$ &  $7.8 \pm 3.8$ &  NA &  NA \\
3 & MOA-2010-BLG-167S $\bigstar$ &  $-0.60 \pm 0.05$ &  $12.9^{+13.5}_{-8.4}$ &  $15.3 \pm 1.7$ &  $15.4 \pm 1.7$ &  $15.4 \pm 3.8$ &  $9.5 \pm 1.5$  &  NA \\
4 & MOA-2010-BLG-311S &  $0.51 \pm 0.19$ &  $3.6^{+7.5}_{-3.1}$ &  $7.6 \pm 2.7$ &  $7.7 \pm 2.7$ &  $7.7 \pm 3.3$ &  $10.5 \pm 1.0$  &  NA \\
5 & MOA-2010-BLG-446S &  $-0.40 \pm 0.08$ &  $3.1^{+4.6}_{-1.4}$ &  $5.1 \pm 1.2$ &  $4.9 \pm 1.2$ &  $4.9 \pm 2.0$ &  NA &  NA \\
6 & MOA-2010-BLG-523S $\bigstar$ &  $0.06 \pm 0.14$ &  $3.3^{+8.2}_{-2.6}$ &  $7.4 \pm 2.5$ &  $7.5 \pm 2.5$ &  $7.5 \pm 3.8$ &  $8.9 \pm 1.8$  &  NA \\
7 & OGLE-2011-BLG-0950S $\bigstar$ {\bf \textdagger } &  $0.33 \pm 0.10$ &  $2.4^{+3.6}_{-1.6}$ &  $3.4 \pm 1.1$ &  $3.3 \pm 1.1$ &  $3.3 \pm 1.6$ &  $7.1 \pm 1.6$  &  $8.3\pm 2.2$  \\
8 & OGLE-2011-BLG-0969S $\bigstar$ &  $-1.57 \pm 0.25$ &  $11.6^{+13.2}_{-6.3}$ &  $13.6 \pm 1.1$ &  $13.4 \pm 1.1$ &  $13.4 \pm 1.5$ &  $10.1 \pm 1.4$  &  NA \\
9 & MOA-2011-BLG-034S &  $0.11 \pm 0.15$ &  $12.4^{+13.4}_{-8.1}$ &  $13.8 \pm 1.9$ &  $14.6 \pm 1.9$ &  $14.6 \pm 2.8$ &  $8.0 \pm 1.8$  &  NA \\
10 & MOA-2011-BLG-058S &  $0.37 \pm 0.25$ &  $10.1^{+12.9}_{-6.1}$ &  $13.6 \pm 1.9$ &  $13.8 \pm 1.9$ &  $13.8 \pm 2.8$ &  $11.5 \pm 1.0$  &  NA \\
11 & OGLE-2011-BLG-1072S &  $0.36 \pm 0.12$ &  $4.8^{+8.4}_{-4.1}$ &  $8.2 \pm 2.3$ &  $8.2 \pm 2.3$ &  $8.2 \pm 3.4$ &  NA &  NA \\
12 & MOA-2011-BLG-090S &  $-0.26 \pm 0.05$ &  $14.3^{+99.9}_{-10.7}$ &  $17.4 \pm 1.1$ &  $17.4 \pm 1.1$ &  $17.4 \pm 2.7$ &  $12.3 \pm 1.1$  &  NA \\
13 & MOA-2011-BLG-104S &  $-0.85 \pm 0.15$ &  $11.6^{+13.0}_{-6.0}$ &  $13.4 \pm 1.2$ &  $13.3 \pm 1.2$ &  $13.3 \pm 1.8$ &  NA &  NA \\
14 & OGLE-2011-BLG-1105S &  $0.00 \pm 0.17$ &  $4.9^{+10.3}_{-2.9}$ &  $9.7 \pm 1.9$ &  $9.9 \pm 1.9$ &  $9.9 \pm 2.9$ &  NA &  NA \\
15 & MOA-2011-BLG-174S &  $-0.18 \pm 0.09$ &  $3.5^{+5.2}_{-1.8}$ &  $5.5 \pm 1.0$ &  $5.0 \pm 1.0$ &  $5.0 \pm 2.5$ &  $9.5 \pm 1.7$  &  NA \\
16 & MOA-2011-BLG-191S &  $0.26 \pm 0.14$ &  $3.8^{+8.6}_{-3.4}$ &  $8.0 \pm 3.1$ &  $8.1 \pm 3.1$ &  $8.1 \pm 4.1$ &  NA &  NA \\
17 & MOA-2011-BLG-234S {\bf \textdagger } &  $-0.02 \pm 0.08$ &  $6.4^{+8.6}_{-2.9}$ &  $8.1 \pm 1.8$ &  $8.3 \pm 1.8$ &  $8.3 \pm 3.0$ &  $6.0 \pm 2.4$  &  $6.8\pm 2.2$  \\
18 & MOA-2011-BLG-278S &  $0.52 \pm 0.39$ &  $10.2^{+12.6}_{-5.0}$ &  $12.3 \pm 1.5$ &  $12.7 \pm 1.5$ &  $12.7 \pm 1.8$ &  NA &  NA \\
19 & OGLE-2011-BLG-1410S &  $0.22 \pm 0.37$ &  $6.5^{+12.1}_{-3.3}$ &  $10.7 \pm 1.6$ &  $10.8 \pm 1.6$ &  $10.8 \pm 2.4$ &  NA &  NA \\
20 & MOA-2011-BLG-445S &  $0.26 \pm 0.31$ &  $8.2^{+12.1}_{-4.0}$ &  $10.7 \pm 1.8$ &  $11.1 \pm 1.8$ &  $11.1 \pm 2.8$ &  $10.5 \pm 1.5$  &  NA \\
21 & MOA-2012-BLG-022S &  $0.42 \pm 0.09$ &  $3.4^{+5.1}_{-1.9}$ &  $5.1 \pm 1.3$ &  $5.0 \pm 1.3$ &  $5.0 \pm 2.2$ &  NA &  NA \\
22 & OGLE-2012-BLG-0026S &  $0.50 \pm 0.44$ &  $6.5^{+12.4}_{-3.6}$ &  $11.0 \pm 1.5$ &  $11.1 \pm 1.5$ &  $11.1 \pm 2.5$ &  NA &  NA \\
23 & MOA-2012-BLG-187S $\bigstar$ &  $-0.74 \pm 0.08$ &  $11.6^{+13.4}_{-9.0}$ &  $15.5 \pm 1.4$ &  $15.5 \pm 1.4$ &  $15.5 \pm 2.2$ &  $9.8 \pm 1.8$  &  NA \\
24 & MOA-2012-BLG-202S $\bigstar$ {\bf \textdagger } &  $-0.15 \pm 0.13$ &  $14.7^{+99.9}_{-8.5}$ &  $14.7 \pm 1.3$ &  $14.6 \pm 1.3$ &  $14.6 \pm 2.1$ &  $9.0 \pm 1.5$  &  $7.4\pm 2.0$  \\
25 & OGLE-2012-BLG-0211S &  $-0.03 \pm 0.08$ &  $7.8^{+11.5}_{-5.9}$ &  $12.5 \pm 2.2$ &  $12.9 \pm 2.2$ &  $12.9 \pm 3.5$ &  $8.7 \pm 1.6$  &  NA \\
26 & OGLE-2012-BLG-0270S &  $-0.84 \pm 0.13$ &  $11.6^{+13.0}_{-5.8}$ &  $13.6 \pm 1.2$ &  $13.5 \pm 1.2$ &  $13.5 \pm 2.0$ &  NA &  NA \\
27 & MOA-2012-BLG-291S &  $0.16 \pm 0.24$ &  $14.7^{+99.9}_{-7.9}$ &  $15.1 \pm 1.2$ &  $15.2 \pm 1.2$ &  $15.2 \pm 2.1$ &  $8.6 \pm 1.8$  &  NA \\
28 & MOA-2012-BLG-391S &  $-0.24 \pm 0.07$ &  $6.7^{+11.8}_{-4.9}$ &  $13.0 \pm 2.7$ &  $12.7 \pm 2.7$ &  $12.7 \pm 3.5$ &  NA &  NA \\
29 & MOA-2012-BLG-410S &  $-0.14 \pm 0.07$ &  $6.1^{+11.2}_{-4.7}$ &  $12.2 \pm 3.0$ &  $12.1 \pm 3.0$ &  $12.1 \pm 3.9$ &  NA &  NA \\
30 & OGLE-2012-BLG-0521S &  $0.09 \pm 0.15$ &  $5.6^{+11.5}_{-4.4}$ &  $10.2 \pm 2.6$ &  $10.7 \pm 2.6$ &  $10.7 \pm 3.7$ &  NA &  NA \\
31 & MOA-2012-BLG-532S $\bigstar$ &  $-0.55 \pm 0.21$ &  $11.6^{+12.6}_{-4.0}$ &  $12.7 \pm 1.3$ &  $12.4 \pm 1.3$ &  $12.4 \pm 1.6$ &  $9.6 \pm 1.6$  &  NA \\
32 & OGLE-2012-BLG-0563S &  $-0.66 \pm 0.07$ &  $8.7^{+11.6}_{-5.1}$ &  $12.9 \pm 1.9$ &  $12.9 \pm 1.9$ &  $12.9 \pm 3.6$ &  $9.0 \pm 1.8$  &  NA \\
33 & OGLE-2012-BLG-0617S $\bigstar$ &  $-0.14 \pm 0.09$ &  $12.6^{+13.3}_{-6.8}$ &  $13.0 \pm 1.9$ &  $12.7 \pm 1.9$ &  $12.7 \pm 3.2$ &  $9.3 \pm 1.5$  &  NA \\
34 & OGLE-2012-BLG-0816S &  $-0.10 \pm 0.14$ &  $1.6^{+6.4}_{-1.6}$ &  $6.3 \pm 3.2$ &  $6.3 \pm 3.2$ &  $6.3 \pm 3.4$ &  $12.1 \pm 1.1$  &  NA \\
35 & OGLE-2012-BLG-1156S &  $-1.89 \pm 0.25$ &  $13.1^{+13.4}_{-6.3}$ &  $13.5 \pm 1.1$ &  $13.4 \pm 1.1$ &  $13.4 \pm 1.6$ &  NA &  NA \\
36 & OGLE-2012-BLG-1217S &  $-0.41 \pm 0.07$ &  $10.2^{+12.4}_{-6.3}$ &  $13.2 \pm 1.8$ &  $13.1 \pm 1.8$ &  $13.1 \pm 3.1$ &  $10.7 \pm 1.2$  &  NA \\
37 & OGLE-2012-BLG-1274S &  $0.07 \pm 0.04$ &  $8.2^{+9.1}_{-6.9}$ &  $9.3 \pm 1.1$ &  $9.4 \pm 1.1$ &  $9.4 \pm 3.2$ &  $9.1 \pm 1.7$  &  NA \\
38 & OGLE-2012-BLG-1279S $\bigstar$ &  $-0.62 \pm 0.06$ &  $11.6^{+12.8}_{-5.5}$ &  $14.2 \pm 1.8$ &  $14.2 \pm 1.8$ &  $14.2 \pm 3.5$ &  $10.0 \pm 1.3$  &  NA \\
39 & OGLE-2012-BLG-1526S $\bigstar$ &  $-0.24 \pm 0.06$ &  $9.5^{+12.6}_{-6.0}$ &  $12.9 \pm 2.4$ &  $12.7 \pm 2.4$ &  $12.7 \pm 3.6$ &  $9.5 \pm 1.7$  &  NA \\
40 & OGLE-2012-BLG-1534S &  $-0.15 \pm 0.04$ &  $6.3^{+7.5}_{-5.5}$ &  $8.6 \pm 0.9$ &  $8.3 \pm 0.9$ &  $8.3 \pm 3.2$ &  NA &  NA \\
41 & MOA-2013-BLG-063S &  $0.47 \pm 0.14$ &  $2.6^{+3.5}_{-1.2}$ &  $3.7 \pm 1.4$ &  $3.6 \pm 1.4$ &  $3.6 \pm 1.8$ &  NA &  NA \\
42 & MOA-2013-BLG-068S &  $-0.27 \pm 0.05$ &  $3.8^{+7.5}_{-3.2}$ &  $8.1 \pm 2.6$ &  $8.1 \pm 2.6$ &  $8.1 \pm 4.5$ &  NA &  NA \\
43 & MOA-2013-BLG-299S &  $-0.48 \pm 0.11$ &  $13.1^{+13.1}_{-4.6}$ &  $14.2 \pm 1.4$ &  $14.0 \pm 1.4$ &  $14.0 \pm 2.1$ &  $9.2 \pm 1.4$  &  $9.4\pm 1.3$  \\
44 & MOA-2013-BLG-402S &  $0.20 \pm 0.25$ &  $5.3^{+11.5}_{-3.8}$ &  $10.2 \pm 1.8$ &  $10.6 \pm 1.8$ &  $10.6 \pm 2.8$ &  $7.9 \pm 2.0$  &  NA \\
45 & OGLE-2013-BLG-0446S &  $0.40 \pm 0.08$ &  $5.5^{+7.5}_{-5.0}$ &  $7.0 \pm 1.2$ &  $7.0 \pm 1.2$ &  $7.0 \pm 2.5$ &  $4.3 \pm 2.5$  &  NA \\
46 & MOA-2013-BLG-517S &  $-1.08 \pm 0.15$ &  $10.2^{+12.7}_{-5.2}$ &  $12.9 \pm 1.4$ &  $12.8 \pm 1.4$ &  $12.8 \pm 1.6$ &  $10.1 \pm 1.6$  &  NA \\
47 & MOA-2013-BLG-524S &  $-1.01 \pm 0.26$ &  $10.2^{+12.7}_{-4.4}$ &  $12.3 \pm 1.2$ &  $12.2 \pm 1.2$ &  $12.2 \pm 1.5$ &  NA &  $10.3\pm 1.0$  \\
48 & MOA-2013-BLG-605S &  $-0.17 \pm 0.09$ &  $8.3^{+12.8}_{-5.3}$ &  $11.9 \pm 2.0$ &  $11.3 \pm 2.0$ &  $11.3 \pm 5.0$ &  $9.6 \pm 1.5$  &  $8.6\pm 1.7$  \\
49 & OGLE-2013-BLG-0692S $\bigstar$ {\bf \textdagger } &  $0.15 \pm 0.19$ &  $4.1^{+9.8}_{-3.3}$ &  $8.6 \pm 2.4$ &  $9.2 \pm 2.4$ &  $9.2 \pm 3.8$ &  $9.6 \pm 1.4$  &  $9.2\pm 2.0$  \\
50 & OGLE-2013-BLG-0835S &  $0.53 \pm 0.25$ &  $5.8^{+12.0}_{-2.7}$ &  $10.5 \pm 1.6$ &  $10.3 \pm 1.6$ &  $10.3 \pm 2.0$ &  NA &  NA \\
\hline 
\end{tabular}
\label{table:mega_age_results}
\end{table*} 
\begin{table*} 
\def\thetable{2 (cont.)}
\centering 
\caption{Age Summary cont.} 
\begin{tabular}{l l c c c ccc  c}  
\hline\hline 
 &	Name &	[Fe/H]	& B17 Age	& MIST Age & .. $\alpha$-enhanced  & ...model err & BDBS Age & Gaia Age  \\ \hline 
51 & OGLE-2013-BLG-0911S &  $0.47 \pm 0.09$ &  $4.0^{+5.7}_{-3.6}$ &  $5.1 \pm 0.9$ &  $5.1 \pm 0.9$ &  $5.1 \pm 2.2$ &  NA &  NA \\
52 & OGLE-2013-BLG-1015S &  $0.36 \pm 0.07$ &  $6.4^{+8.2}_{-5.3}$ &  $8.0 \pm 1.4$ &  $8.0 \pm 1.4$ &  $8.0 \pm 3.5$ &  NA &  NA \\
53 & OGLE-2013-BLG-1114S &  $0.42 \pm 0.19$ &  $1.6^{+3.3}_{-1.0}$ &  $3.3 \pm 1.8$ &  $3.5 \pm 1.8$ &  $3.5 \pm 1.7$ &  NA &  NA \\
54 & OGLE-2013-BLG-1125S &  $-0.12 \pm 0.07$ &  $10.3^{+12.6}_{-4.2}$ &  $11.3 \pm 1.4$ &  $10.9 \pm 1.4$ &  $10.9 \pm 3.8$ &  NA &  NA \\
55 & OGLE-2013-BLG-1147S &  $0.29 \pm 0.09$ &  $4.1^{+5.7}_{-2.8}$ &  $5.4 \pm 1.9$ &  $5.4 \pm 1.9$ &  $5.4 \pm 2.6$ &  $10.7 \pm 1.4$  &  $9.9\pm 1.4$  \\
56 & OGLE-2013-BLG-1259S &  $-0.21 \pm 0.06$ &  $9.5^{+12.6}_{-5.9}$ &  $13.0 \pm 2.4$ &  $12.8 \pm 2.4$ &  $12.8 \pm 4.1$ &  $7.1 \pm 2.2$  &  $8.1\pm 1.8$  \\
57 & OGLE-2013-BLG-1768S &  $0.11 \pm 0.25$ &  $11.0^{+12.7}_{-4.0}$ &  $11.2 \pm 1.3$ &  $11.2 \pm 1.3$ &  $11.2 \pm 1.6$ &  NA &  NA \\
58 & OGLE-2013-BLG-1793S &  $0.32 \pm 0.07$ &  $5.4^{+7.6}_{-4.3}$ &  $7.4 \pm 2.0$ &  $7.4 \pm 2.0$ &  $7.4 \pm 3.3$ &  NA &  NA \\
59 & OGLE-2013-BLG-1868S $\bigstar$ &  $0.38 \pm 0.07$ &  $5.3^{+6.8}_{-4.3}$ &  $6.0 \pm 1.1$ &  $6.0 \pm 1.1$ &  $6.0 \pm 2.5$ &  $9.6 \pm 1.4$  &  $9.5\pm 1.6$  \\
60 & OGLE-2013-BLG-1938S {\bf \textdagger } &  $-0.41 \pm 0.09$ &  $14.6^{+99.9}_{-6.5}$ &  $12.9 \pm 1.8$ &  $12.9 \pm 1.8$ &  $12.9 \pm 4.6$ &  $9.0 \pm 1.8$  &  $8.7\pm 1.3$  \\
61 & MOA-2014-BLG-131S &  $0.29 \pm 0.34$ &  $13.1^{+99.9}_{-3.8}$ &  $13.2 \pm 0.9$ &  $13.2 \pm 0.9$ &  $13.2 \pm 1.3$ &  $9.1 \pm 1.9$  &  NA \\
62 & OGLE-2014-BLG-0157S $\bigstar$ {\bf \textdagger } &  $-0.52 \pm 0.10$ &  $14.0^{+99.9}_{-7.4}$ &  $13.7 \pm 1.9$ &  $13.7 \pm 1.9$ &  $13.7 \pm 3.8$ &  $10.0 \pm 1.1$  &  $10.3\pm 1.2$  \\
63 & OGLE-2014-BLG-0801S &  $0.55 \pm 0.34$ &  $1.2^{+11.5}_{-99.9}$ &  $10.1 \pm 1.4$ &  $9.7 \pm 1.4$ &  $9.7 \pm 2.6$ &  $9.4 \pm 1.6$  &  NA \\
64 & OGLE-2014-BLG-0953S &  $-0.13 \pm 0.12$ &  $6.2^{+12.3}_{-4.6}$ &  $13.3 \pm 2.4$ &  $13.4 \pm 2.4$ &  $13.4 \pm 3.0$ &  NA &  NA \\
65 & OGLE-2014-BLG-0987S &  $0.23 \pm 0.16$ &  $10.1^{+12.8}_{-6.0}$ &  $13.1 \pm 2.1$ &  $13.4 \pm 2.1$ &  $13.4 \pm 2.6$ &  $2.4 \pm 1.6$  &  NA \\
66 & OGLE-2014-BLG-1122S $\bigstar$ &  $-0.32 \pm 0.06$ &  $14.6^{+99.9}_{-9.3}$ &  $16.9 \pm 1.3$ &  $16.9 \pm 1.3$ &  $16.9 \pm 2.8$ &  $9.3 \pm 1.6$  &  NA \\
67 & OGLE-2014-BLG-1370S &  $0.43 \pm 0.29$ &  $11.6^{+13.3}_{-7.2}$ &  $13.8 \pm 1.2$ &  $14.1 \pm 1.2$ &  $14.1 \pm 2.6$ &  NA &  NA \\
68 & OGLE-2014-BLG-1418S &  $0.44 \pm 0.14$ &  $8.3^{+10.7}_{-6.2}$ &  $10.1 \pm 1.8$ &  $10.1 \pm 1.8$ &  $10.1 \pm 3.5$ &  $8.9 \pm 1.9$  &  $9.8\pm 0.9$  \\
69 & OGLE-2014-BLG-1469S &  $-0.70 \pm 0.08$ &  $7.9^{+12.2}_{-5.0}$ &  $12.7 \pm 2.2$ &  $12.7 \pm 2.2$ &  $12.7 \pm 3.5$ &  NA &  NA \\
70 & OGLE-2014-BLG-2040S &  $0.19 \pm 0.17$ &  $6.4^{+9.8}_{-4.1}$ &  $8.2 \pm 1.9$ &  $8.9 \pm 1.9$ &  $8.9 \pm 2.8$ &  NA &  NA \\
71 & OGLE-2015-BLG-0078S &  $-0.55 \pm 0.15$ &  $4.3^{+11.6}_{-2.7}$ &  $10.5 \pm 1.6$ &  $10.2 \pm 1.6$ &  $10.2 \pm 2.9$ &  NA &  NA \\
72 & MOA-2015-BLG-111S &  $0.07 \pm 0.14$ &  $6.6^{+11.7}_{-4.8}$ &  $11.9 \pm 2.8$ &  $12.2 \pm 2.8$ &  $12.2 \pm 3.6$ &  NA &  NA \\
73 & OGLE-2015-BLG-0159S $\bigstar$ &  $0.36 \pm 0.13$ &  $3.7^{+7.3}_{-3.1}$ &  $7.1 \pm 2.4$ &  $7.1 \pm 2.4$ &  $7.1 \pm 3.6$ &  $7.9 \pm 2.1$  &  NA \\
74 & MOA-2010-BLG-037S &  $0.55 \pm 0.20$ &  $4.2^{+6.2}_{-3.1}$ &  $5.8 \pm 1.8$ &  $5.9 \pm 1.8$ &  $5.9 \pm 2.1$ &  NA &  NA \\
75 & MOA-2010-BLG-049S &  $-0.40 \pm 0.07$ &  $12.3^{+13.4}_{-8.8}$ &  $15.9 \pm 1.6$ &  $15.9 \pm 1.6$ &  $15.9 \pm 3.2$ &  $9.6 \pm 1.3$  &  NA \\
76 & OGLE-2009-BLG-076S &  $-0.72 \pm 0.11$ &  $11.6^{+13.3}_{-6.9}$ &  $14.4 \pm 1.5$ &  $14.4 \pm 1.5$ &  $14.4 \pm 2.3$ &  $10.2 \pm 1.5$  &  NA \\
77 & MOA-2010-BLG-078S &  $-1.00 \pm 0.34$ &  $14.6^{+99.9}_{-4.8}$ &  $12.0 \pm 1.3$ &  $11.9 \pm 1.3$ &  $11.9 \pm 1.5$ &  NA &  NA \\
78 & MOA-2006-BLG-099S &  $0.44 \pm 0.21$ &  $3.5^{+7.0}_{-1.7}$ &  $6.5 \pm 1.9$ &  $6.7 \pm 1.9$ &  $6.7 \pm 2.7$ &  NA &  NA \\
79 & MOA-2009-BLG-133S $\bigstar$ &  $-0.69 \pm 0.07$ &  $14.6^{+99.9}_{-6.1}$ &  $15.2 \pm 1.0$ &  $15.1 \pm 1.0$ &  $15.1 \pm 1.9$ &  $11.3 \pm 1.4$  &  NA \\
80 & OGLE-2008-BLG-209S $\bigstar$ {\bf \textdagger } &  $-0.31 \pm 0.09$ &  $7.2^{+12.2}_{-5.0}$ &  $12.0 \pm 2.2$ &  $11.7 \pm 2.2$ &  $11.7 \pm 3.4$ &  $9.1 \pm 1.6$  &  $11.0\pm 1.3$  \\
81 & OGLE-2006-BLG-265S &  $0.46 \pm 0.19$ &  $8.7^{+12.1}_{-5.8}$ &  $10.7 \pm 2.0$ &  $11.0 \pm 2.0$ &  $11.0 \pm 3.1$ &  $7.9 \pm 1.9$  &  NA \\
82 & MOA-2010-BLG-285S $\bigstar$ {\bf \textdagger } &  $-1.21 \pm 0.10$ &  $11.6^{+13.4}_{-8.1}$ &  $15.1 \pm 1.2$ &  $15.1 \pm 1.2$ &  $15.1 \pm 1.7$ &  $11.2 \pm 1.4$  &  $9.3\pm 1.1$  \\
83 & MOA-2008-BLG-310S &  $0.41 \pm 0.11$ &  $4.9^{+8.2}_{-4.2}$ &  $7.0 \pm 1.5$ &  $7.0 \pm 1.5$ &  $7.0 \pm 2.8$ &  $1.4 \pm 1.2$  &  NA \\
84 & MOA-2008-BLG-311S &  $0.35 \pm 0.08$ &  $0.9^{+2.9}_{-0.8}$ &  $2.9 \pm 0.8$ &  $2.6 \pm 0.8$ &  $2.6 \pm 1.7$ &  $12.1 \pm 1.1$  &  NA \\
85 & OGLE-2007-BLG-349S &  $0.42 \pm 0.26$ &  $13.1^{+13.5}_{-6.3}$ &  $13.8 \pm 1.1$ &  $14.0 \pm 1.1$ &  $14.0 \pm 2.2$ &  NA &  NA \\
86 & MOA-2009-BLG-456S &  $0.13 \pm 0.10$ &  $8.7^{+11.1}_{-6.3}$ &  $9.7 \pm 2.1$ &  $10.4 \pm 2.1$ &  $10.4 \pm 3.4$ &  $5.3 \pm 2.8$  &  NA \\
87 & MOA-2009-BLG-475S &  $-0.52 \pm 0.20$ &  $8.7^{+12.3}_{-4.3}$ &  $11.9 \pm 1.5$ &  $11.4 \pm 1.5$ &  $11.4 \pm 2.1$ &  NA &  NA \\
88 & MOA-2009-BLG-489S &  $-0.21 \pm 0.07$ &  $13.6^{+13.8}_{-9.1}$ &  $16.3 \pm 1.4$ &  $15.8 \pm 1.4$ &  $15.8 \pm 3.8$ &  NA &  NA \\
89 & MOA-2009-BLG-493S &  $-0.74 \pm 0.15$ &  $13.1^{+99.9}_{-4.3}$ &  $13.7 \pm 0.7$ &  $13.4 \pm 0.7$ &  $13.4 \pm 1.2$ &  NA &  NA \\
90 & OGLE-2007-BLG-514S &  $0.29 \pm 0.23$ &  $6.4^{+10.3}_{-4.6}$ &  $8.9 \pm 1.8$ &  $9.8 \pm 1.8$ &  $9.8 \pm 2.5$ &  NA &  NA \\
91 & MACHO-1999-BLG-022S &  $-0.49 \pm 0.17$ &  $13.1^{+13.5}_{-7.6}$ &  $14.9 \pm 1.5$ &  $14.5 \pm 1.5$ &  $14.5 \pm 2.1$ &  $11.0 \pm 1.5$  &  NA \\
\hline 
\end{tabular}
\begin{flushleft}
\small{All ages shown in Gyr. ``MIST Age'' refers to the re-fitting of \citet{Bensby17}'s own physical coordinates (Teff, logg) in the present analysis. The $\bigstar$ symbol indicates membership in the BDBS gold sample; \textdagger indicates membership in the Gaia gold sample. We strictly emphasize that the uncertainties reported for the photometric results (BDBS, \textit{Gaia}) are not physical (see discussion in Appendix); they are simply the results of simulations performed with insufficiently precise observational constraints. Only the uncertainties associated with spectroscopic age determinations should be interpreted as physical, and only those that further account for modeling uncertainties should be taken as meaningful lower limits on real age uncertainties.}
\end{flushleft}
\end{table*} }

\end{center}


\section{Modeling Uncertainties}
\label{sec:modeling_uncertainty}
\begin{figure*}
\begin{center}
\includegraphics[width=\columnwidth]{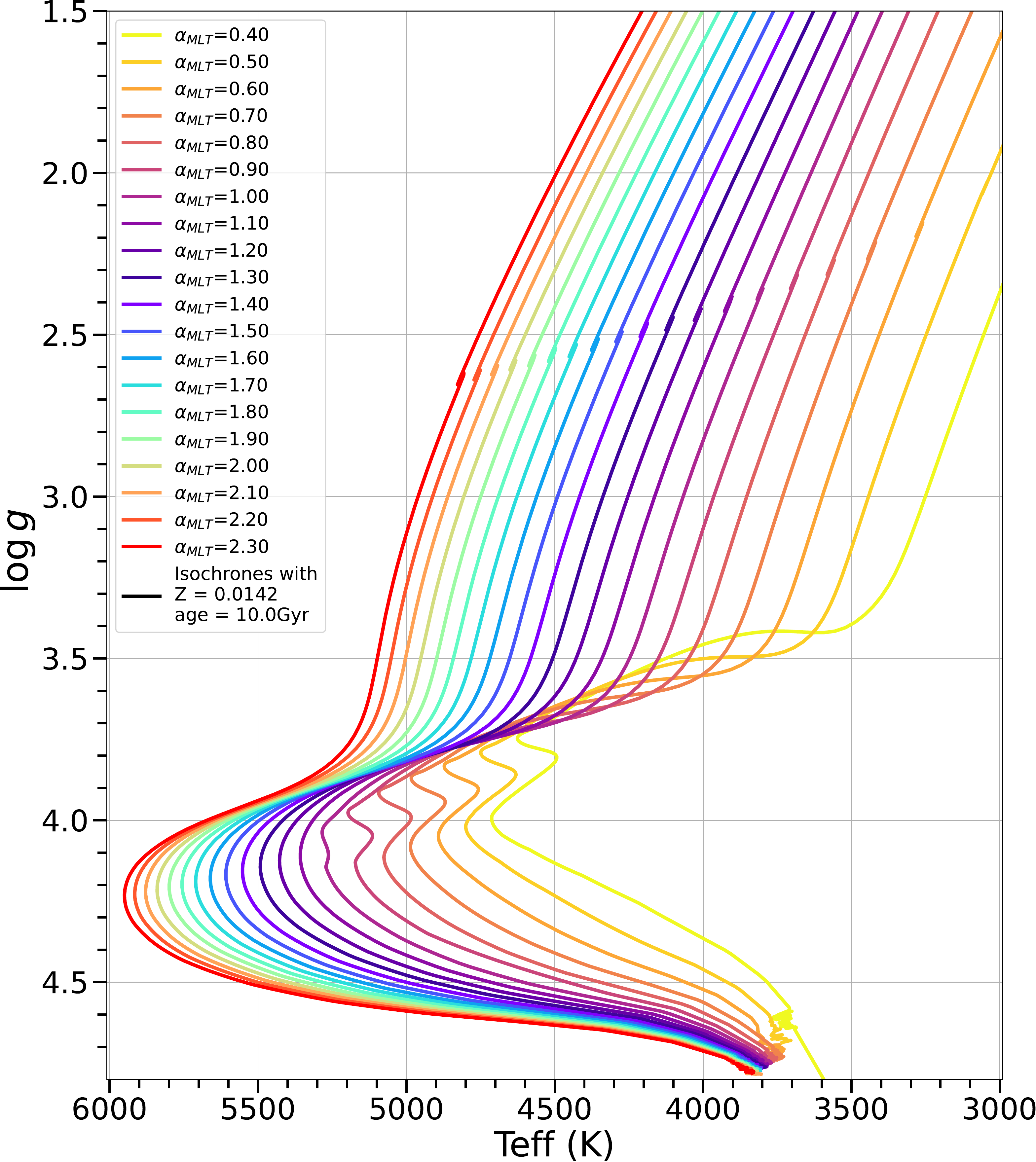}
\includegraphics[width=\columnwidth]{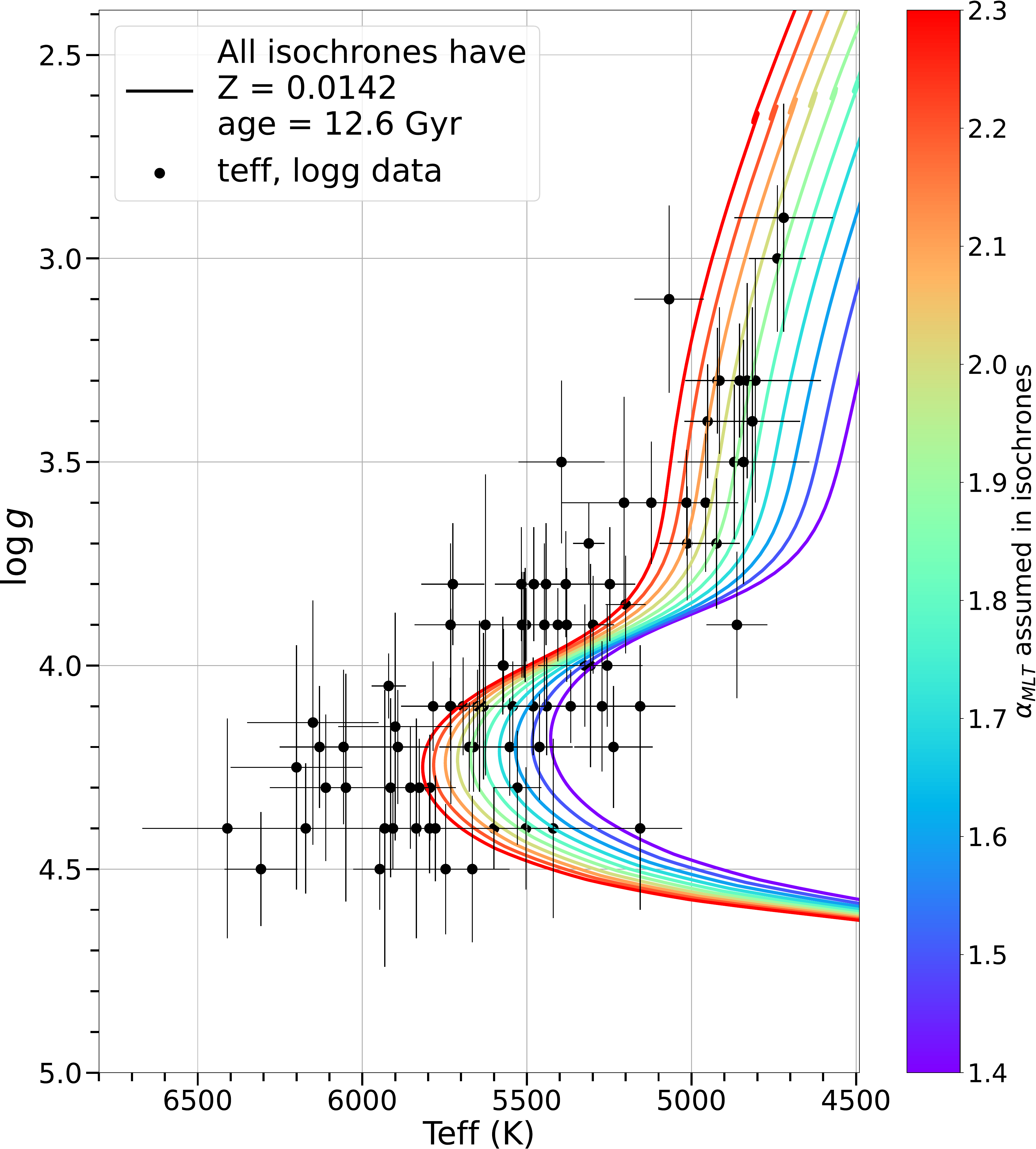}
\includegraphics[width=\columnwidth]{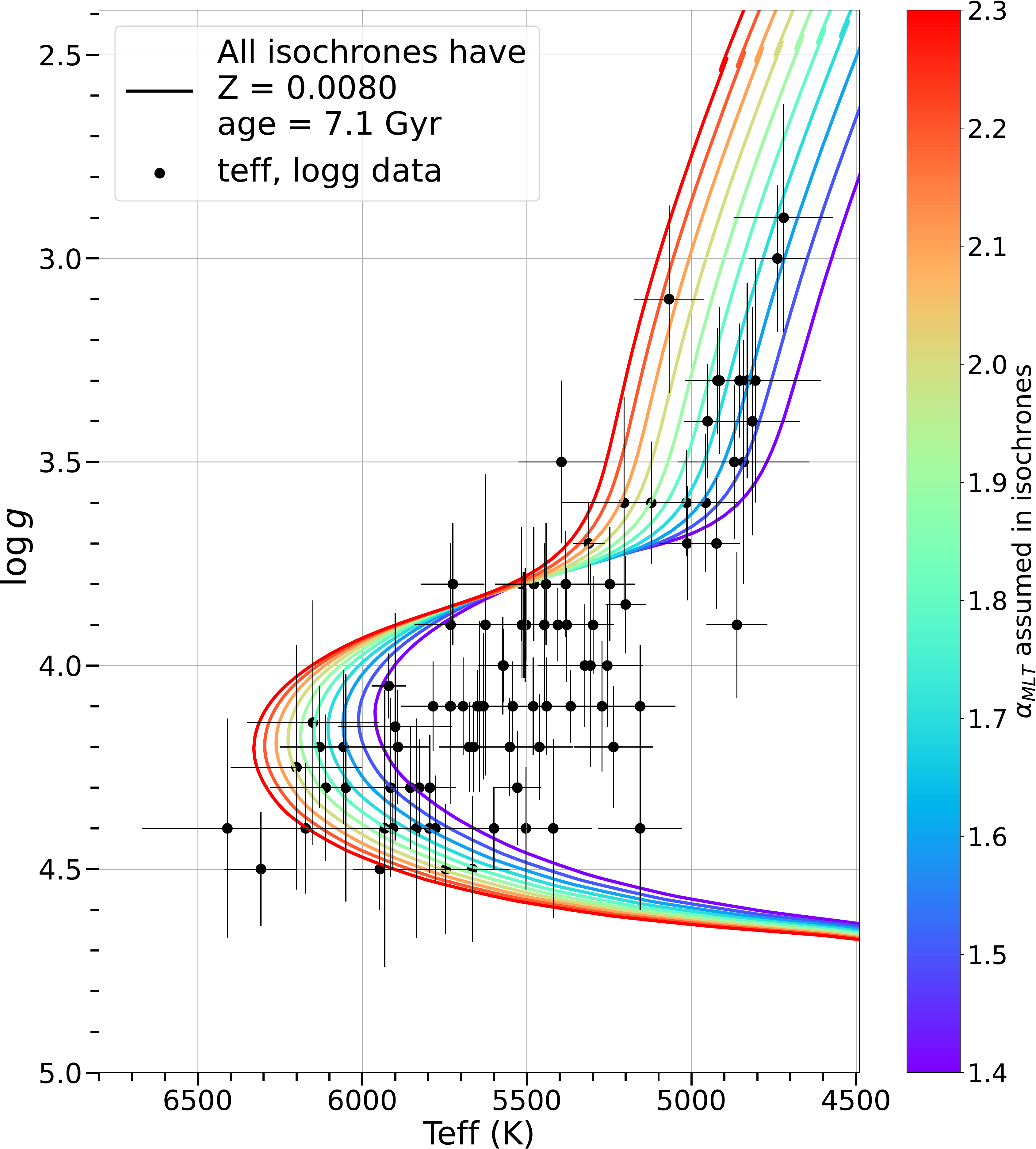}
\includegraphics[width=\columnwidth]{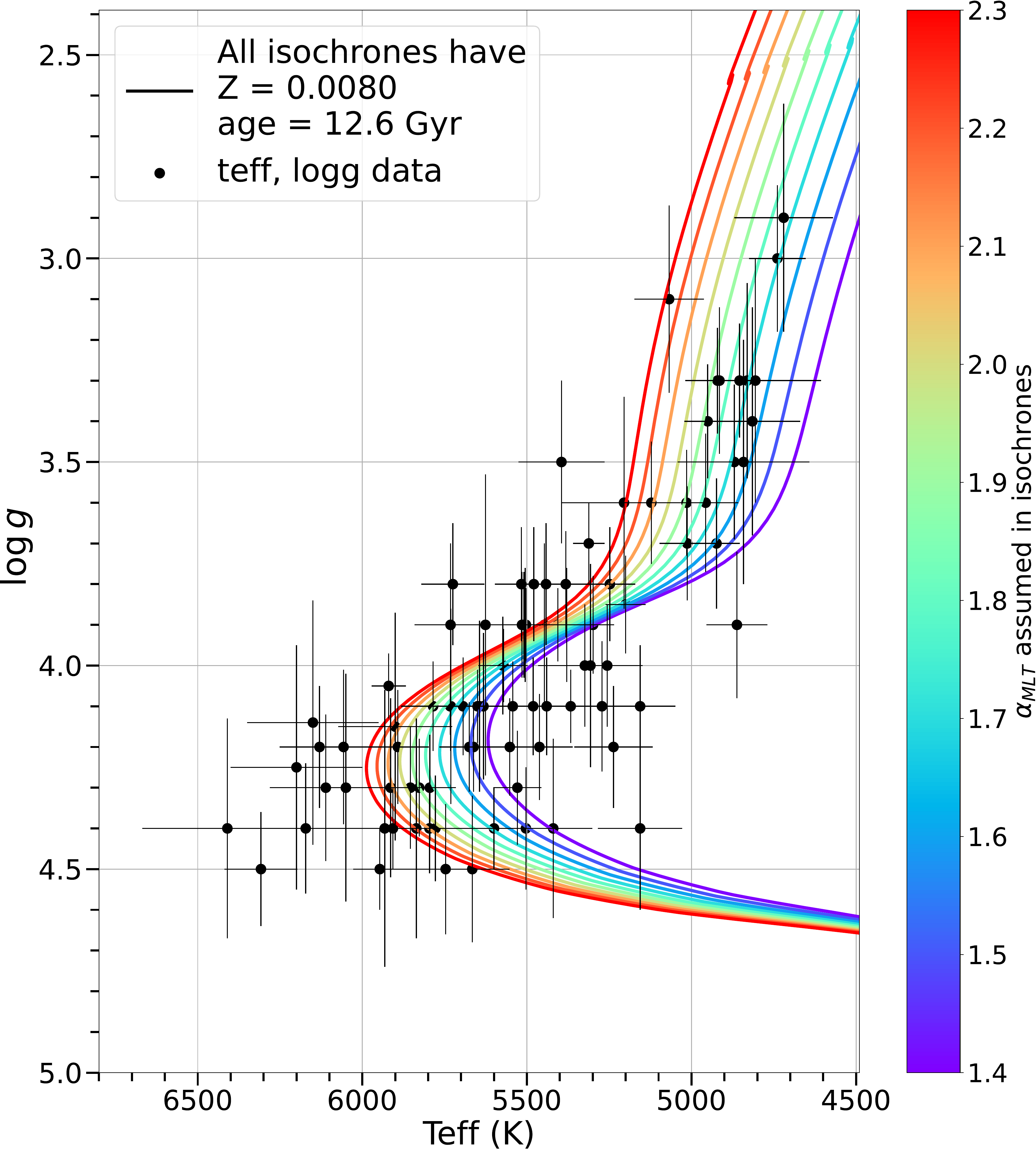}
\caption{As an example, the intrinsic temperature scatter among custom isochrones (computed with MESA models) due to \textit{ad hoc} assumptions about convection (choice of mixing length, $\alpha_{\text{MLT}}$) is at least half of the nearly $2000$ K temperature span shown here near the main sequence turn--off region and base of the subgiant branch.}
\label{fig:mixing_length}
\end{center}
\end{figure*}

While a comprehensive analysis of modeling systematics in MESA and MIST is beyond the scope of the current study, we must make an effort to avoid the most prominent flaw in all stellar age determinations: incomplete, or completely absent, consideration of theoretical uncertainties.

As mentioned in Section \ref{sec:statistical_analysis}, the isochrones themselves adopt a complex network of physical assumptions, each of which has its own associated uncertainty. This patchwork of parameters is highly covariant, and the sum of their effects is difficult to quantify. While some sources of modeling uncertainty have been studied extensively, such as metallicity and abundance assumptions, atmospheric boundary conditions, convective overshoot, and treatment of diffusion (e.g. \citealt{Feiden2015, Choi2018, Joyce2018aNotAll, Tayar2022}), these considerations are not typically applied to more than one idealized observational target at a time. Further, there are many additional sources of modeling uncertainty that are neglected entirely; in many cases, this is due to poor or non-existent constraints on the values of free parameters. A prime example of such a case is the convective mixing length, $\alpha_{\text{MLT}}$. More than a decade of research shows that (1) inappropriate choices for $\alpha_{\text{MLT}}$ can impart huge inaccuracies to model--derived fundamental stellar parameters, and (2) that the appropriate choice mixing length is dependent on metallicity, evolutionary stage, and probably mass \citep{Trampedach14, Tayar, Joyce2018aNotAll, Joyce2018balphaCen, Viani, TangJoyce2021, CinquegranaJoyce2022}. 

Yet, the \textit{de facto} approach in stellar modeling and isochrone construction is to adopt a fixed, solar value with no further exploration. As Figure \ref{fig:mixing_length} makes clear, this is not acceptable. 

Figure \ref{fig:mixing_length} shows a suite of isochrones computed with the same MESA version and physical settings used in MIST, but with variations in $\alpha_{\text{MLT}}$ permitted. In the upper left panel, the metallicity ($Z=0.0142$; roughly solar) and age (10 Gyr) are identical for every isochrone; the only difference is in the value of the mixing length adopted in its constituent stellar tracks, indicated by color. The mixing length values range from 0.4 to 2.3. These choices are physically justified, as some magnetically active M dwarfs have been found to be fit best by mixing lengths as low as $\alpha_{\text{MLT}}= 0.4$ H$_P$ (e.g. \citealt{Mann2015,IrelandBrowning2018}), and values above $2.0$ H$_P$ are common in fits to solar-like stars. 

The effects are striking, particularly in terms of effective temperature: in the evolutionary region most pertinent for these observations---the MSTO and subgiant branch---the range in Teff induced by varying $\alpha_{\text{MLT}}$ alone is more than 1500 K. Just above this, on the early RGB, it balloons to 2000 K (the danger of using the red giant branch for age determinations for precisely this reason has, at least, been noted in \citealt{Joyce2018aNotAll}). In terms of $\log g$, the variation is at least $\Delta \log g = 0.5$. In the lowest--valued cases, below about $\alpha_{\text{MLT}} = 1.2$, the models show the development of a convective core, indicated by the hooked feature at the MSTO.   

In the remaining three panels, the span in mixing length is truncated at a lower value of $\alpha_{\text{MLT}} = 1.4$ to more realistically capture the range of values appropriate for subgiant and early RGB stars (which are not fully convective). The B17 data set is shown overlaid.
Even with a more conservative estimate of the variance in this parameter, the spread in the HR diagram cannot be ignored. The two right--hand panels show the same age assumption but different metallicities: [Fe/H]$=0$ (solar) on top versus [Fe/H] $\approx -0.5$ underneath. Here we can see clearly that adjustments to the mixing length (that are well within reason) will comparably or more significantly impact age determinations than metallicity variations of 0.5 dex. Comparing instead the lower two panels, which have the same metallicity assumption but different ages, we find that a difference of $\alpha_{\text{MLT}} = 1.4$ vs $\alpha_{\text{MLT}} = 2.3$ is nearly equivalent, in terms of shifts in Teff, to an age difference of 5 Gyr.

Based on these order--of--impact arguments, we perform a simplified estimate of the \textit{total} age uncertainty through the inclusion of theoretical terms motivated by Figure \ref{fig:mixing_length}. In a new set of Monte Carlo simulations, we adopt the same definitions for $\chi^2$ and $t_S$ given in Equations \ref{eq:chisq_Bensby} and \ref{eq:weighted_age} and sample from the same distributions in observational parameters given in Equation \ref{eq:normal_dist}, but we introduce an additional term accounting for variation in the isochrones. In each Monte Carlo simulation, we sample normal distributions given by
\begin{subequations}
\begin{alignat}{2}
\label{eq:error_normal_dista}
\mu &= 0, &\quad \sigma &= \sigma_{T_{\text{eff},\text{th} }}  
\\
\label{eq:error_normal_distb}
\mu &= 0, &\quad \sigma &= \sigma_{\log g_\text{th}},
\end{alignat}
\end{subequations}
uniformly shifting the entire basis of isochrones horizontally by a random sample of Equation \ref{eq:error_normal_dista} and vertically by a random sample from Equation \ref{eq:error_normal_distb}. The rest of the algorithm proceeds as normal. The values adopted for the order--of--magnitude theoretical uncertainties in each direction are $\sigma_{T_{\text{eff},\text{th} }} = 200$ K and $\sigma_{\log g_\text{th}} = 0.17$ dex, representing approximately one third of the spans\footnote{The span is taken to represent a $3\sigma$ spread, as a crude approximation.} shown in the HR diagrams of Figure \ref{fig:mixing_length}. As these choices are based on mixing length arguments alone, these numbers should be taken to represent lower limits on the true degree of model variation. While a rigorous estimate of the total model uncertainty is beyond the scope of this paper, uncertainty in $\alpha_{\text{MLT}}$ is likely to dominate, so this is not an unreasonable first--order approximation.

The results of these simulations are given in the ``model err'' column of Table \ref{table:mega_age_results}. For a given star, the typical increase from the observational age uncertainty to the global age uncertainty is $1-2$ Gyr, thus putting the true age uncertainties anywhere between 2 and 5 Gyr. These are not only more realistic, they are realistic \textit{lower limits.} 

\section{Comparison with HST Data}
\label{sec:hst_comparison}
As discussed in Section \ref{sec:intro}, most HST analyses have concluded that the bulge is uniformly old with a small age spread, typically finding a mean age of $\sim$ 10-11 Gyr and a spread of $\sim$ 1-2 Gyr (e.g.\ \citealt{Kuijken02,Clarkson08,Brown10,Clarkson11, Gennaro15, Renzini18}).
In contrast, \citet{Haywood16} note that the main-sequence turn--offs of HST CMDs are too narrow in color to be compatible with a population that simultaneously possesses a large metallicity spread and a small age spread.  \citet{Haywood16} instead claim that incorporating an age--metallicity relation that forces metal--rich stars to be progressively younger reduces the main-sequence turn--off color spread and resolves the tension between photometric and spectroscopic studies finding discrepant age distributions.

\begin{figure*}
\begin{center}
\includegraphics[width=0.95\columnwidth]{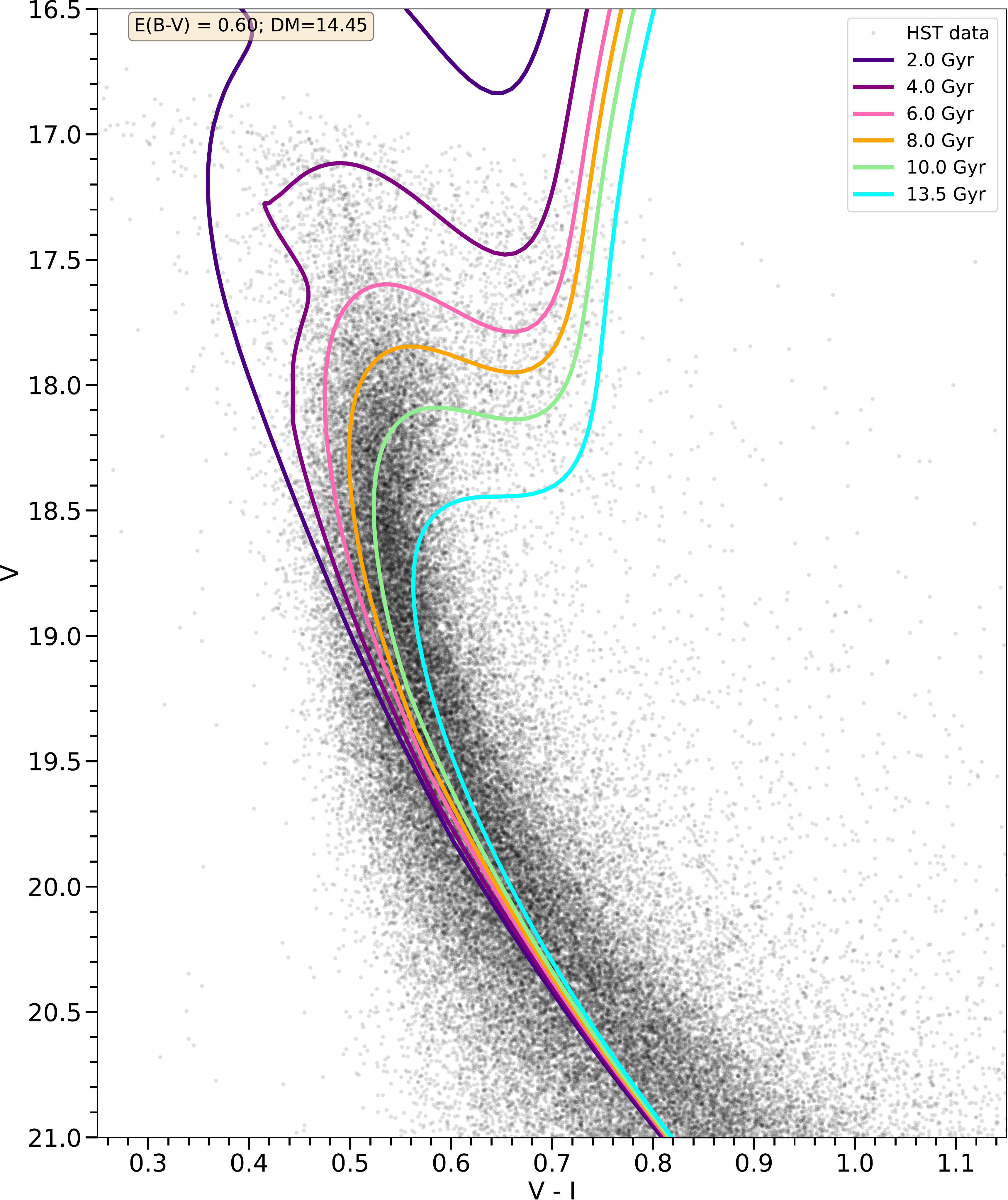}
\hfil
\includegraphics[width=0.95\columnwidth]{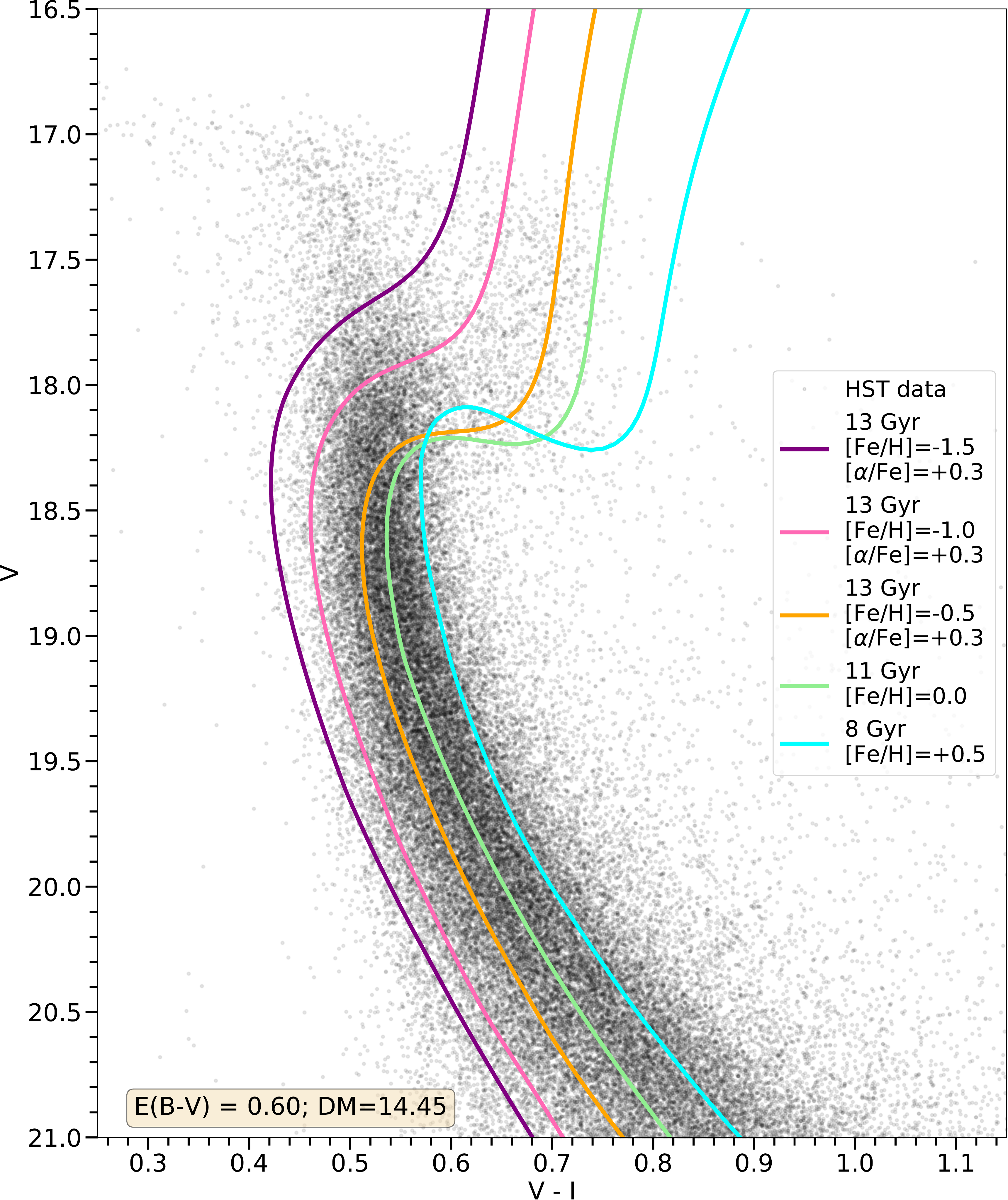}
\caption{\textbf{LEFT:} A proper motion cleaned HST CMD of the low latitude SWEEPS field using data from \citet{Calamida2015} is shown with a set of MIST isochrones spanning ages of 2-13 Gyr and with [Fe/H] = 0 over plotted.  We find that the bulge main-sequence turn--off is bracketed by ages of $\sim$ 6-13 Gyr, and that stars younger than 4 Gyr are too blue and bright to be compatible with the data. Note that most of the stars brighter and bluer than the nominal main-sequence turn--off are known to be blue stragglers \citep{Clarkson11}.  \textbf{RIGHT:} Similar isochrones are shown that represent our derived age--metallicity relation (Figure \ref{fig:age_dispersions}).  Since our age--metallicity relation restricts the youngest stars to those with [Fe/H] $>$ 0, the younger ages, which push stars bluer, are offset by the increased metallicity, which drives stars redder, such that the breadth of the expected main-sequence turn--off is reduced without producing the large contingent of bright, blue main-sequence stars seen in \citet{Haywood16}.
}
\label{fig:HST_isochrones}
\end{center}
\end{figure*}

However, the age--metallicity relation adopted by \citet{Haywood16}, which follows \citet{bensby13}, includes a significant number of stars younger than 5 Gyr.  As a result, \citet{Barbuy18} pointed out that the \citet{Haywood16} age--metallicity relation produces too many bright, blue main--sequence turn--off stars to be compatible with the observed CMDs.  Their age--metallicity relation also contrasts with ours presented in Figure \ref{fig:age_dispersions}, which is much flatter, contains only a small number of stars $<$ 6 Gyr in age, and has a median age of $\sim$ 8 Gyr for the most metal--rich bin.  Both \citet{Haywood16} and the present study agree that the bulge is ``predominantly old," but our work strongly removes the youngest age bins.  As a result, Figure \ref{fig:HST_isochrones}, which shows a proper motioned cleaned HST CMD of the low latitude SWEEPS field described in \citet{Calamida2015}, illustrates that our derived age--metallicity relation is more compatible with the observations. The mild slope of the age--metallicity relation found here retains the benefit of reducing the main-sequence turn--off color spread but does not include such young ages as to generate a large contingent of unobserved bright, blue stars.  Therefore, Figure \ref{fig:HST_isochrones} reinforces our earlier conclusion that the bulge is bracketed by ages of $\sim$ 6-13 Gyr, and that no appreciable fraction of stars have ages $<$ 4-5 Gyr.

\section{Implications of our derived Ages}
\label{sec:implications}

The application of multiple approaches to age determination has not relieved the tension between isochrone--derived ages and ages derived from analysis of HST color--magnitude diagrams with proper motion cleaning \citep{Clarkson08,Renzini18} or foreground subtraction \citep{Zoccali03}, though our new age--metallicity relation may help improve compatibility between some of these methods (see Section \ref{sec:hst_comparison}). Figure \ref{fig:age_vs_met} illustrates a sharp break in the derived age dispersion at [Fe/H]$\sim -0.5$, with more metal rich stars spanning a range from 12 to 3 Gyr in age, and being in the mean $\sim3$ Gyr younger than their metal poor counterparts.  
In agreement with \citet{Bernard18}, we find that most bulge stars are old ($>$ 10 Gyr) or of intermediate ($>$ 5-8 Gyr) age; however, we do not find any stars $<$3 Gyr old and the prevalence of intermediate-age stars rises sharply for [Fe/H] $>$ 0.3.
We cannot find any aspect of our analysis or the isochrone physics that would lead to a systematic derivation of younger ages at high metallicity.   In contrast to \citet{Bensby17}, our analysis gives an increase in intermediate age stars that is clearly pronounced at high metallicity; we conclude this is an actual physical aspect of our sample.

We may find some support for the age range at high metallicity from other observations.  \citet{catchpole16} find that long period Mira variables (attributed to intermdiate age populations) tend to follow the bar structure, while short period (older) Miras do not. \citet{Hasselquist20} use a large sample of APOGEE data to argue that a large fraction of the bulge---at least half their sample---is old ($>$ 8 Gyr), and that the population of young (2-5 Gyr) stars they do find is strongly constrained to the Galactic plane.  The present work, along with \citet{Marchetti22}, do not confirm the presence of such young stars.  However, we note that \citet{Marchetti22} does not sample inside $b \sim -3.5^{\degr}$ where young stars may be more prevalent.  On the other hand, the \citet{Bensby17} sample analyzed here shows no significant correlation between age and Galactic latitude.

It is difficult to age date populations of late--type stars such as SiO masers from their bolometric luminosities (uncertain also due to distance), but there are indications of a widespread population of luminous SiO masers that follow the bulge/bar population kinematics \citep{trapp18}. However, the most luminous late--type stars are extremely concentrated to the Galactic center \citep{blom07}. This work and our foreground cleaning of disk stars from the Galactic bulge (Marchetti et al. 2022, \textit{subm.}) argue strongly against a significant population of main sequence progenitors younger than 3 Gyr. Considering the lack of main sequence progenitor candidates, we conclude that no substantial population of massive, evolved AGB stars is present in large numbers in the bulge.

Studies of galaxy morphology \citep{sheth08} find that the fraction of barred galaxies decreases strongly with redshift, dropping from 0.6 at $z=0.1$ to 0.3 at $z=0.8$.  The evolution of bar fraction is striking in that it occurs at relatively low redshift.  This suggests that the formation of bars in spiral galaxies may be a relatively recent phenomenon that took place within the last 5-10 Gyr.

Dynamical models \citep[e.g.,][]{Shen10} have long found that the formation of bars from a massive stellar disk occurs rapidly in just a few rotation times. Such a scenario would appear at first to be in stark disagreement with an age range for stars in the bar, and would especially challenge a scenario in which stars of higher metallicity are more concentrated to the plane. It is possible that once the bar forms, gas flows support additional star formation from the enriched gas, with the younger, more metal rich population observed closer to the plane. \citet{val-gerhard13} advance a scenario in which the initial disk has a radial metallicity gradient that, after the bar forms, is mapped into a comparable vertical metallicity gradient. Their model preserves the basic premise of a rapid buckling of the bar while allowing for an observed metallicity gradient. The present-day region of the bar closest to the Galactic center, however, shows negligible evidence of a metallicity gradient \citep{Johnson20}.

The chemistry of the bulge offers some additional support for this age/metallicity relationship.  The trend of $\rm [\alpha/Fe] $ vs [Fe/H] is clearly toward scaled Solar with increasing metallicity \citep[e.g.,][]{Gonzalez11,Johnson14}, consistent with Type Ia SNe having the time to contribute enough iron to force this trend.  However, similar trends are seen in the thick disk, for which the evidence for an extended formation history is less compelling, and the long timescales needed for Type I SNe to pull $\rm [alpha/Fe]$ toward Solar scaled composition \citep{tinsley79,gregg08,maoz10} may not be available.

In conclusion, our new ages do not find support for a population of young, massive stars in the bulge (age $<$ 3 Gyr). However, the tension between the ages of micro-lensed dwarfs derived from spectroscopic analysis and isochrone fitting (and also via photometric isochrone fits) and proper--motion cleaned HST-based color--magnitude diagrams remains unresolved. The new, more carefully computed ages point toward a growing likelihood that the most metal rich stars in the bar are younger than their more metal poor counterparts. This conclusion looks increasingly attractive as new evidence accumulates for a progressive concentration of metal rich stars toward the plane.  However, we emphasize that while a number of disparate approaches point toward support for modestly younger ages or a wider age dispersion in the bar, all age determination methods face challenges in precise measurements for stars older than $\sim 5$ Gyr; additional robust measurements are needed.  While challenging, it would be desirable to construct proper--motion separated HST-based color--magnitude diagrams that are also segregated by metallicity, to confirm whether the metal rich population shows evidence of a greater age dispersion. The work by \citet{Renzini18} is a good first step in this direction, and we envision building on their methods to provide more granular examinations of the age-metallicity relation in the bulge. It will be important to find new approaches to test the age dispersion of the bulge/bar population, as there are important implications for the formation and evolution of the bulge/bar.  

\section{Summary}
\label{sec:summary}
We present new age determinations of the Milky Way bulge micro-lensed dwarf sample presented in \citet{Bensby17} by combining their spectroscopic stellar parameters with out-of-the-box and custom modern MIST isochrones and carefully considered statistical analysis techniques. We also include an analysis of the impact various error sources have on the derived ages, such as: the choice of isochrone grid (MIST versus Yale-Yonsei), including (or not) metallicity dependent $\alpha$-element enhancement, and reasonable variations in the mixing length parameter. The newly derived ages are then combined with the previously published [Fe/H] values to construct an updated age--metallicity relation for the bulge.

The full micro-lensed dwarf sample analyzed here has a mean age of about 11 Gyr and a spread of $\sim$3.5 Gyr. We find a dominant old peak near 13 Gyr and a secondary peak near 8 Gyr, but we do not find any younger local maxima in the bulge age distribution.  Our newly derived mean age is nearly 3 Gyr older than the mean age presented in \citet{Bensby17}, and unlike the \citet{Bensby17} results, we do not find any appreciable population of stars $<$ 5 Gyr in age.  The strong reduction in the number of young bulge stars is most clearly illustrated when comparing our results against those in \citet{Bensby17} for stars $<$ 7 Gyr in age.  The latter finds that nearly 50$\%$ of their sample is younger than 7 Gyr while we find only 16$\%$. Similarly, we find 66$\%$ of the micro-lensed sample have ages $>$ 10 Gyr while \citet{Bensby17} find only 40$\%$.  Therefore, our derived age distribution is both more strongly weighted toward older stars and has a sharper cut-off on the young end compared to the \citet{Bensby17} distribution.  These results are robust against variations in assumed [$\alpha$/Fe] ratios and choice of stellar isochrone grid as we find nearly the same distributions, modulo some small scale zero point shifts, when purely solar--scaled compositions are used and when adopting the Yale-Yonsei grid used by \citet{Bensby17}.

In addition to deriving ages using the spectroscopic parameters, we cross-matched the \citet{Bensby17} sample against stars observed in the g and i-bands with the Blanco DECam Bulge Survey and with the Gaia EDR3 database. In the former, we found 51 stars in common, while in the latter, we found 17 matches.  In both photometric cases, we included all dominant error sources in the Monte Carlo fitting procedure and found mean ages of $\sim$ 9-10 Gyr and spreads of $\sim$ 1-2 Gyr, which is about 1-2 Gyr younger than the spectroscopic sample but with a much more narrow distribution.  Interestingly, our photometric age distribution for individual stars closely matches those found in the literature when fitting cleaned main-sequence turn--offs in bulge CMDs.  However, further analysis showed that the combined effects of photon noise, distance, bulge membership, and small scale reddening uncertainties were too large to provide any meaningful results from isochrone fitting. Instead, when using only photometry, the fitting procedure simply trended toward the median of the isochrone grid, thus providing no useful information on the age distribution (see Appendix \ref{sec:photometric_fits}). 

We investigated the age sensitivity to the MIST isochrone's adopted mixing length, which is one of the most significant sources of uncertainty in stellar models, especially near the MSTO.  Simply changing $\alpha_{MLT}$ between 1.4 and 2.3, which are all plausible values for low mass stars in the parameter space analyzed here, added at least 1-2 Gyr to the global age uncertainty.  Modifying this parameter alone increases the lower limit of the age uncertainty of an individual star to between 2 and 5 Gyr, thus showing that the age uncertainty values from many previous works are likely severely underestimated.

In order to further investigate the discrepancies between photometric and spectroscopic age distributions of bulge main-sequence turn-off stars, we applied the age--metallicity relation derived here from spectroscopy and overlaid the resulting isochrones onto a proper motioned cleaned HST CMD of the bulge SWEEPS field. A previous analysis by \citet{Haywood16} suggested that the bulge must have a significant age dispersion, rather than being uniformly old, because a purely old bulge that also has a large metallicity spread would yield too broad of main-sequence turn--off to match the observations. However, the age--metallicity relation adopted by \citet{Haywood16}, which follows the results of \citet{bensby13}, includes a substantial number of stars younger than 5 Gyr in age. As a result, their simulated CMD produces a prominent plume of bright, blue stars that are not observed in the cleaned CMDs. However, we found that applying our new age--metallicity relation, which has very few stars $<$ 5 Gyr in age and is on average older, is able to reduce the color spread at the main-sequence turn--off to the expected levels while avoiding the production of much younger, brighter stars that are clearly not present in the data.  Therefore, we conclude that the bulge is bracketed by ages of about 6-13 Gyr with little room for a significant contingent of younger stars.

\section*{Acknowledgements}
M.J.\ was supported by the Lasker Data Science Fellowship, awarded by the Space Telescope Science Institute. M.J.\ acknowledges the Kavli Institute for Theoretical Physics at the University of California, Santa Barbara, whose collaborative residency program TRANSTAR21 was supported by the National Science Foundation under Grant No. NSF PHY-1748958. M.J.\ wishes to thank L\'aszl\'o Moln\'ar, Jamie Tayar, Aaron Dotter, and Brian Chaboyer for helpful discussion. 
C.I.J. gratefully acknowledges Annalisa Calamida and Will Clarkson for sharing data and details of their HST reductions for the HST SWEEPS field.
T.M.\ acknowledges an ESO Fellowship.

\vspace{5mm}
\facilities{HST(STIS), Swift(XRT and UVOT), AAVSO, CTIO:1.3m,
CTIO:1.5m,CXO}

%
%
\appendix

\section{Age fits using photometric systems}
\label{sec:photometric_fits}
As an independent check on our age determinations from \citet{Bensby17}'s own measurements, we attempt to derive ages from colors and magnitudes for those B17 sources which also have photometry from the Blanco DECam Bulge Survey \citep[BDBS;][]{Rich20,Johnson20} and/or \textit{Gaia} EDR3 \citep{GaiaEDR3}. Among \citet{Bensby17}'s 91 stars, we find 51 that have reliable BDBS $g$- and $i$-band photometry and 16 that have reliable $G, B_P, R_P$ Gaia photometry. Within these, especially well-determined targets are selected for ``gold samples,'' described below. 

\subsection{BDBS Photometry}
\label{sec:BDBS_photometry}
The BDBS survey is a deep, UV-optical Dark Energy Camera (DECam) imaging campaign spanning $>$ 200 contiguous square degrees of the southern Milky Way bulge.  The survey reaches below the main-sequence turn-off along most sight lines, and includes photometry in the DECam $ugrizY$ bands that are calibrated onto the SDSS ($u$) and PanSTARRS ($grizy$) systems \citep{Tonry2012}.  Nearly 250 million unique point sources are included in the BDBS catalog, along with a 1$\arcmin$ $\times$ 1$\arcmin$ reddening map spanning $|$l$|$ $<$ 10$^{\circ}$ and $b > -10^{\circ}$.  Details regarding the project goals, performance, and early science results are presented in \citet{Rich20} and \citet{Johnson20}.

Cross-matching the BDBS catalog with the micro-lensed sample presented in \citet{Bensby17}, we find 60 stars in common.  Stars that were not cross-matched either reside outside the BDBS footprint or were in a very crowded region where we could not be certain that the correct star was identified.  Among the 60 cross-matched targets, 9 were removed due to poor photometry or unavailability of distances. Among the remaining 51 candidates, we further identified a set of 18 stars as belonging to a ``gold sample".  These stars were selected as having higher quality photometric measurements.  Specifically, the gold sample BDBS stars were observed a minimum of 3 times, were not located in high background regions, and did not include any measurements with peculiar image shape parameters.

\begin{figure*}
\def\figwid{0.41\textwidth}
\begin{center}
\includegraphics[width=\figwid]{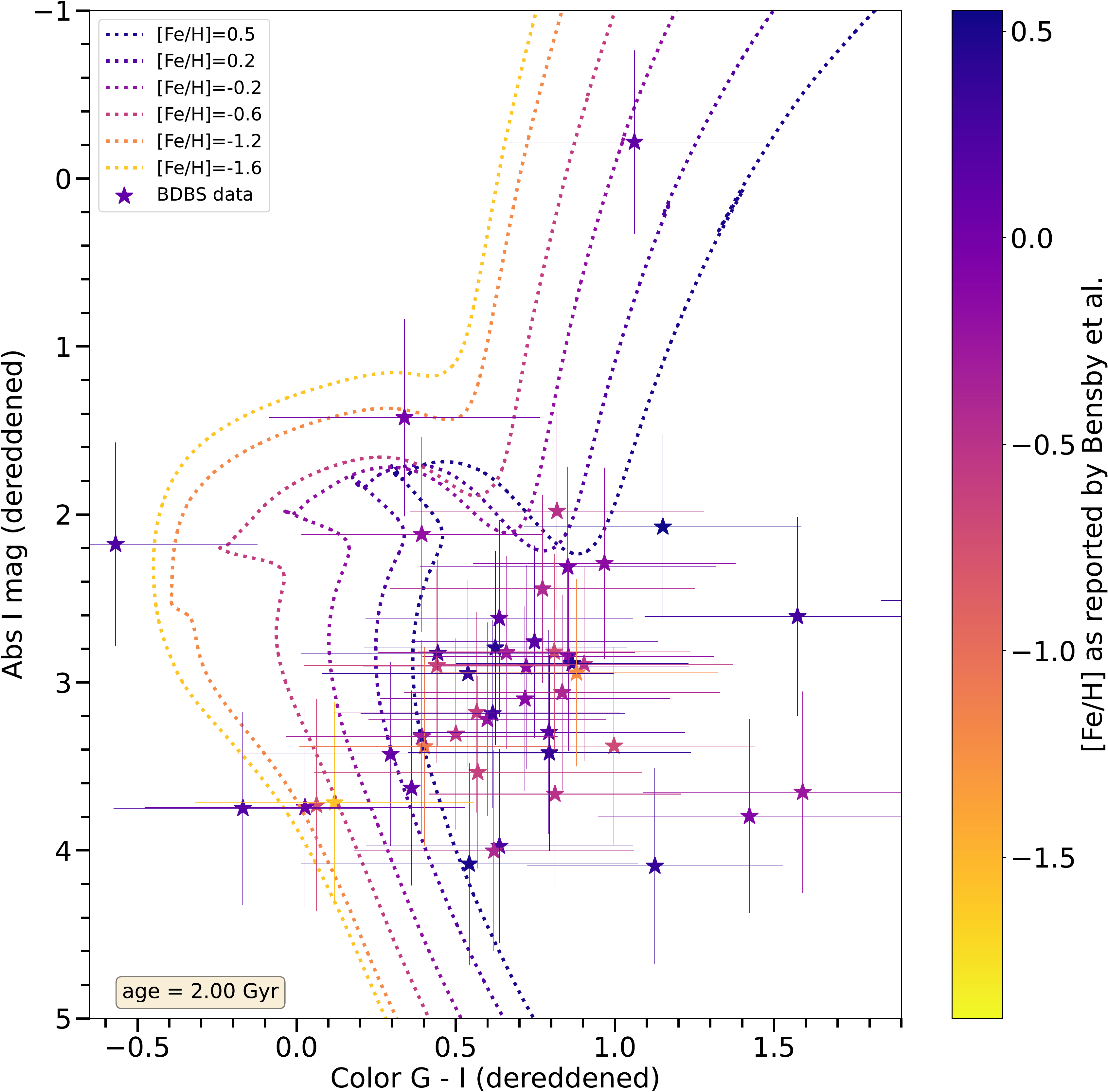}
\hfil
\includegraphics[width=\figwid]{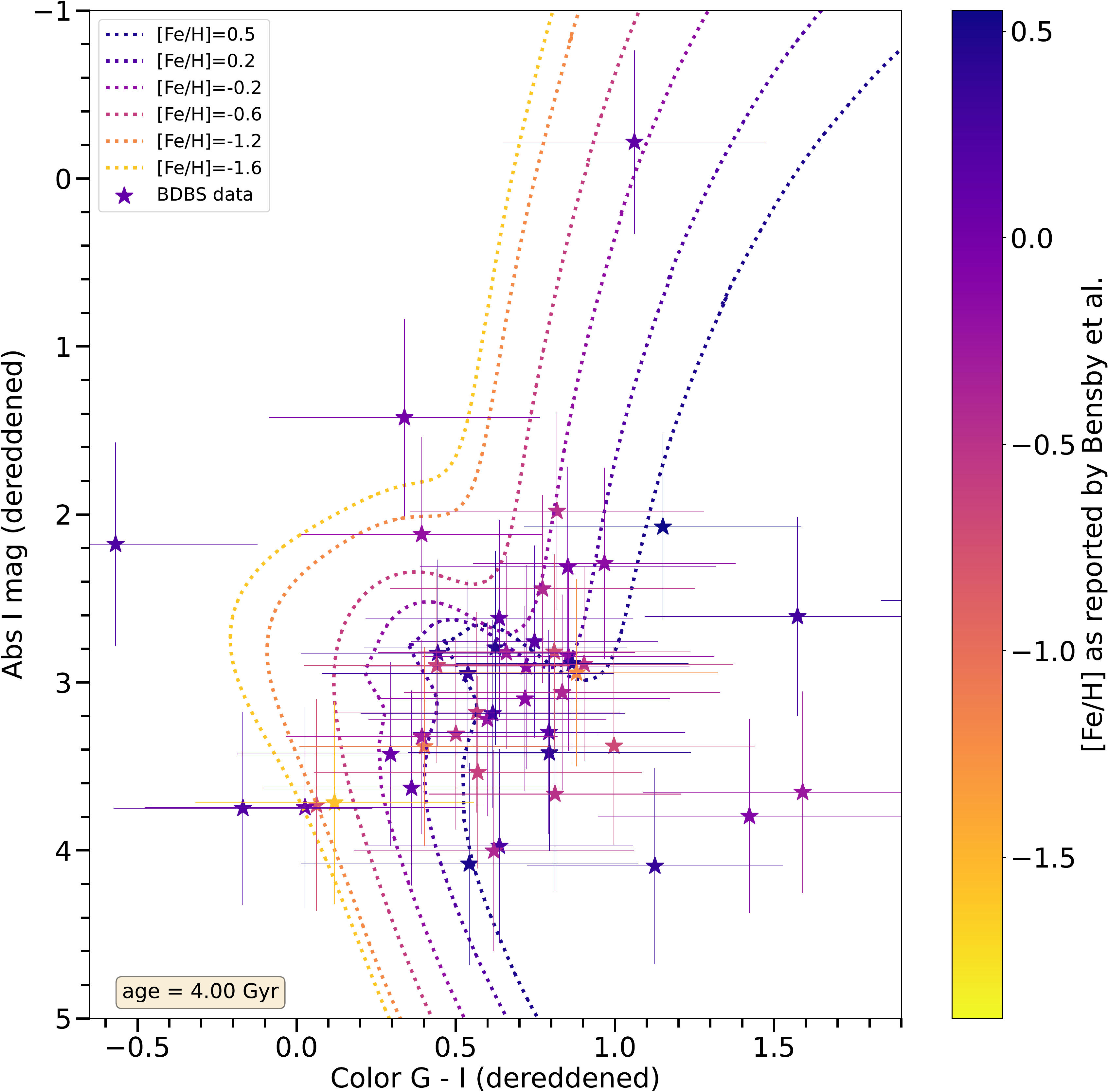}
\vskip8pt   
\includegraphics[width=\figwid]{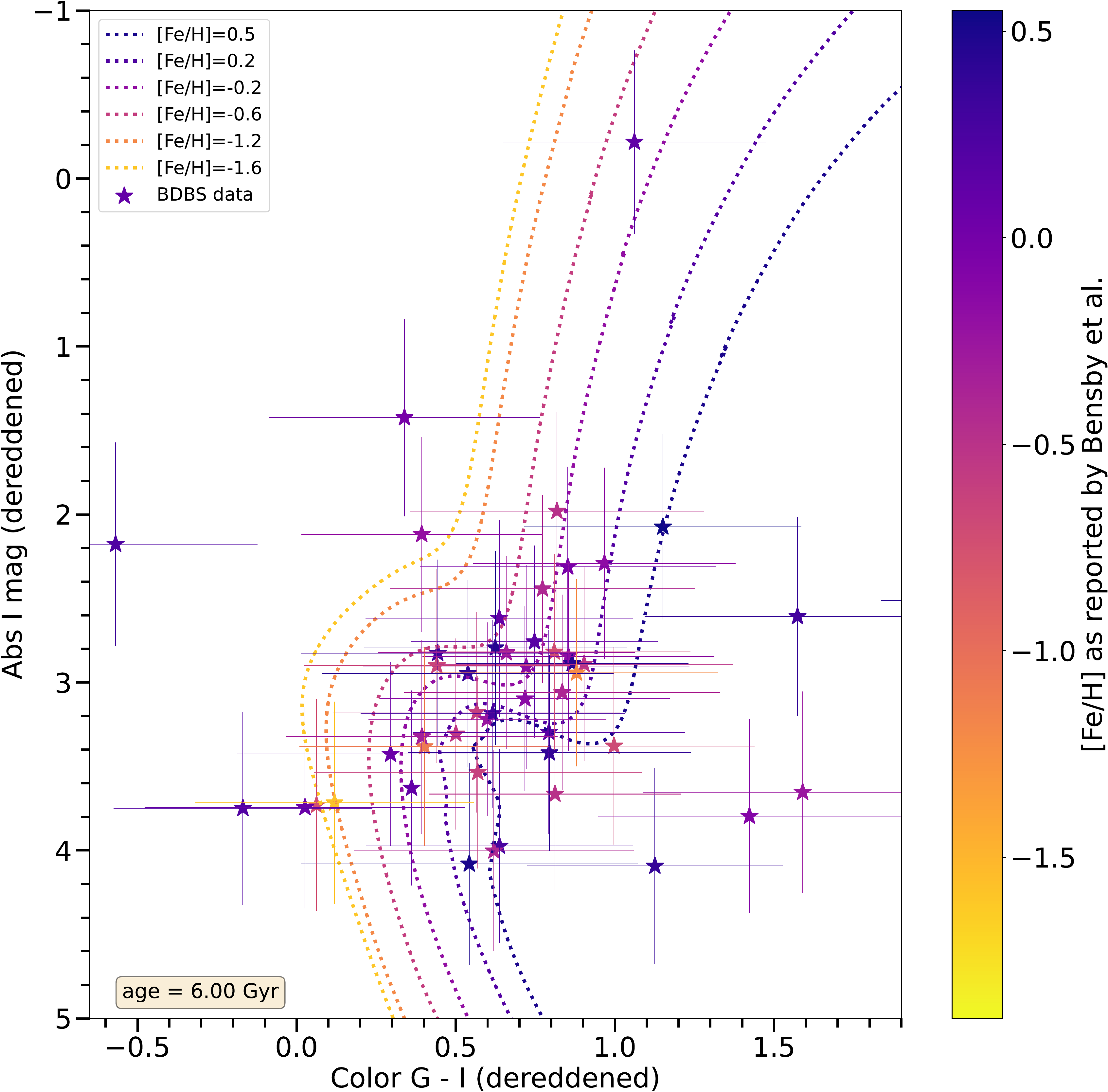}
\hfil
\includegraphics[width=\figwid]{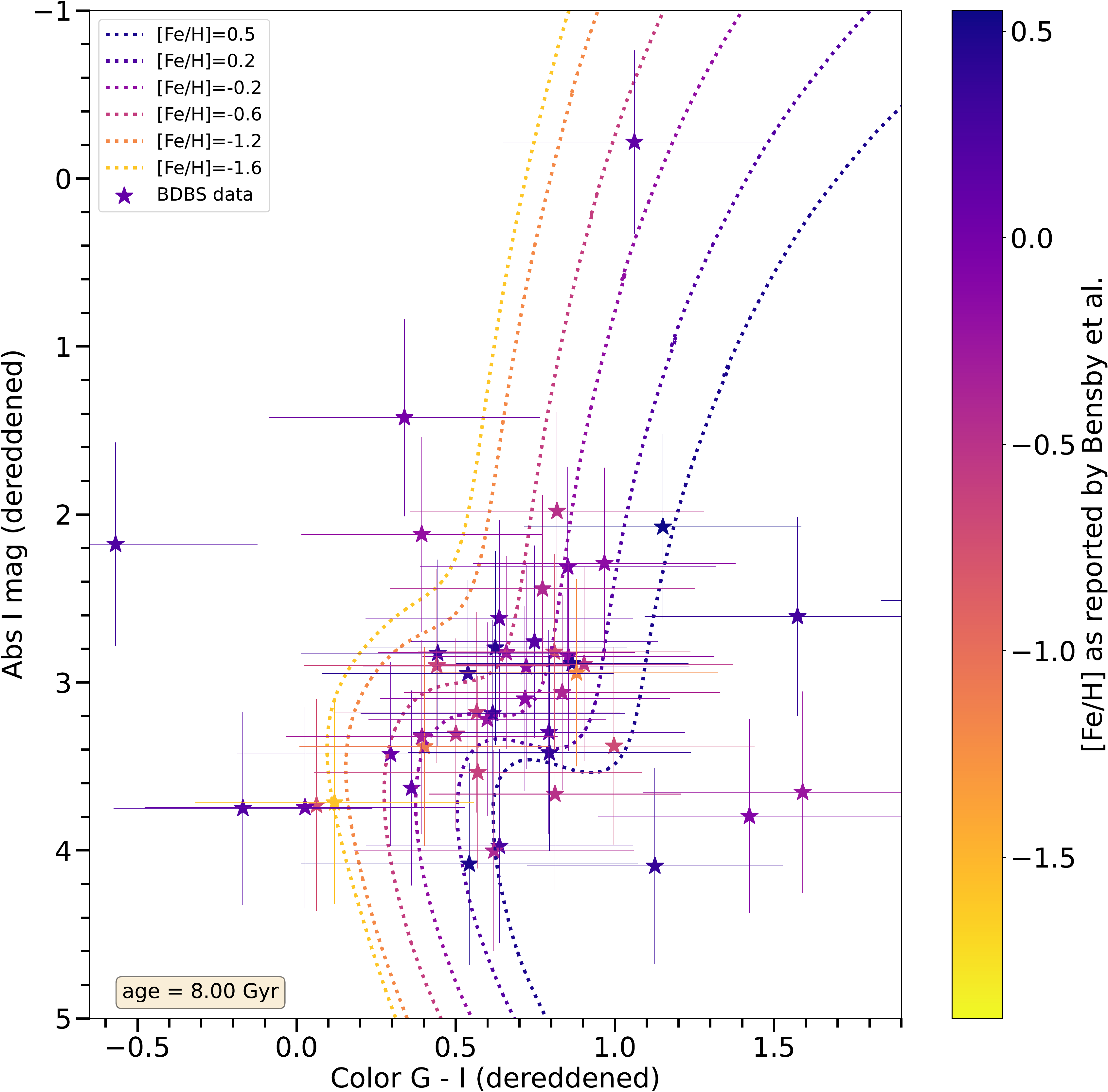}
\vskip8pt
\includegraphics[width=\figwid]{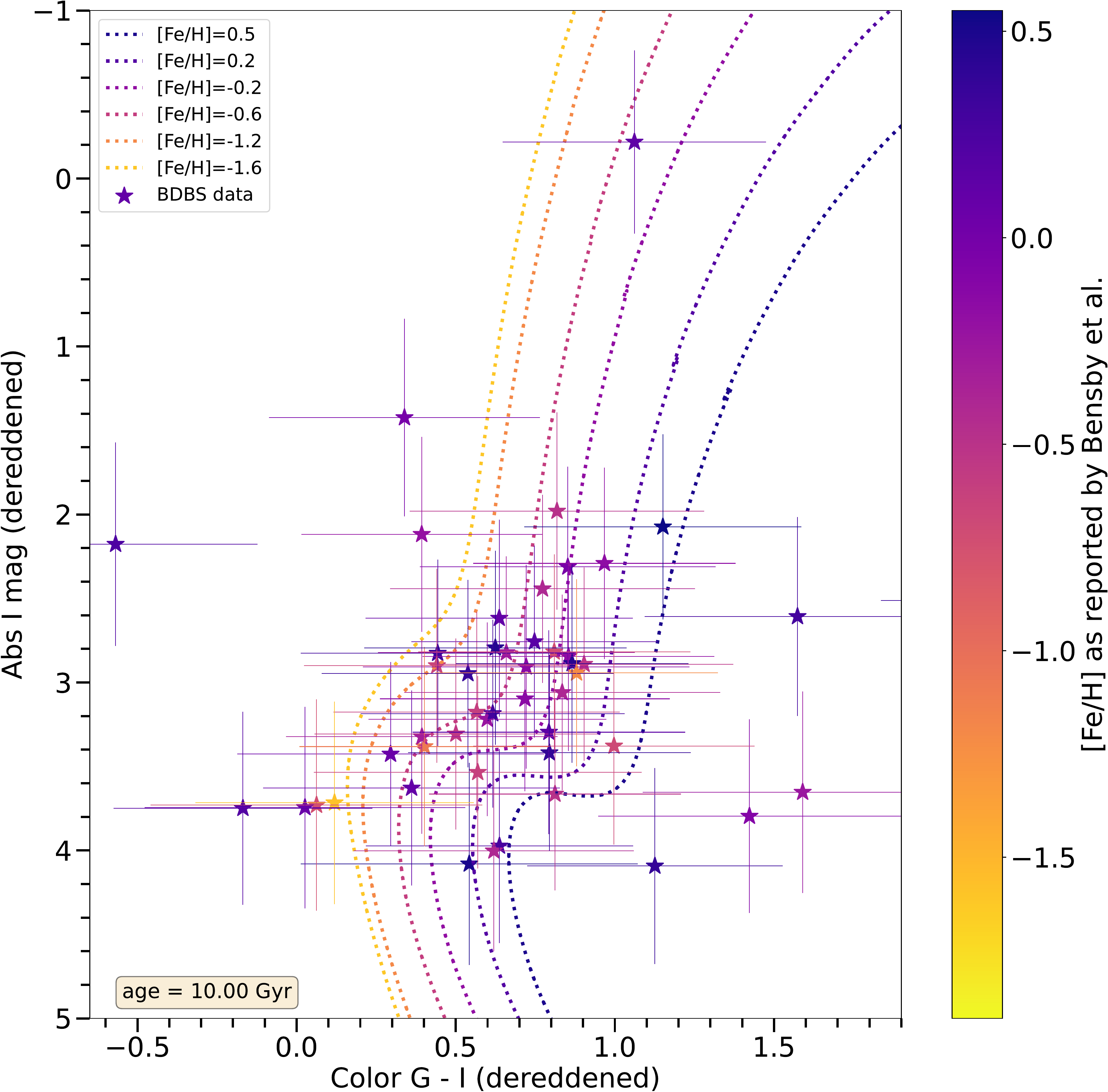}
\hfil
\includegraphics[width=\figwid]{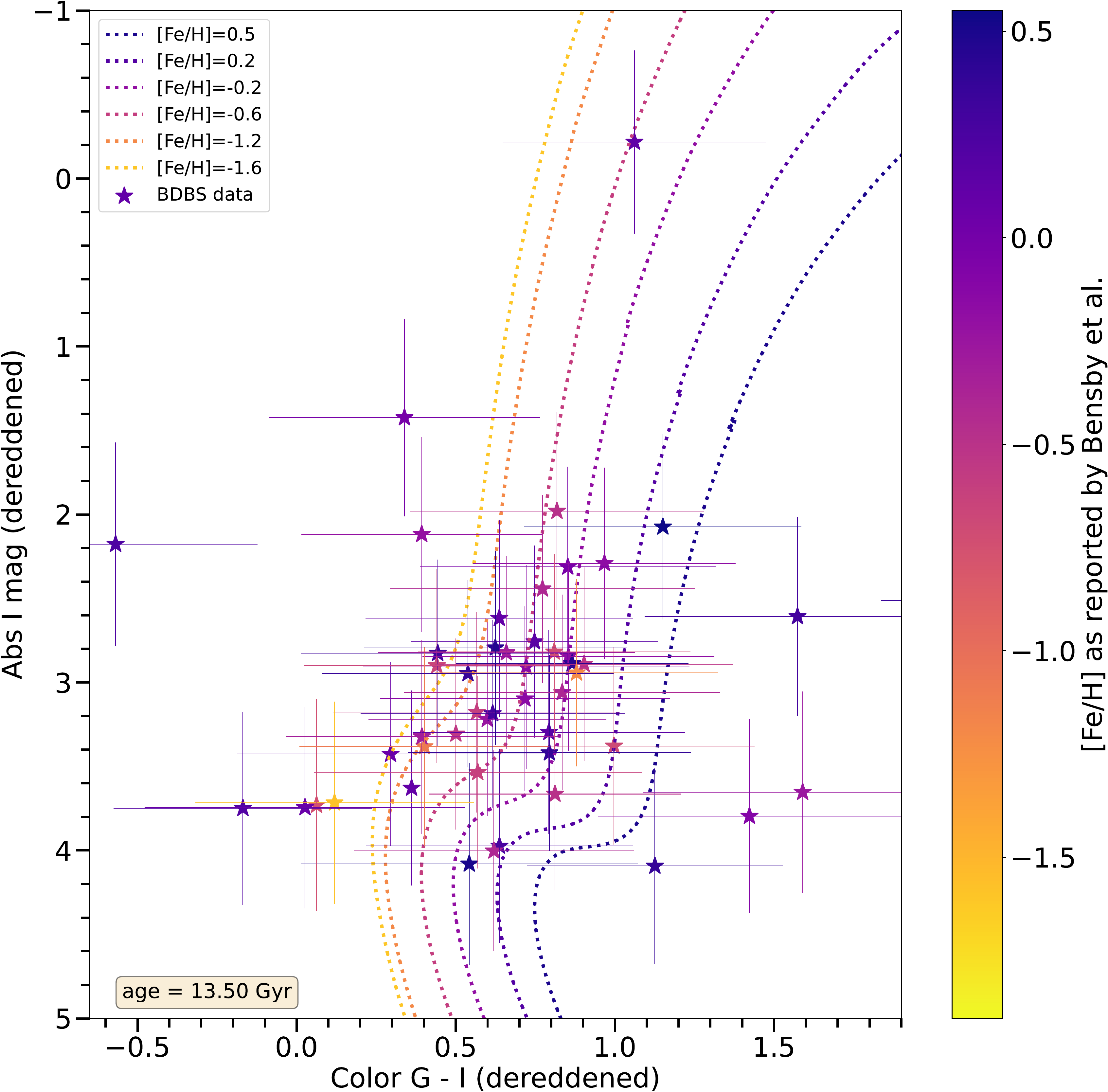}
\caption{MIST isochrones in the PanSTARRS color basis are shown against BDBS photometry in $I$ vs $G - I$.}
\label{fig:BDBS_cmd}
\end{center}
\end{figure*}

CMDs showing the absolute, de-reddened magnitudes and colors for 51 BDBS targets against MIST isochrones are presented in Figure \ref{fig:BDBS_cmd}. Panels show ages ranging from 2 to 13.5 Gyr. Metallicities are assigned to the stars from B17 directly and indicated for both data and isochrones according to the color bar. 

To fit models to BDBS photometry, the MIST isochrones must be converted to the PanSTARRS color--magnitude system for consistency with the BDBS $i$- and $g$-band filters, and the observations must be shifted into an absolute magnitude system, which requires knowledge of the distance and extinction. Where these transformations are possible (i.e., where sufficiently reliable distance and reddening information is available), equation \ref{eq:chisq} still holds as the appropriate definition of $\chi^2$. The analog of Equation \ref{eq:chisq_Bensby} for BDBS, Gaia, or any photometric fit is then
\begin{equation}
\chi^2_{\text{BDBS}} 
= \frac{ (\text{color}_o - \text{color}_t)^2 }{\sigma_{\text{color},o}^2 }  + \frac{ (\text{mag}_o - \text{mag}_t)^2 }{\sigma_{\text{mag},o}^2 }  
+ \frac{ (Z_{o}  - Z_t)^2 }{\sigma_{Z,o}^2}. 
\label{eq:chisq_BDBS}
\end{equation}
However, unlike in the fits to B17's physical coordinates, the terms $\sigma_{\text{color},o}$ and $\sigma_{\text{mag},o}$ are composite uncertainties. The full definition of color, $G - I$, is actually $(g - A_g) - (i - A_i)$, where $A_g$ and $A_i$ are the extinctions in those bands, respectively. However, we also note that the extinction estimates are based on a reddening map that aimed to minimize the infrared color/magnitude spread of red clump stars inside $1\arcmin \times 1\arcmin$ bins on the sky; this is not the same as measuring or fitting the extinction for each star individually in the B17 sample.  We also do not account for possible variations or errors in the reddening laws, such as changes in the total-to-selective extinction ratio.  Further, there is an uncertainty in both $g$ and $i$ due to instrumental systematics. Therefore, the uncertainty in color as a whole is given by
\begin{equation}
    \sigma_{\text{color},o}^2
    = \sqrt{ \sigma_{i\text{-band}}^2 + \sigma_{g\text{-band}}^2
    + \sigma_{A_g}^2 + \sigma_{A_i}^2 + \sigma_{i, \text{sys}}^2 + \sigma_{g, \text{sys}}^2 },
\label{eq:sigma_color_BDBS}
\end{equation}
%
with $\sigma_{i, \text{sys}} = 0.03$ mag and  $\sigma_{g, \text{sys}} = 0.02$ mag \citep{Johnson20}.
Similarly, $\sigma_{\text{mag}}$ must incorporate uncertainty in the $i$-band magnitude, $A_i$, instrumental systematics in $i$, and uncertainty in the distance modulus, $DM$: 
\begin{equation}
    \sigma_{\text{mag},o} = \sqrt{ \sigma_{i\text{-band}}^2 + \sigma_{A_I}^2 +  \sigma_{i, \text{sys}}^2 + \sigma_{\text{DM}}^2}.
\label{eq:sigma_mag_BDBS}
\end{equation}

For each star’s line of sight, we calculated the distance at which the bulge density distribution reaches a maximum. We adopt the bulge parametric model described in \citet[Model E;][]{Simion17} which provided the best fit to the infrared Vista Variables in the Via Lactea (VVV; \citealt{Minniti10}) survey. The dispersion encompasses the 0.16--0.84 density fractions and is typically $\sim 1.5$ kpc.

Since the Gaia EDR3 parallaxes are insufficient for deriving accurate distances for faint bulge stars, the second set of distances is estimated from a sample of 2.6 million red clump stars described in \citet{Johnson2022}.
The red clump distance map covers sight lines between $|l|$ = $\pm$10$^\circ$ and $b < -3.25^\circ$ and encompasses a large fraction of the micro-lensed dwarf sample.  For micro-lensed stars residing inside the red clump map, a mean distance and dispersion were calculated from all red clump stars that had a similar ($\Delta$ [Fe/H] $<$ 0.1 dex) metallicity and were within a 0.5$^\circ$ radius.  micro-lensed stars that reside outside the red clump map were excluded from further analysis due to insufficient distance information.

The presence of both data sets allows us to fit as many BDBS targets as possible. Due to the effects on the uncertainty terms, and therefore on the $\chi^2$ and $p$ scores, stars that do not have distances according to either map cannot be fit and are removed from the sample. 

One of the largest error sources for a given star is the uncertainty in the extinction correction factors for the BDBS $g$-band (A$_{g}$) and $i$-band (A$_{i}$). We deredden the BDBS photometry following the approach described in \citet{Johnson20}. We employ a high resolution E(J-Ks) dust map computed using the VVV survey. The (J-Ks) color dispersion $\sigma_\text{E(J-Ks)}$ caused by the extinction correction was propagated to the PanSTARRS photometry using the photometric transformations outlined in \citet{Johnson20}.

\begin{figure}
\begin{center}
\includegraphics[width=0.45\columnwidth]{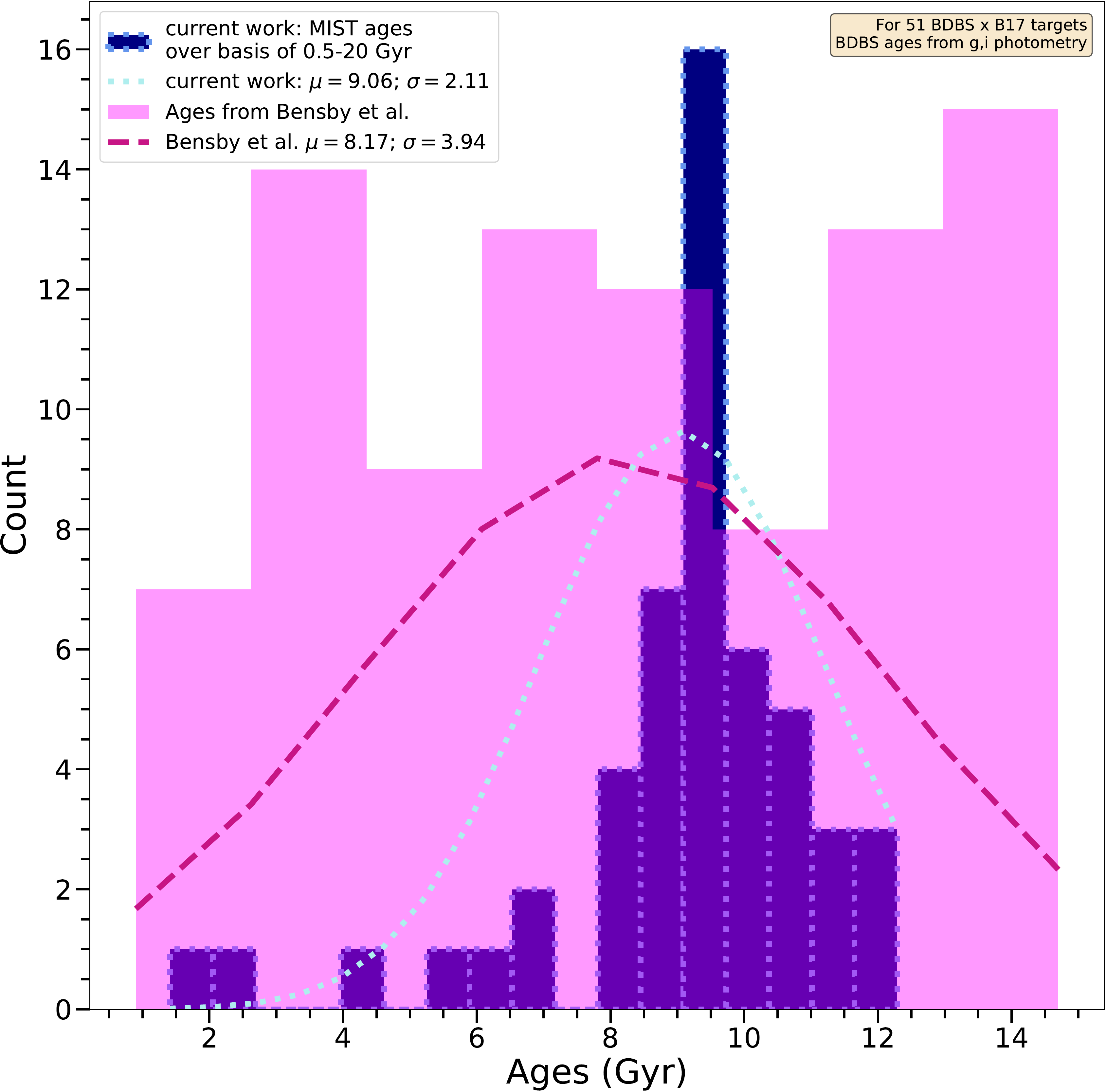}
\hfil
\includegraphics[width=0.45\columnwidth]{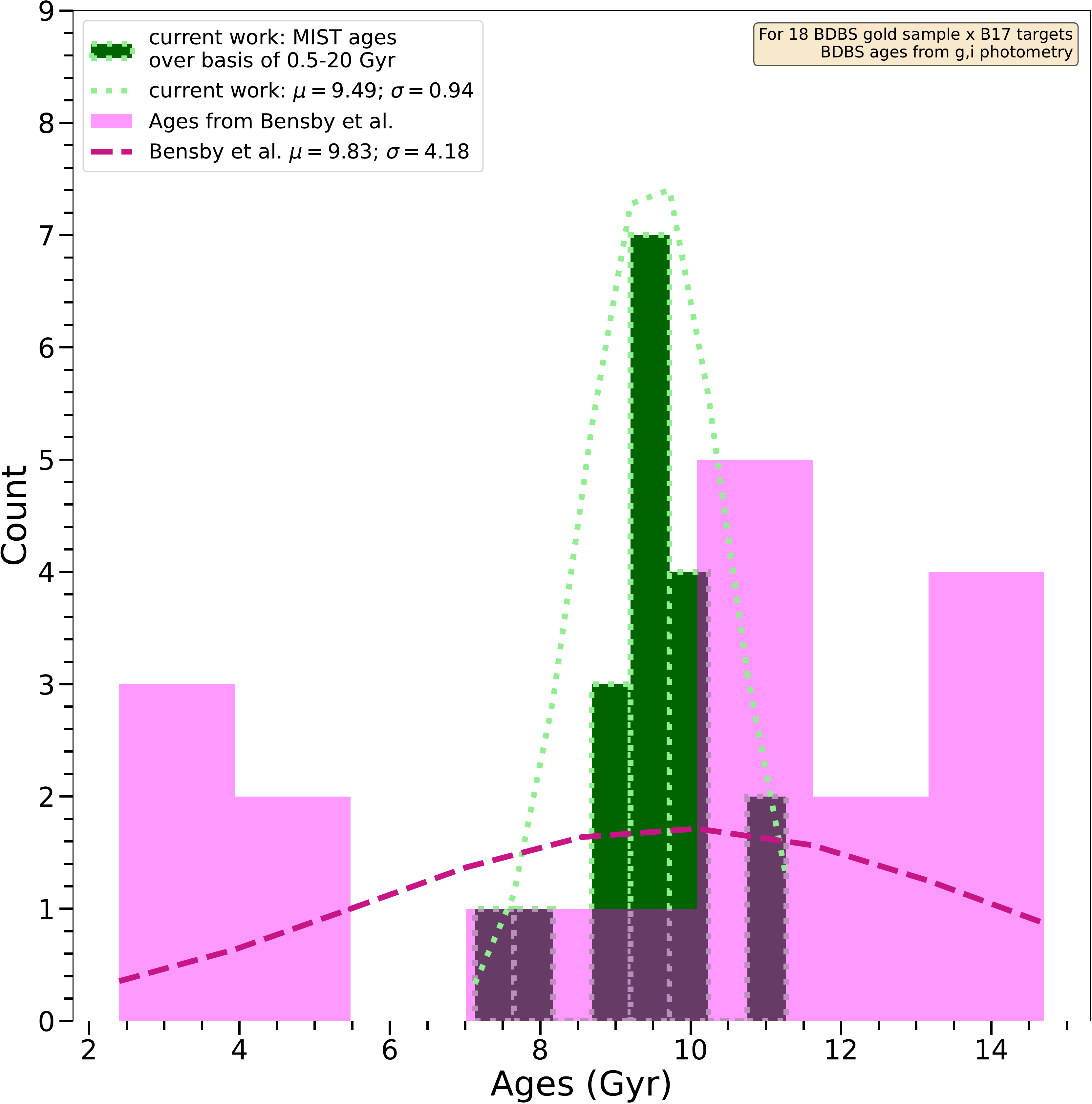}
\caption{ Age histograms in the style of Figures \ref{fig:age_histogram} and \ref{fig:age_histogram_alpha} but with ages derived from BDBS photometry. 
\textbf{LEFT:} The full intersection of the B17 and BDBS target lists for which complete photometric and distance information was available, totalling 51. The age distribution according to \citet{Bensby17} for the same subset of stars is overlaid in pink. 
\textbf{RIGHT:} Same as left panel, but for the gold sample containing 18 members. The ages for the same 18 stars according to B17 is overlaid in pink. }
\label{fig:age_histogram_BDBS}
\end{center}
\end{figure}

Using definitions \ref{eq:chisq_BDBS}, \ref{eq:sigma_color_BDBS}, and \ref{eq:sigma_mag_BDBS} in the calculation of $\chi^2$ and $p$, we compute ages for 51 BDBS targets.
The resulting distribution is shown in Figure \ref{fig:age_histogram_BDBS}.  

To understand this distribution---in particular, its narrowness---let us return briefly to the CMDs in Figure \ref{fig:BDBS_cmd}, where we see that the observational error bars from BDBS photometry are much larger relative to the spacing of the isochrones than in analogous Figures \ref{fig:6panel} and \ref{fig:young_alpha_shift}. Given the larger number of constituent uncertainty terms embedded in the photometric $\sigma_{\text{color}}$ and $\sigma_{\text{mag}}$ definitions (and the well-established difficulty of making photometric age determinations for this reason), it is not surprising that this is the case. It follows, then, that a star adopting these $1 \sigma$ color and magnitude uncertainty definitions will intersect with a much larger number of isochrones, therefore achieving good (or over-fit) $\chi^2$ scores for several hundreds of evolutionary points spanning a large range of ages and metallicities. The resulting age distribution should therefore reflect the weakly constrained star's plausible consistency with nearly any model, thus yielding a sharply peaked distribution near the median of the trial ages (10.25 Gyr for ages 0.5 Gyr to 20 Gyr) and low variance, as shown. 

What the sensitivity of the photometric ages to the set of input age candidates (i.e.\ grid of isochrones) really tells us is that the stars are not well enough constrained through photometry alone to be discerning among models. Much more precise brightness, distance, and reddening measurements would be necessary to produce truly valid, input--agnostic age determinations using the method presented here.\footnote{This statement is likely true for any similar statistical method as well.} A further consequence of insufficient observational precision is unreliable age uncertainties. The Monte Carlo calculations re-determine a star's age across variations in its observed parameters, but if the observational uncertainties are so forgiving that good fits are found across all isochrones regardless of changes to the central values of color and magnitude (the values substituted for color$_o$ and mag$_o$ in Equation \ref{eq:chisq_BDBS}), the age determination will not change much from trial to trial. This produces an output distribution with low variance because the limiting factor on age variation is the maximum and minimum of the isochrone grid rather than the observational error bars. As such, age determinations made using this photometry should be approached with caution. 

The distribution clearly bears no resemblance to the near-uniform distribution of the B17 ages for the same sub-sample (reduced from 91 to the 51 stars for which we have BDBS measurements), shown overlaid in pink, but neither does it resemble the MIST age determinations for the micro-lensed coordinates shown in Figure \ref{fig:age_histogram_alpha}. However, this distribution is typical of other photometrically derived age distributions for this region in the literature (e.g.\ \citealt{Gennaro15}).

When the age determination algorithm is applied to the BDBS gold sample only (lower panel of Figure \ref{fig:age_histogram_BDBS}), we find an even sharper distribution, with lower variance, and starker disagreement with the B17 age determinations for the equivalent sub-sample. Though the problems associated with small number statistics should not be discounted when the sample is reduced to 18 candidates, it is worth noting that the mean of the distribution does increase when the only most reliable BDBS targets are fit. The lower panel of Figure \ref{fig:age_histogram_BDBS} also makes clear that the stars being classified as young or intermediate-age by MIST and BDBS photometry are not the same as those being classified as young by \citet{Bensby17}.

However, we wish to make clear that a direct comparison between the spectroscopic and photometric age distributions is not strictly valid  given the caveats regarding the photometric uncertainties discussed above. It is not surprising, then, that there is tension between age distributions derived from photometry and those derived from micro-lensing and spectroscopy; the impact of uncertainties cannot be ignored. 

\begin{figure}
\begin{center}
\includegraphics[width=0.5\columnwidth]{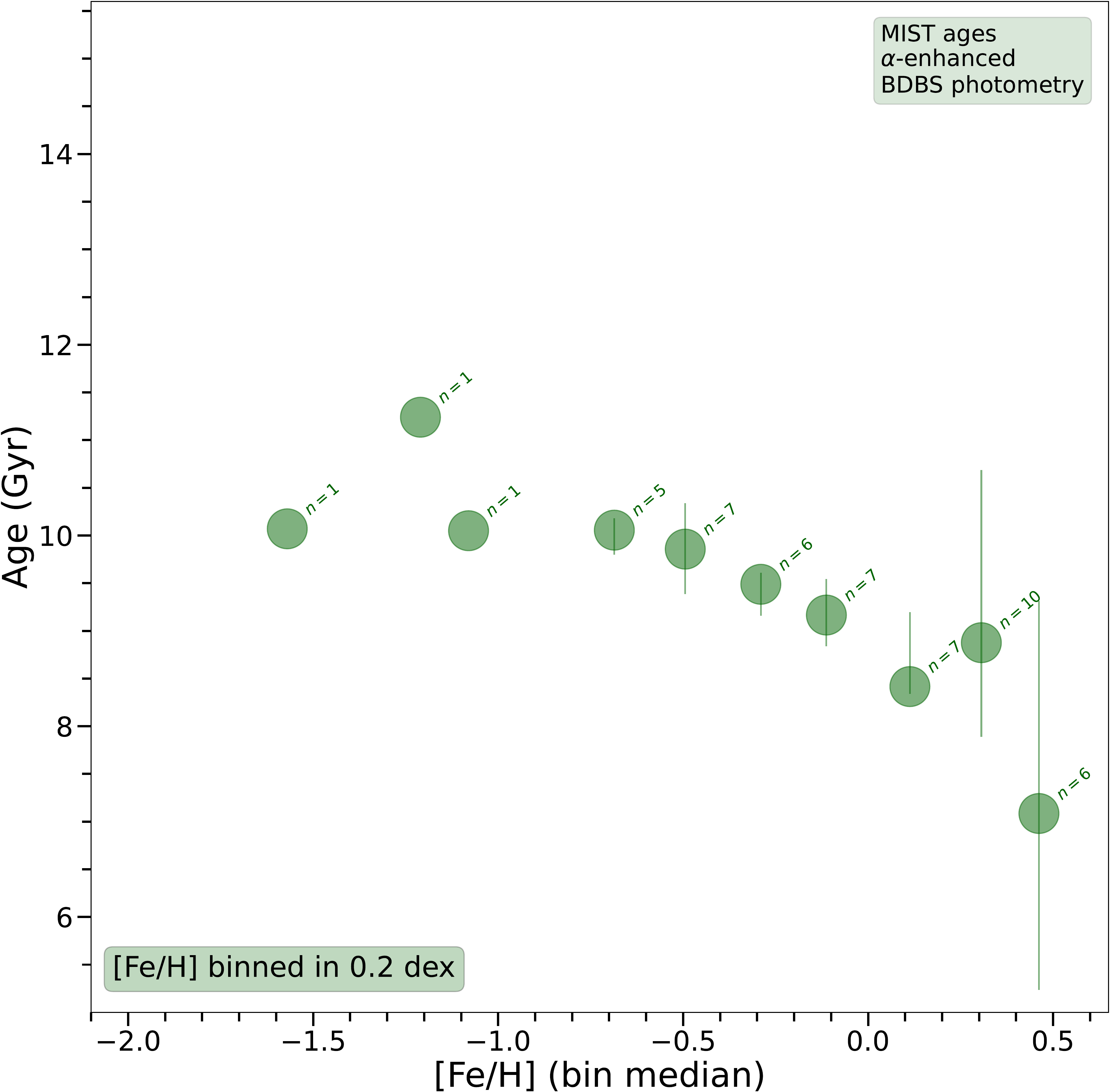}
\caption{
Same as Figure \ref{fig:age_dispersions}, but using the ages derived from BDBS photometry. Axes are on the same scale in both panels for direct comparison. In both cases, there is a trend towards younger ages at higher metallicities, though the age spread is significantly smaller in the photometric case. We emphasize once again that the age spread and age dispersions are smaller in this case not because the photometric ages are more reliable, but because the photometric uncertainties are not precise enough to discern between age hypotheses. }
\label{fig:age_dispersions_BDBS}
\end{center}
\end{figure}

A comparison between Figure \ref{fig:age_dispersions_BDBS} and Figure \ref{fig:age_dispersions} underscores this point. Binned average age is shown as a function of metallicity, with number of stars in each bin annotated, for the BDBS photometry. While both data sets show an increase in age dispersion with increasing metallicity, the photometric data are consistent with the claim that the stellar ages display no metallicity dependence. This is clearly not the same story told by the spectroscopic data. A na\"ive interpretation of this discrepancy is that there is some fundamental tension between the data collection methods; but, in reality, this is once again explained by photometric imprecision in absolute magnitudes, driven particularly by distance and small scale reddening variations along with the extreme crowding of low latitude bulge fields. The age spread and age dispersions are smaller in the photometric case because the photometric uncertainties are of the same order as the entire ranges in color and magnitude covered by the isochrones. 

\subsection{Gaia photometry}
\label{sec:Gaia_photometry}

\begin{table*} 
\centering 
\caption{Sources with reliable {\it Gaia} photometry}
\begin{tabular}{c ll c  c c c c}  
\hline\hline
& Name & \textit{Gaia} source ID &	[Fe/H] &	[Fe/H] error	& B17 Age	& B17 lower age limit &	B17 upper age limit \\
\hline
1 & OGLE-2013-BLG-0692S & 4052447161630182400	& 0.15	& 0.19	& 4.10	& 3.30	& 9.80 \\
2 & MOA-2012-BLG-202S	& 4064990283016815744  & $-0.15$	& 0.13	& 14.70	& 8.50	& 99.90 \\
3 & MOA-2011-BLG-234S	& 4064581887363003520 & $-0.02$	& 0.08	& 6.40	& 2.90	& 8.60 \\
4 & OGLE-2011-BLG-0950S	& 4043354643038544256  & 0.33	& 0.10	& 2.40	& 1.60	& 3.60 \\
5 & OGLE-2014-BLG-0157S	& 4042474694255127168  & $-0.52$	& 0.10	& 14.00	& 7.40	& 99.90 \\
6 & OGLE-2013-BLG-1938S	& 4041757847027545984  & $-0.41$	& 0.09	& 14.60	& 6.50	& 99.90 \\
7 & OGLE-2008-BLG-209S	& 4050303698056121728  & $-0.31$	& 0.09	& 7.20	& 5.00	& 12.20 \\
8 & MOA-2010-BLG-285S	& 4056277031255698304  & $-1.21$	& 0.10	& 11.60	& 8.10	& 13.40 \\
\hline
\end{tabular}
\begin{flushleft}
\small{The \textit{Gaia} source IDs and fundamental parameters as reported by \citet{Bensby17} are shown for 8 cross-matched \textit{Gaia} targets determined to have reliable photometry as described in Section \ref{sec:Gaia_photometry}.}
\end{flushleft}
\label{table:gaia_targets}
\end{table*}

\begin{figure*}
\def\figwid{0.41\textwidth}
\begin{center}
\includegraphics[width=\figwid]{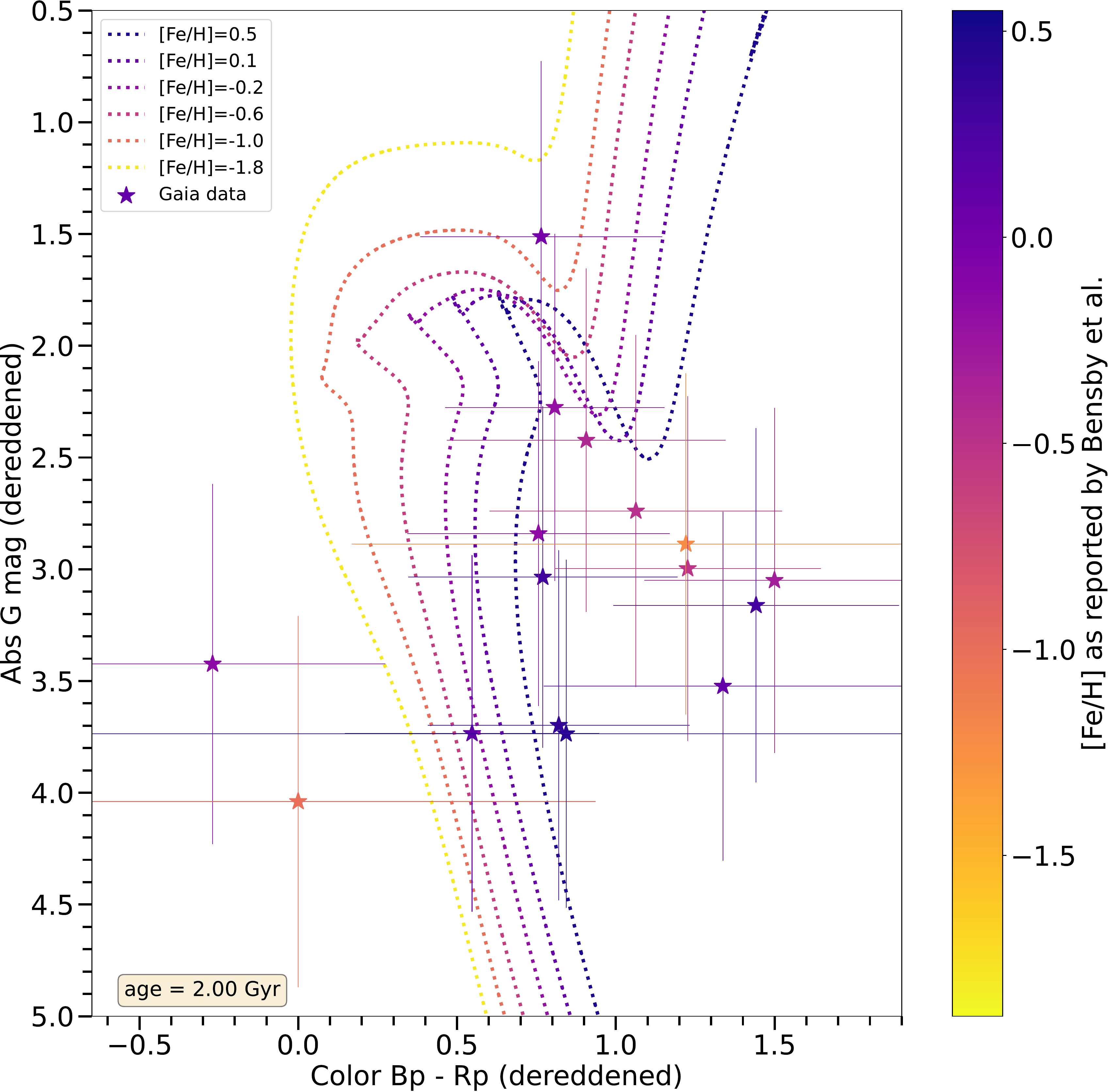}
\hfil
\includegraphics[width=\figwid]{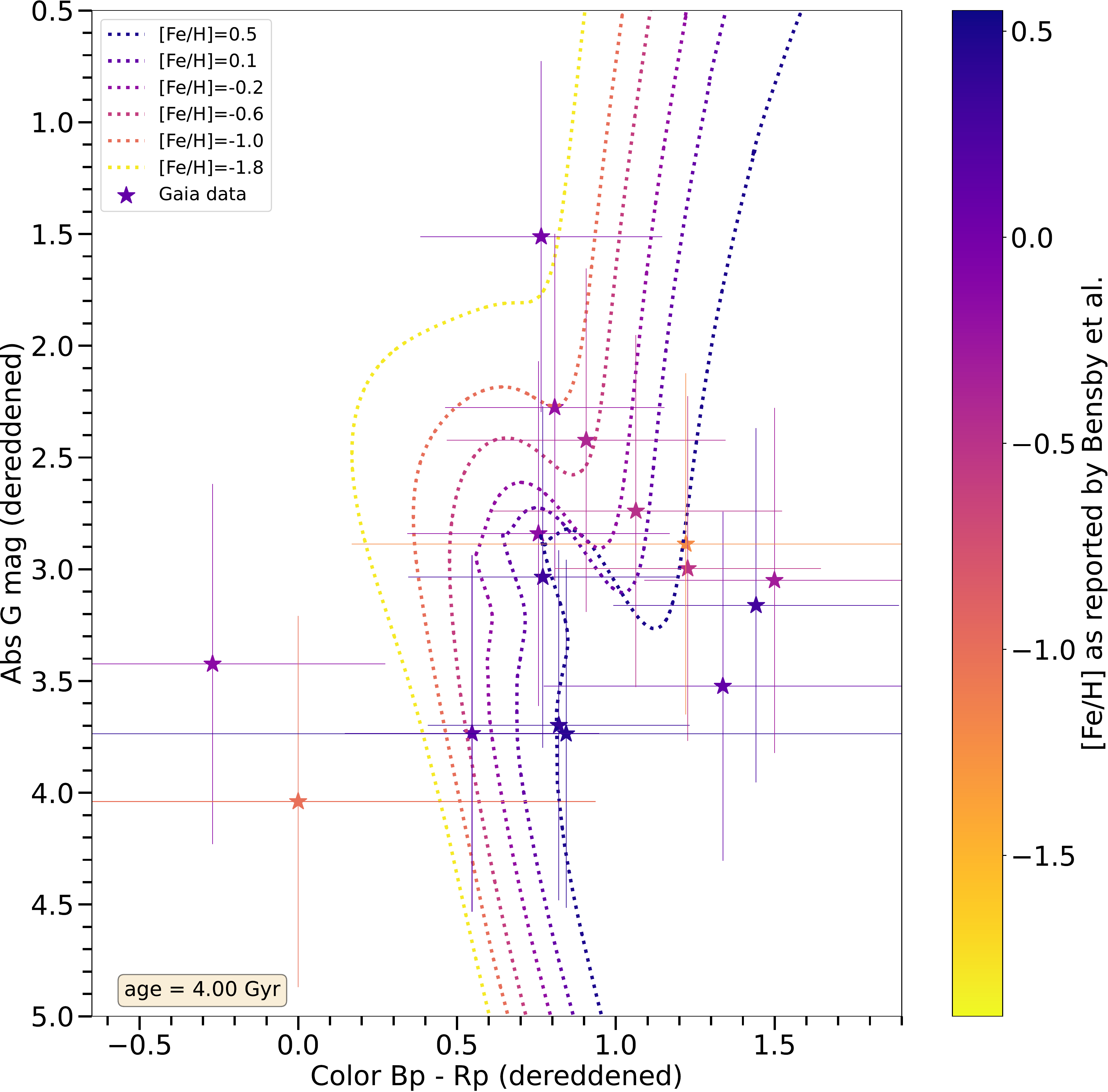}
\vskip8pt   
\includegraphics[width=\figwid]{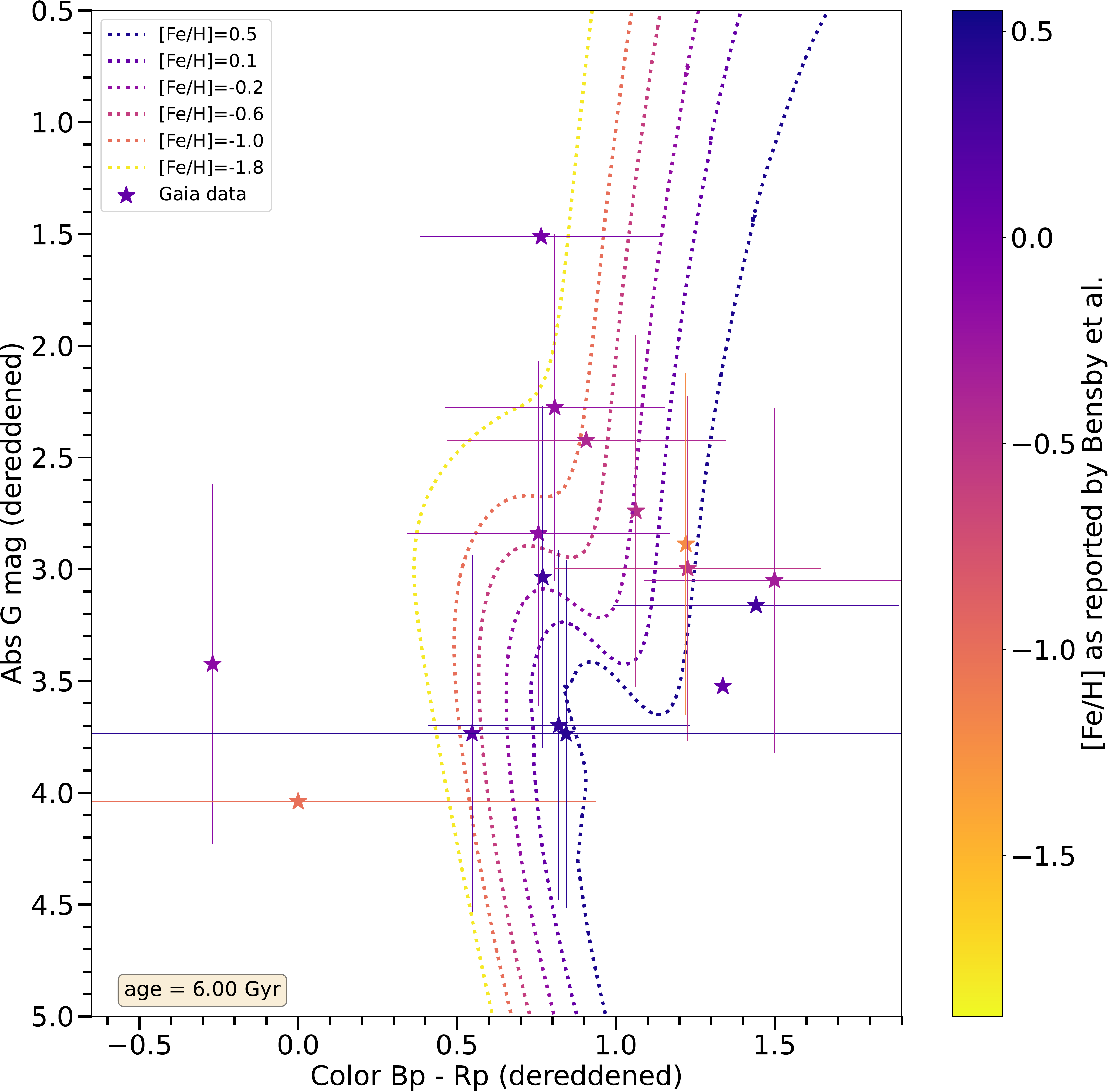}
\hfil
\includegraphics[width=\figwid]{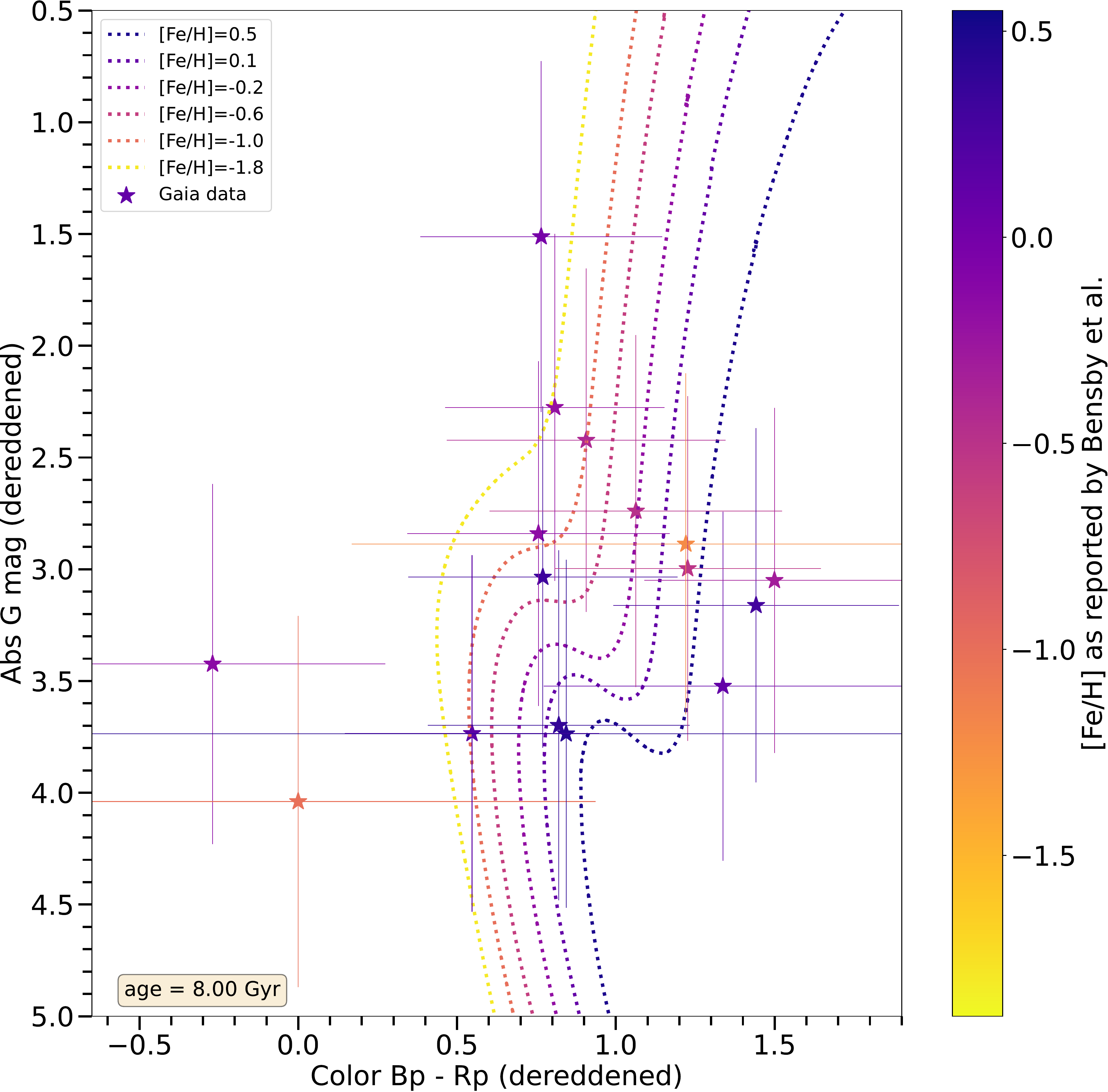}
\vskip8pt
\includegraphics[width=\figwid]{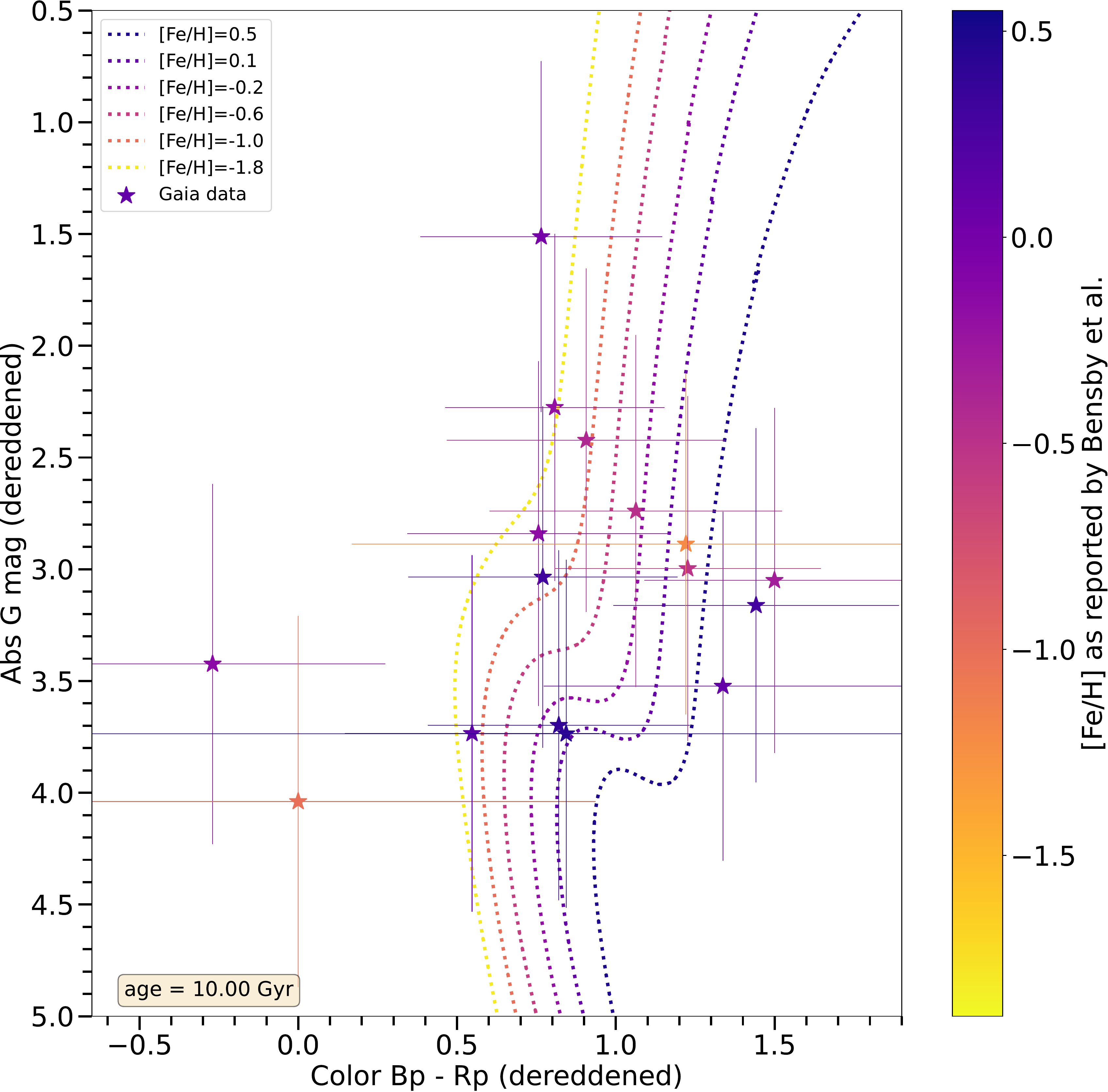}
\hfil
\includegraphics[width=\figwid]{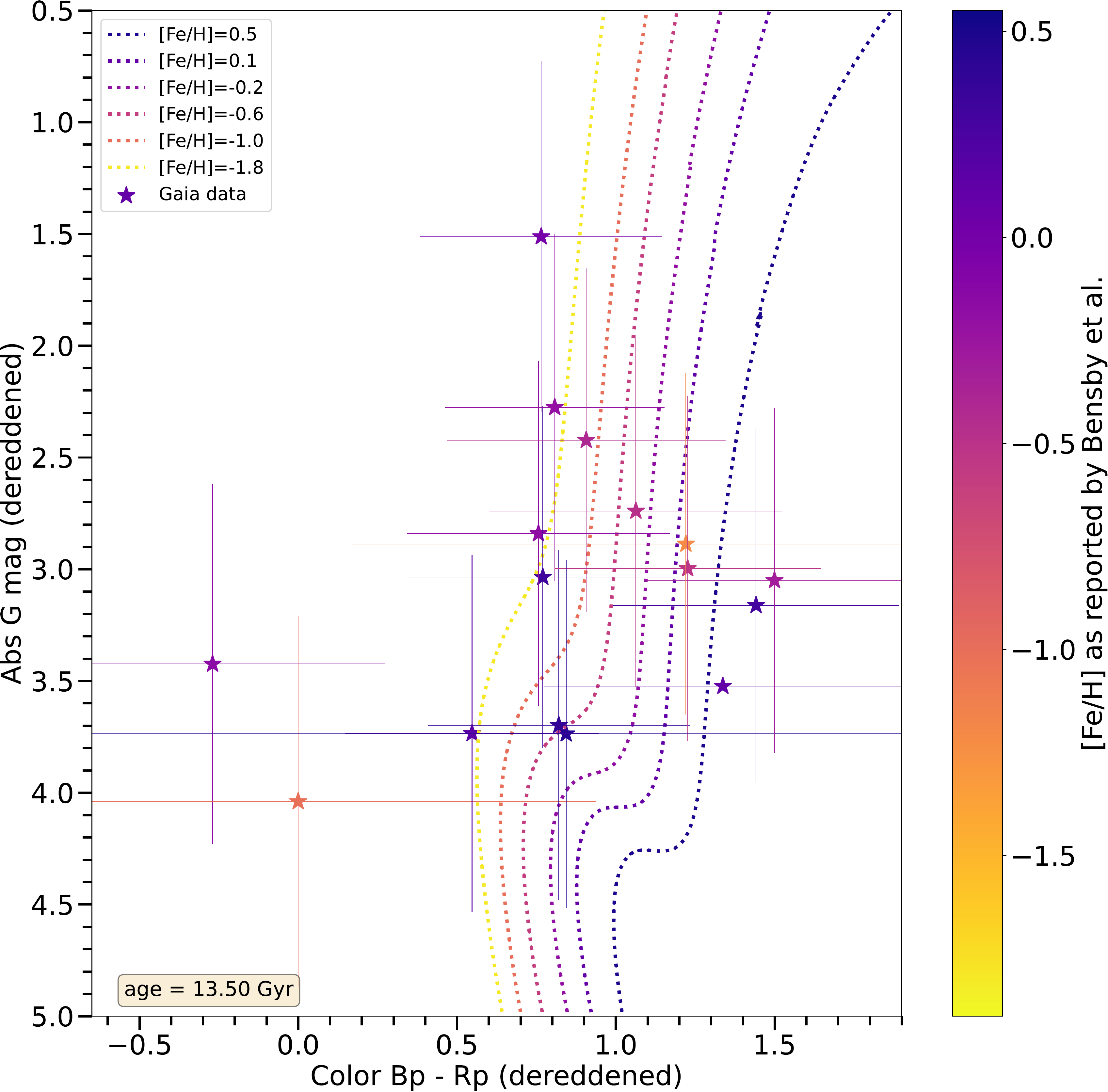}
\caption{Same as Figure \ref{fig:BDBS_cmd}, but for {\it Gaia} photometry.}
\label{fig:Gaia_cmd}
\end{center}
\end{figure*}

\begin{figure}
\begin{center}
\includegraphics[width=0.45\columnwidth]{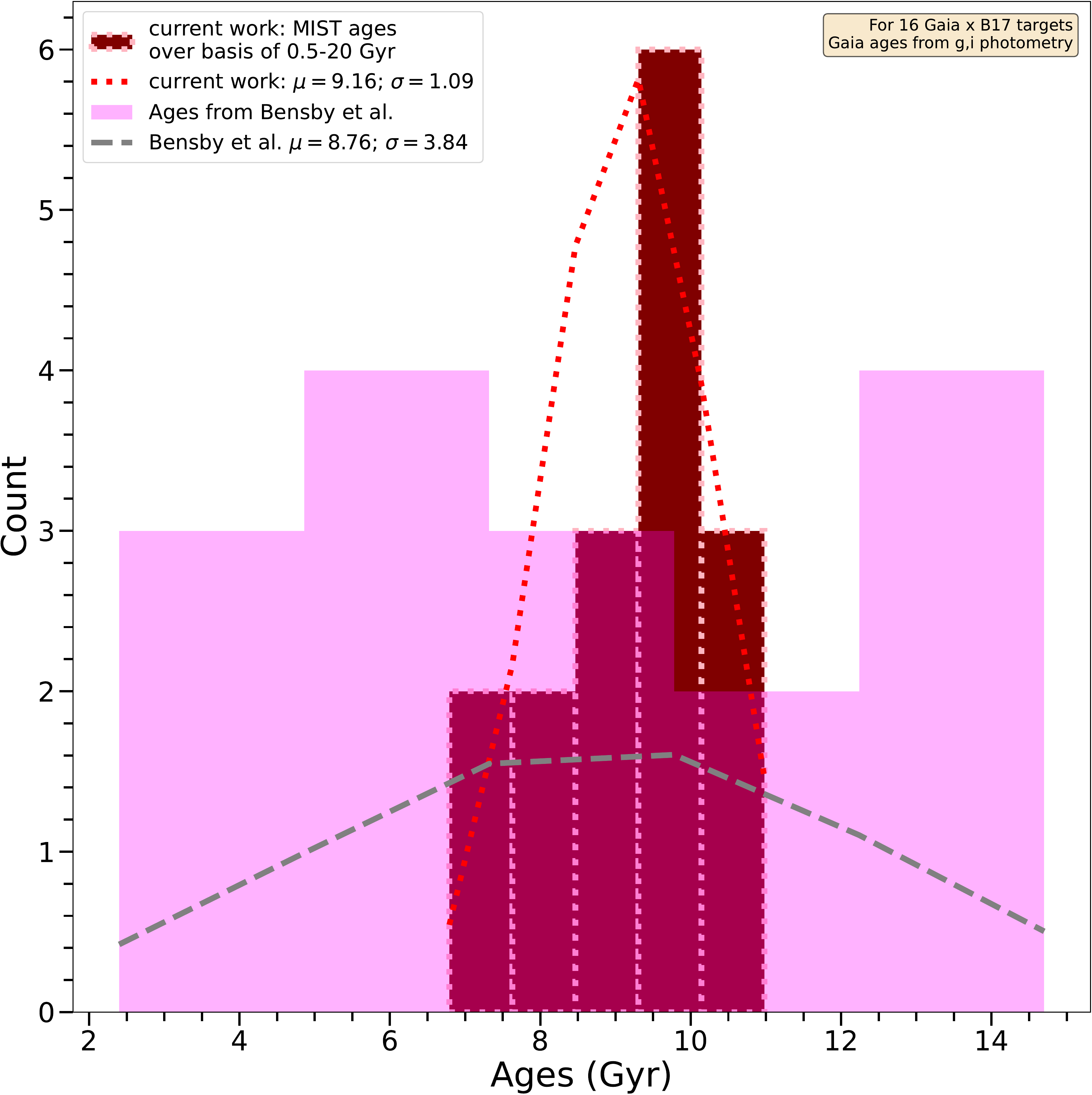}
\hfil
\includegraphics[width=0.45\columnwidth]{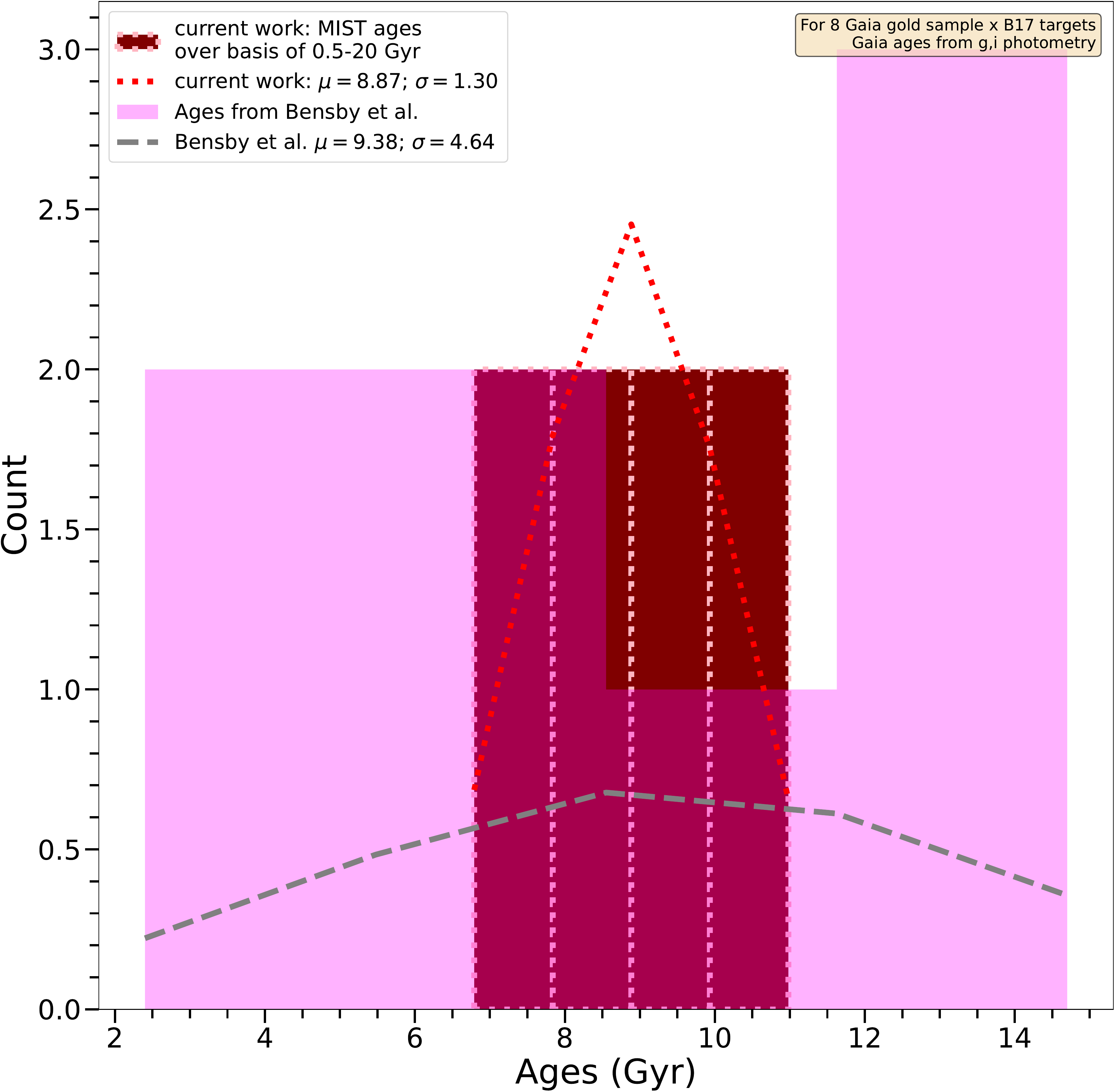}
\caption{ 
\textbf{LEFT:} Same as Figure \ref{fig:age_histogram_BDBS}, but for the intersection of the B17 and {\it Gaia} target lists examining only those stars selected according to the description in Section \ref{sec:Gaia_photometry}. This is the ``{\it Gaia} gold sample.''
\textbf{RIGHT:} Same as left panel, but for the entire intersection of the B17 and {\it Gaia} target lists. This totals 16 stars after the removal of {\it Gaia} stars with either (1) bad photometry or (2) for which two independent distance determinations with uncertainties were not available. 
}
\label{fig:age_histogram_Gaia}
\end{center}
\end{figure}

%
%
\begin{figure*}
\maxrows=6
\figgrid{alpha_histos/histogram_ages_alpha-basis_MCMC_OGLE-2015-BLG-0159S_iter100.pdf,
alpha_histos/histogram_ages_alpha-basis_MCMC_OGLE-2015-BLG-0078S_iter100.pdf,
alpha_histos/histogram_ages_alpha-basis_MCMC_OGLE-2014-BLG-2040S_iter100.pdf,
alpha_histos/histogram_ages_alpha-basis_MCMC_OGLE-2014-BLG-1469S_iter100.pdf,
alpha_histos/histogram_ages_alpha-basis_MCMC_OGLE-2014-BLG-1418S_iter100.pdf,
alpha_histos/histogram_ages_alpha-basis_MCMC_OGLE-2014-BLG-1370S_iter100.pdf,
alpha_histos/histogram_ages_alpha-basis_MCMC_OGLE-2014-BLG-1122S_iter100.pdf,
alpha_histos/histogram_ages_alpha-basis_MCMC_OGLE-2014-BLG-0987S_iter100.pdf,
alpha_histos/histogram_ages_alpha-basis_MCMC_OGLE-2014-BLG-0953S_iter100.pdf,
alpha_histos/histogram_ages_alpha-basis_MCMC_OGLE-2014-BLG-0801S_iter100.pdf,
alpha_histos/histogram_ages_alpha-basis_MCMC_OGLE-2014-BLG-0157S_iter100.pdf,
alpha_histos/histogram_ages_alpha-basis_MCMC_OGLE-2013-BLG-1938S_iter100.pdf,
alpha_histos/histogram_ages_alpha-basis_MCMC_OGLE-2013-BLG-1868S_iter100.pdf,
alpha_histos/histogram_ages_alpha-basis_MCMC_OGLE-2013-BLG-1793S_iter100.pdf,
alpha_histos/histogram_ages_alpha-basis_MCMC_OGLE-2013-BLG-1768S_iter100.pdf,
alpha_histos/histogram_ages_alpha-basis_MCMC_OGLE-2013-BLG-1259S_iter100.pdf,
alpha_histos/histogram_ages_alpha-basis_MCMC_OGLE-2013-BLG-1147S_iter100.pdf,
alpha_histos/histogram_ages_alpha-basis_MCMC_OGLE-2013-BLG-1125S_iter100.pdf,
alpha_histos/histogram_ages_alpha-basis_MCMC_OGLE-2013-BLG-1114S_iter100.pdf,
alpha_histos/histogram_ages_alpha-basis_MCMC_OGLE-2013-BLG-1015S_iter100.pdf,
alpha_histos/histogram_ages_alpha-basis_MCMC_OGLE-2013-BLG-0911S_iter100.pdf,
alpha_histos/histogram_ages_alpha-basis_MCMC_OGLE-2013-BLG-0835S_iter100.pdf,
alpha_histos/histogram_ages_alpha-basis_MCMC_OGLE-2013-BLG-0692S_iter100.pdf,
alpha_histos/histogram_ages_alpha-basis_MCMC_OGLE-2013-BLG-0446S_iter100.pdf,
alpha_histos/histogram_ages_alpha-basis_MCMC_OGLE-2012-BLG-1534S_iter100.pdf,
alpha_histos/histogram_ages_alpha-basis_MCMC_OGLE-2012-BLG-1526S_iter100.pdf,
alpha_histos/histogram_ages_alpha-basis_MCMC_OGLE-2012-BLG-1279S_iter100.pdf,
alpha_histos/histogram_ages_alpha-basis_MCMC_OGLE-2012-BLG-1274S_iter100.pdf,
alpha_histos/histogram_ages_alpha-basis_MCMC_OGLE-2012-BLG-1217S_iter100.pdf,
alpha_histos/histogram_ages_alpha-basis_MCMC_OGLE-2012-BLG-1156S_iter100.pdf,
alpha_histos/histogram_ages_alpha-basis_MCMC_OGLE-2012-BLG-0816S_iter100.pdf,
alpha_histos/histogram_ages_alpha-basis_MCMC_OGLE-2012-BLG-0617S_iter100.pdf,
alpha_histos/histogram_ages_alpha-basis_MCMC_OGLE-2012-BLG-0563S_iter100.pdf,
alpha_histos/histogram_ages_alpha-basis_MCMC_OGLE-2012-BLG-0521S_iter100.pdf,
alpha_histos/histogram_ages_alpha-basis_MCMC_OGLE-2012-BLG-0270S_iter100.pdf,
alpha_histos/histogram_ages_alpha-basis_MCMC_OGLE-2012-BLG-0211S_iter100.pdf,
alpha_histos/histogram_ages_alpha-basis_MCMC_OGLE-2012-BLG-0026S_iter100.pdf,
alpha_histos/histogram_ages_alpha-basis_MCMC_OGLE-2011-BLG-1410S_iter100.pdf,
alpha_histos/histogram_ages_alpha-basis_MCMC_OGLE-2011-BLG-1105S_iter100.pdf,
alpha_histos/histogram_ages_alpha-basis_MCMC_OGLE-2011-BLG-1072S_iter100.pdf,
alpha_histos/histogram_ages_alpha-basis_MCMC_OGLE-2011-BLG-0969S_iter100.pdf,
alpha_histos/histogram_ages_alpha-basis_MCMC_OGLE-2011-BLG-0950S_iter100.pdf,
alpha_histos/histogram_ages_alpha-basis_MCMC_OGLE-2009-BLG-076S_iter100.pdf,
alpha_histos/histogram_ages_alpha-basis_MCMC_OGLE-2008-BLG-209S_iter100.pdf,
alpha_histos/histogram_ages_alpha-basis_MCMC_OGLE-2007-BLG-514S_iter100.pdf,
alpha_histos/histogram_ages_alpha-basis_MCMC_OGLE-2007-BLG-349S_iter100.pdf,
alpha_histos/histogram_ages_alpha-basis_MCMC_OGLE-2006-BLG-265S_iter100.pdf,
alpha_histos/histogram_ages_alpha-basis_MCMC_MOA-2015-BLG-111S_iter100.pdf,
alpha_histos/histogram_ages_alpha-basis_MCMC_MOA-2014-BLG-131S_iter100.pdf,
alpha_histos/histogram_ages_alpha-basis_MCMC_MOA-2013-BLG-605S_iter100.pdf,
alpha_histos/histogram_ages_alpha-basis_MCMC_MOA-2013-BLG-524S_iter100.pdf,
alpha_histos/histogram_ages_alpha-basis_MCMC_MOA-2013-BLG-517S_iter100.pdf,
alpha_histos/histogram_ages_alpha-basis_MCMC_MOA-2013-BLG-402S_iter100.pdf,
alpha_histos/histogram_ages_alpha-basis_MCMC_MOA-2013-BLG-299S_iter100.pdf,
alpha_histos/histogram_ages_alpha-basis_MCMC_MOA-2013-BLG-068S_iter100.pdf,
alpha_histos/histogram_ages_alpha-basis_MCMC_MOA-2013-BLG-063S_iter100.pdf,
alpha_histos/histogram_ages_alpha-basis_MCMC_MOA-2012-BLG-532S_iter100.pdf,
alpha_histos/histogram_ages_alpha-basis_MCMC_MOA-2012-BLG-410S_iter100.pdf,
alpha_histos/histogram_ages_alpha-basis_MCMC_MOA-2012-BLG-391S_iter100.pdf,
alpha_histos/histogram_ages_alpha-basis_MCMC_MOA-2012-BLG-291S_iter100.pdf,
alpha_histos/histogram_ages_alpha-basis_MCMC_MOA-2012-BLG-202S_iter100.pdf,
alpha_histos/histogram_ages_alpha-basis_MCMC_MOA-2012-BLG-187S_iter100.pdf,
alpha_histos/histogram_ages_alpha-basis_MCMC_MOA-2012-BLG-022S_iter100.pdf,
alpha_histos/histogram_ages_alpha-basis_MCMC_MOA-2011-BLG-445S_iter100.pdf,
alpha_histos/histogram_ages_alpha-basis_MCMC_MOA-2011-BLG-278S_iter100.pdf,
alpha_histos/histogram_ages_alpha-basis_MCMC_MOA-2011-BLG-234S_iter100.pdf,
alpha_histos/histogram_ages_alpha-basis_MCMC_MOA-2011-BLG-191S_iter100.pdf,
alpha_histos/histogram_ages_alpha-basis_MCMC_MOA-2011-BLG-174S_iter100.pdf,
alpha_histos/histogram_ages_alpha-basis_MCMC_MOA-2011-BLG-104S_iter100.pdf,
alpha_histos/histogram_ages_alpha-basis_MCMC_MOA-2011-BLG-090S_iter100.pdf,
alpha_histos/histogram_ages_alpha-basis_MCMC_MOA-2011-BLG-058S_iter100.pdf,
alpha_histos/histogram_ages_alpha-basis_MCMC_MOA-2011-BLG-034S_iter100.pdf,
alpha_histos/histogram_ages_alpha-basis_MCMC_MOA-2010-BLG-523S_iter100.pdf,
alpha_histos/histogram_ages_alpha-basis_MCMC_MOA-2010-BLG-446S_iter100.pdf,
alpha_histos/histogram_ages_alpha-basis_MCMC_MOA-2010-BLG-311S_iter100.pdf,
alpha_histos/histogram_ages_alpha-basis_MCMC_MOA-2010-BLG-285S_iter100.pdf,
alpha_histos/histogram_ages_alpha-basis_MCMC_MOA-2010-BLG-167S_iter100.pdf,
alpha_histos/histogram_ages_alpha-basis_MCMC_MOA-2010-BLG-078S_iter100.pdf,
alpha_histos/histogram_ages_alpha-basis_MCMC_MOA-2010-BLG-049S_iter100.pdf,
alpha_histos/histogram_ages_alpha-basis_MCMC_MOA-2010-BLG-037S_iter100.pdf,
alpha_histos/histogram_ages_alpha-basis_MCMC_MOA-2009-BLG-493S_iter100.pdf,
alpha_histos/histogram_ages_alpha-basis_MCMC_MOA-2009-BLG-489S_iter100.pdf,
alpha_histos/histogram_ages_alpha-basis_MCMC_MOA-2009-BLG-475S_iter100.pdf,
alpha_histos/histogram_ages_alpha-basis_MCMC_MOA-2009-BLG-456S_iter100.pdf,
alpha_histos/histogram_ages_alpha-basis_MCMC_MOA-2009-BLG-259S_iter100.pdf,
alpha_histos/histogram_ages_alpha-basis_MCMC_MOA-2009-BLG-174S_iter100.pdf,
alpha_histos/histogram_ages_alpha-basis_MCMC_MOA-2009-BLG-133S_iter100.pdf,
alpha_histos/histogram_ages_alpha-basis_MCMC_MOA-2008-BLG-311S_iter100.pdf,
alpha_histos/histogram_ages_alpha-basis_MCMC_MOA-2008-BLG-310S_iter100.pdf,
alpha_histos/histogram_ages_alpha-basis_MCMC_MOA-2006-BLG-099S_iter100.pdf,
alpha_histos/histogram_ages_alpha-basis_MCMC_MACHO-1999-BLG-022S_iter100.pdf}
\caption{Error bar simulations for MIST-based re-fit to physical coordinates: Teff, logg.}
\end{figure*}
\begin{figure*}
\maxrows=6
\contfiggrid
\caption{Error bar simulations for MIST-based re-fit to physical coordinates: Teff, logg.}
\end{figure*}
\begin{figure*}
\maxrows=6
\contfiggrid
\caption{Error bar simulations for MIST-based re-fit to physical coordinates: Teff, logg.}
\end{figure*}
\begin{figure*}
\contfiggrid
\caption{Error bar simulations for MIST-based re-fit to physical coordinates: Teff, logg.}
\end{figure*}

\newpage

%
%
\begin{figure*}
\maxrows=6
\figgrid{BDBS_histos/histogram_ages_BDBS_MCMC_log-err-defn_OGLE-2015-BLG-0159S_iter100.pdf,
BDBS_histos/histogram_ages_BDBS_MCMC_log-err-defn_OGLE-2014-BLG-1418S_iter100.pdf,
BDBS_histos/histogram_ages_BDBS_MCMC_log-err-defn_OGLE-2014-BLG-1122S_iter100.pdf,
BDBS_histos/histogram_ages_BDBS_MCMC_log-err-defn_OGLE-2014-BLG-0987S_iter100.pdf,
BDBS_histos/histogram_ages_BDBS_MCMC_log-err-defn_OGLE-2014-BLG-0801S_iter100.pdf,
BDBS_histos/histogram_ages_BDBS_MCMC_log-err-defn_OGLE-2014-BLG-0157S_iter100.pdf,
BDBS_histos/histogram_ages_BDBS_MCMC_log-err-defn_OGLE-2013-BLG-1938S_iter100.pdf,
BDBS_histos/histogram_ages_BDBS_MCMC_log-err-defn_OGLE-2013-BLG-1868S_iter100.pdf,
BDBS_histos/histogram_ages_BDBS_MCMC_log-err-defn_OGLE-2013-BLG-1259S_iter100.pdf,
BDBS_histos/histogram_ages_BDBS_MCMC_log-err-defn_OGLE-2013-BLG-1147S_iter100.pdf,
BDBS_histos/histogram_ages_BDBS_MCMC_log-err-defn_OGLE-2013-BLG-0692S_iter100.pdf,
BDBS_histos/histogram_ages_BDBS_MCMC_log-err-defn_OGLE-2012-BLG-1526S_iter100.pdf,
BDBS_histos/histogram_ages_BDBS_MCMC_log-err-defn_OGLE-2012-BLG-1279S_iter100.pdf,
BDBS_histos/histogram_ages_BDBS_MCMC_log-err-defn_OGLE-2012-BLG-1274S_iter100.pdf,
BDBS_histos/histogram_ages_BDBS_MCMC_log-err-defn_OGLE-2012-BLG-1217S_iter100.pdf,
BDBS_histos/histogram_ages_BDBS_MCMC_log-err-defn_OGLE-2012-BLG-0816S_iter100.pdf,
BDBS_histos/histogram_ages_BDBS_MCMC_log-err-defn_OGLE-2012-BLG-0617S_iter100.pdf,
BDBS_histos/histogram_ages_BDBS_MCMC_log-err-defn_OGLE-2012-BLG-0563S_iter100.pdf,
BDBS_histos/histogram_ages_BDBS_MCMC_log-err-defn_OGLE-2012-BLG-0211S_iter100.pdf,
BDBS_histos/histogram_ages_BDBS_MCMC_log-err-defn_OGLE-2011-BLG-0969S_iter100.pdf,
BDBS_histos/histogram_ages_BDBS_MCMC_log-err-defn_OGLE-2011-BLG-0950S_iter100.pdf,
BDBS_histos/histogram_ages_BDBS_MCMC_log-err-defn_OGLE-2009-BLG-076S_iter100.pdf,
BDBS_histos/histogram_ages_BDBS_MCMC_log-err-defn_OGLE-2008-BLG-209S_iter100.pdf,
BDBS_histos/histogram_ages_BDBS_MCMC_log-err-defn_OGLE-2006-BLG-265S_iter100.pdf,
BDBS_histos/histogram_ages_BDBS_MCMC_log-err-defn_MOA-2014-BLG-131S_iter100.pdf,
BDBS_histos/histogram_ages_BDBS_MCMC_log-err-defn_MOA-2013-BLG-605S_iter100.pdf,
BDBS_histos/histogram_ages_BDBS_MCMC_log-err-defn_MOA-2013-BLG-517S_iter100.pdf,
BDBS_histos/histogram_ages_BDBS_MCMC_log-err-defn_MOA-2013-BLG-402S_iter100.pdf,
BDBS_histos/histogram_ages_BDBS_MCMC_log-err-defn_MOA-2013-BLG-299S_iter100.pdf,
BDBS_histos/histogram_ages_BDBS_MCMC_log-err-defn_MOA-2012-BLG-532S_iter100.pdf,
BDBS_histos/histogram_ages_BDBS_MCMC_log-err-defn_MOA-2012-BLG-291S_iter100.pdf,
BDBS_histos/histogram_ages_BDBS_MCMC_log-err-defn_MOA-2012-BLG-202S_iter100.pdf,
BDBS_histos/histogram_ages_BDBS_MCMC_log-err-defn_MOA-2012-BLG-187S_iter100.pdf,
BDBS_histos/histogram_ages_BDBS_MCMC_log-err-defn_MOA-2011-BLG-445S_iter100.pdf,
BDBS_histos/histogram_ages_BDBS_MCMC_log-err-defn_MOA-2011-BLG-234S_iter100.pdf,
BDBS_histos/histogram_ages_BDBS_MCMC_log-err-defn_MOA-2011-BLG-174S_iter100.pdf,
BDBS_histos/histogram_ages_BDBS_MCMC_log-err-defn_MOA-2011-BLG-090S_iter100.pdf,
BDBS_histos/histogram_ages_BDBS_MCMC_log-err-defn_MOA-2011-BLG-058S_iter100.pdf,
BDBS_histos/histogram_ages_BDBS_MCMC_log-err-defn_MOA-2011-BLG-034S_iter100.pdf,
BDBS_histos/histogram_ages_BDBS_MCMC_log-err-defn_MOA-2010-BLG-523S_iter100.pdf,
BDBS_histos/histogram_ages_BDBS_MCMC_log-err-defn_MOA-2010-BLG-311S_iter100.pdf,
BDBS_histos/histogram_ages_BDBS_MCMC_log-err-defn_MOA-2010-BLG-285S_iter100.pdf,
BDBS_histos/histogram_ages_BDBS_MCMC_log-err-defn_MOA-2010-BLG-167S_iter100.pdf,
BDBS_histos/histogram_ages_BDBS_MCMC_log-err-defn_MOA-2010-BLG-049S_iter100.pdf,
BDBS_histos/histogram_ages_BDBS_MCMC_log-err-defn_MOA-2009-BLG-456S_iter100.pdf,
BDBS_histos/histogram_ages_BDBS_MCMC_log-err-defn_MOA-2009-BLG-174S_iter100.pdf,
BDBS_histos/histogram_ages_BDBS_MCMC_log-err-defn_MOA-2009-BLG-133S_iter100.pdf,
BDBS_histos/histogram_ages_BDBS_MCMC_log-err-defn_MOA-2008-BLG-311S_iter100.pdf,
BDBS_histos/histogram_ages_BDBS_MCMC_log-err-defn_MOA-2008-BLG-310S_iter100.pdf,
BDBS_histos/histogram_ages_BDBS_MCMC_log-err-defn_MACHO-1999-BLG-022S_iter100.pdf,
BDBS_histos/histogram_ages_BDBS_MCMC_full_OGLE-2013-BLG-0446S_iter100.pdf}
\caption{Error bar simulation for BDBS targets.}
\end{figure*}
\begin{figure*}
\maxrows=6
\contfiggrid
\caption{Error bar simulation for BDBS targets.}
\end{figure*}

\newpage
%
%
\begin{figure*}
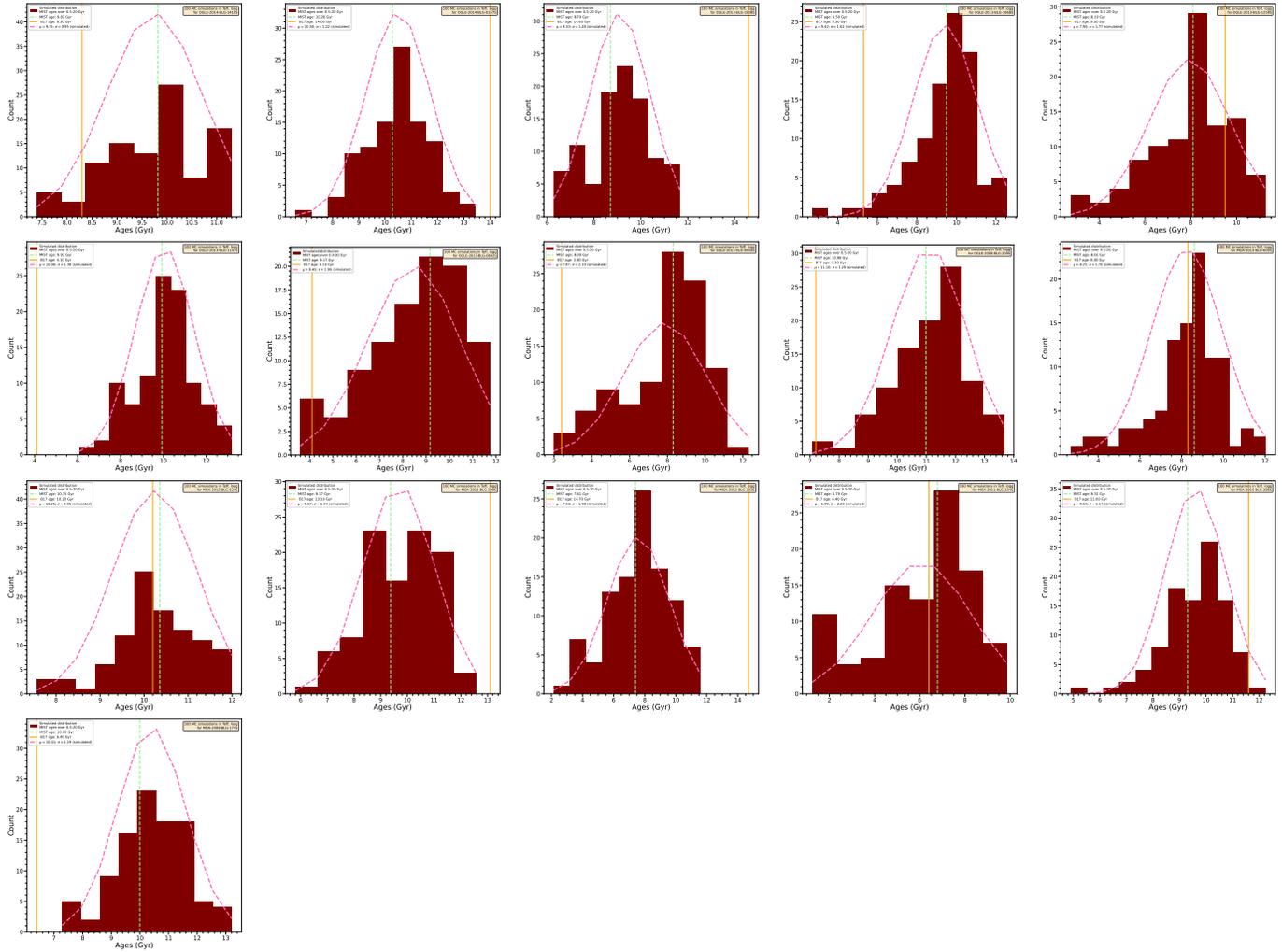

\maxrows=6
\figgrid[5]{Gaia_histos/histogram_ages_Gaia_MCMC_full_OGLE-2014-BLG-1418S_iter100.pdf,
Gaia_histos/histogram_ages_Gaia_MCMC_full_OGLE-2014-BLG-0157S_iter100.pdf,
Gaia_histos/histogram_ages_Gaia_MCMC_full_OGLE-2013-BLG-1938S_iter100.pdf,
Gaia_histos/histogram_ages_Gaia_MCMC_full_OGLE-2013-BLG-1868S_iter100.pdf,
Gaia_histos/histogram_ages_Gaia_MCMC_full_OGLE-2013-BLG-1259S_iter100.pdf,
Gaia_histos/histogram_ages_Gaia_MCMC_full_OGLE-2013-BLG-1147S_iter100.pdf,
Gaia_histos/histogram_ages_Gaia_MCMC_full_OGLE-2013-BLG-0692S_iter100.pdf,
Gaia_histos/histogram_ages_Gaia_MCMC_full_OGLE-2011-BLG-0950S_iter100.pdf,
Gaia_histos/histogram_ages_Gaia_MCMC_full_OGLE-2008-BLG-209S_iter100.pdf,
Gaia_histos/histogram_ages_Gaia_MCMC_full_MOA-2013-BLG-605S_iter100.pdf,
Gaia_histos/histogram_ages_Gaia_MCMC_full_MOA-2013-BLG-524S_iter100.pdf,
Gaia_histos/histogram_ages_Gaia_MCMC_full_MOA-2013-BLG-299S_iter100.pdf,
Gaia_histos/histogram_ages_Gaia_MCMC_full_MOA-2012-BLG-202S_iter100.pdf,
Gaia_histos/histogram_ages_Gaia_MCMC_full_MOA-2011-BLG-234S_iter100.pdf,
Gaia_histos/histogram_ages_Gaia_MCMC_full_MOA-2010-BLG-285S_iter100.pdf,
Gaia_histos/histogram_ages_Gaia_MCMC_full_MOA-2009-BLG-174S_iter100.pdf}
\caption{Error bar simulation for Gaia targets.}
\end{figure*}

Nonetheless, it is worthwhile to repeat this exercise with an orthogonal photometric data set. We crossmatch the B17 sample with {\it Gaia} EDR3, finding a total of $17$ stars with {\it Gaia} photometry and colors. We correct apparent $G$-band magnitudes using the formula provided by \citet{Riello21}. To select a sample of stars with reliable {\it Gaia} photometry, we also correct the photometric flux excess factor following the approach described in  \citep{Riello21}. The corrected excess factor $C^*$ is computed comparing the integrated flux in the $G_\mathrm{BP}$ and $G_\mathrm{RP}$ bands to the one in the $G$ band, and it is therefore a measure of the consistency of the photometry in the different bands. We then estimate the corresponding uncertainty $\sigma_{C^*}$ as a function of (corrected) $G$ magnitude using equation 18 in \citet{Riello21}. We find a total of $8$ stars that satisfy the condition $C^* \leq 5 \sigma_{C^*}$, and their identifiers are reported in Table \ref{table:gaia_targets}. {\it Gaia} EDR3 parallaxes and proper motions are available for a subset of $5$ stars. Even though {\it Gaia} EDR3 parallaxes for these faint stars are not precise enough to determine reliable individual distances, for one of the targets, OGLE-2013-BLG-0692S, the reported parallax $\varpi = (2.75\pm 0.84)$ mas might indicate that the star is a foreground star and therefore does not belong to the bulge. This is further confirmed by {\it Gaia} EDR3 proper motions. For each of the $5$ stars, we compute its probability to belong kinematically to the bulge based on its Galactic proper motions $(\mu_{\ell *}, \mu_b)$, which we determine by fitting the distribution in proper motion for all the stars within $0.5^\circ$ from the centre of the fields presented in Marchetti et al. (2022, \textit{subm.}). By this metric, we find OGLE-2013-BLG-0692S to be the only star with a probability marginally lower than $50\%$ of belonging to the bulge, $P_\mathrm{bulge} = 0.48$. We include it our gold sample and fits regardless. 

Distances are once again provided by \citet{Johnson2022}
and \citet{Simion17}, with extinction uncertainties in the BP, RP filters calculated as described in Section \ref{sec:BDBS_photometry}. A lack of reliable distance reduces the full cross-matched sample from 17 to 16 targets, while the gold sample includes 8 targets. The color and magnitude uncertainty definitions for the {\it Gaia} targets are given by 
\begin{equation}
    \sigma_{\text{color},o}^2
    = \sqrt{ \sigma_\mathrm{BP}^2 + \sigma_\mathrm{RP}^2
    + \sigma_{A_{\text{BP}}}^2 + \sigma_{A_\text{RP}}^2 }
\label{eq:sigma_color_Gaia}
\end{equation}
and
\begin{equation}
    \sigma_{\text{mag},o} = \sqrt{ \sigma_{G\text{-band}}^2 + \sigma_{A_G}^2 + \sigma_{\text{DM}}^2},
\label{eq:sigma_mag_Gaia}
\end{equation}
respectively. 
In this case, the analogs of the instrumental systematics terms (e.g.\ $\sigma_{i,\text{sys}}$) used in Equations \ref{eq:sigma_color_BDBS} and \ref{eq:sigma_mag_BDBS} are the \textit{Gaia} zero--point offsets per filter ($G, B_P, R_P$), which are already folded into the definitions of $\sigma_{G\text{-band}}$, $\sigma_\mathrm{BP}$ and $\sigma_\mathrm{RP}$. 

Figure \ref{fig:Gaia_cmd} shows the CMDs for all 16 available {\it Gaia} targets with uncertainties defined according to Equations \ref{eq:sigma_color_Gaia} and \ref{eq:sigma_mag_Gaia}. The MIST isochrones are transformed into the UBV(RI)c/2MASS/Kepler/Hipparcos/{\it Gaia} (DR2/EDR3)/Tess photometric system. As in the BDBS CMDs of Figure \ref{fig:BDBS_cmd}, we see that the $1\sigma$ observational uncertainties for most stars overlap with nearly the entire set of isochrones (horizontal error bars) and cover a similarly wide expanse of evolutionary time and theoretical brightness (vertical error bars). 

We perform the age determinations as in previous sections. The resulting distributions for the entire sample (left panel) and gold sample (right panel) are shown in Figure \ref{fig:age_histogram_Gaia}, with the {\it Gaia}/MIST ages shown in dark red and the histograms built from B17 overlaid. Qualitatively, we observe once again the narrowness of the MIST-based distributions compared to the equivalent sub-samples in B17, though we reiterate the point made in Appendix \ref{sec:BDBS_photometry} regarding the impropriety of claiming robust photometric ages for stars with such large uncertainties---and whose age determinations display such clear model-dependence---in the first place.
Further, the small number of targets in both cases makes meaningful statistical inferences difficult: there is no apparent relationship between the {\it Gaia}/MIST ages and equivalent B17 ages across the two samples.

\section{Comparison of photometric age determinations}
\label{sec:comparison_of_photometric_age_determinations}
While the age uncertainties quoted in Table \ref{table:mega_age_results} for the photometric fits are the true, numerical results of Monte Carlo simulations, they should not be interpreted as meaning anything beyond this. Neither BDBS nor \textit{Gaia} photometric data are precise enough that the age determination algorithm can distinguish between candidate ages. 

The two photometric ages are typically inconsistent with their spectroscopic counterparts but consistent with each other. Both of these features speak once again to the unreliability of these photometric age determinations and, in particular, their sensitivity to the input age grid. The fact that the photometric age ``uncertainties'' are on par with, or lower than, the spectroscopic age uncertainties for the same star should not be misinterpreted either: the photometric age variance over 100 MC simulations is low because the trial age determinations are always near the median of the input grid, which is always the same, regardless of parameter variations. 

\section{Results of MC Error Bar Simulations}
For each star fit according to each of the three sets of observational parameters, we provide the age histograms generated from the MC simulations of their uncertainties.

\bibliography{isochrones}
\bibliographystyle{aasjournal}

\end{document}